\documentclass[twocolumn]{aastex62}
\usepackage{amssymb,amsmath}
\usepackage{wasysym}

\usepackage{rotating}
\usepackage{xcolor}
\usepackage{sidecap}

\usepackage{apjfonts}

\newcommand{\muv}{$\mu_{1,V}$}
\newcommand{\mus}{$\mu_{1,\sigma}$}
\newcommand{\sigv}{$\mu_{2,V}$}
\newcommand{\sigs}{$\mu_{2,\sigma}$}
\newcommand{\hthreev}{$|\mu_{3,V}|$}
\newcommand{\hthrees}{$|\mu_{3,\sigma}|$}
\newcommand{\hfourv}{$\mu_{4,V}$}
\newcommand{\hfours}{$\mu_{4,\sigma}$}

\newcommand{\vasym}{$\mathrm{v}_{\mathrm{asym}}$}
\newcommand{\sigasym}{$\sigma_{\mathrm{asym}}$}
\newcommand{\sasym}{$\sigma_{\mathrm{asym}}$}
\newcommand{\lambdare}{$\lambda_{R_e}$}

\newcommand{\sunrise}{\texttt{SUNRISE}}

\usepackage{graphics}
\usepackage{relsize}
\usepackage{natbib}

\accepted{February 2, 2021}

\begin{document}

\title{Accurate Identification of Galaxy Mergers with Stellar Kinematics}

\author[0000-0002-2397-206X]{R. Nevin}
\affil{Center for Astrophysics | Harvard \& Smithsonian, 60 Garden St., Cambridge, MA 02138, USA }

\author[0000-0002-2183-1087]{L. Blecha}
\affil{ Department of Physics, University of Florida, Gainesville, FL 32611, USA}

\author[0000-0001-8627-4907]{J. Comerford}
\affil{Department of Astrophysical and Planetary Sciences, University of Colorado, Boulder, CO 80309, USA}

\author[0000-0002-5612-3427]{J. E. Greene}
\affil{Department of Astrophysical Sciences, Princeton University, Princeton, NJ 08544, USA}

\author[0000-0002-9402-186X]{D. R. Law}
\affil{Space Telescope Science Institute, 3700 San Martin Drive, Baltimore, MD 21218, USA}

\author{D. V. Stark}
\affil{Department of Physics and Astronomy, Haverford College, 370 Lancaster Ave, Haverford, PA 19041, USA}

\author{K. B. Westfall}
\affil{University of California Observatories, University of California,
Santa Cruz, 1156 High St., Santa Cruz, CA 95064, USA}

\author[0000-0001-8694-1204]{J. A. Vazquez-Mata}
\affil{Instituto de Astronom\'{i}a, Universidad Nacional Aut\'{o}noma de M\'{e}xico, A.P. 70-264, 04510, CDMX, M\'{e}xico}

\author[0000-0001-6417-7196]{R. Smethurst}
\affil{Oxford Astrophysics, Department of Physics, University of Oxford, Denys Wilkinson Building, Keble Road, Oxford, OX1 3RH, UK}

\author[0000-0002-0789-2326]{M. Argudo-Fern{\'a}ndez} 
\affil{Instituto de F{\'i}sica, Pontificia Universidad Cat{\'o}lica de Valpara{\'i}so, Casilla 4059, Valpara{\'i}so, Chile}

\author{J. R. Brownstein}
\affil{Department of Physics and Astronomy, University of Utah, 115 S. 1400 E., Salt Lake City, UT 84112, USA}

\author{N. Drory}
\affil{McDonald Observatory, The University of Texas at Austin, 1 University Station, Austin, TX 78712, USA}

\correspondingauthor{R. Nevin}
\email{rebecca.nevin@cfa.harvard.edu}

\begin{abstract}
To determine the importance of merging galaxies to galaxy evolution, it is necessary to design classification tools that can identify different types and stages of merging galaxies. Previously, using \texttt{GADGET-3/SUNRISE} simulations of merging galaxies and linear discriminant analysis (LDA), we created an accurate merging galaxy classifier from imaging predictors. Here, we develop a complementary tool based on stellar kinematic predictors derived from the same simulation suite. We design mock stellar velocity and velocity dispersion maps to mimic the specifications of the Mapping Nearby Galaxies at Apache Point (MaNGA) integral field spectroscopy (IFS) survey and utilize an LDA to create a classification based on a linear combination of 11 kinematic predictors.  The classification varies significantly with mass ratio; the major (minor) merger classifications have a mean statistical accuracy of 80\% (70\%), a precision of 90\% (85\%), and a recall of 75\% (60\%). The major mergers are best identified by predictors that trace global kinematic features, while the minor mergers rely on local features that trace a secondary stellar component. While the kinematic classification is less accurate than the imaging classification, the kinematic predictors are better at identifying post-coalescence mergers. A combined imaging + kinematic classification has the potential to reveal more complete merger samples from imaging and IFS surveys like MaNGA. We note that since the suite of simulations used to train the classifier covers a limited range of galaxy properties (i.e., the galaxies are intermediate mass and disk-dominated), the results may not be applicable to all MaNGA galaxies.

\end{abstract}


\section{Introduction}

Observations of galaxies reveal that they evolve over cosmic time from smaller, bluer, more irregular star-forming galaxies at higher redshifts to larger, redder, more elliptical galaxies in the local universe (e.g., \citealt{Glazebrook1995,Lilly1995,Giavalisco1996}). Additionally, the bimodality of galaxy properties such as color, mass, and star formation rate at low redshift implies that galaxies are quenching, or shutting down their star formation, in the local universe as well (e.g., \citealt{Schawinski2009,Masters2010,Weigel2017}). Galaxy evolution, or changes in the size, structures, and star formation properties of galaxies, in lower mass (log$(M_*/M_{\odot}) < 10.5$) galaxies is largely driven by the accretion of gas and/or the prevention of this gas from forming stars (\citealt{Robotham2014}). Many different processes can drive this evolution, from internal processes that are dependent on the galaxy properties, to external processes, which are related to the surroundings of the galaxy. Examples of internal processes include feedback from active galactic nuclei (AGNs; \citealt{Croton2006,fabian12,Heckman2014}), star formation driven outflows (\citealt{Rupke2018} and references therein), and morphological quenching due to structures such as bars or stellar bulges (\citealt{Sheth2005}). External processes include galaxy interactions with the hot intracluster medium that remove or heat gas (\citealt{Gunn1972}), `cold flow' accretion from the cosmic web (\citealt{Dekel2009}), and galaxy mergers (\citealt{Silk1998,DiMatteo2005,Kaviraj2013}). 

While the current $\Lambda$CDM framework for structural formation in the universe points to the importance of mergers for assembling dark matter halos (\citealt{White1978,White1991,Cole2008}),  the relative contribution of mergers to galaxy evolution through processes such as star formation, AGN activity, and/or morphological transformation remains unclear. This disagreement stems largely from the difficulty of building large, unambiguous samples of merging galaxies. Galaxy mergers are inherently difficult to identify; they persist for $\sim$few Gyr and they have a diversity of identifying characteristics that vary with merger stage, mass ratio, gas fraction, orbital parameters, and other merger initial conditions. 

The difficulty of identifying merging galaxies also contributes to the uncertainty in the merger rate ($R_{\mathrm{merg}}$), which is a key measurement for quantifying the role of mergers in galaxy evolution and comparing observations to simulations (e.g., \citealt{Lopez-Sanjuan2008}). The merger rate can either be measured directly from simulations or empirically, using the observed merger fraction and assuming a merger `observability' timescale (\citealt{Lotz2011}). Both techniques show large scatter between different estimates of the merger rate; semi-analytic models and hydrodynamical simulations have discrepancies of about an order of magnitude (see \citet{Hopkins2010} and references therein), while observations have also not converged, due to uncertainties in merger timescales and the completeness of the different methodologies. Recently, however, \citet{Mantha2018} have demonstrated that different empirical estimates of the merger rate can be brought into agreement. This work demonstrates the importance of careful calibration of the completeness of the merger identification methodology and the observability timescale.

Clean and complete samples of merging galaxies are therefore needed to address the contributions of mergers to evolutionary processes in galaxies and to reduce systematic uncertainties in the galaxy merger rate. This in turn necessitates a thorough understanding of the limitations and observability timescale of the technique used to identify merging galaxies. A variety of imaging techniques exist to identify merging galaxies, all of which are susceptible to their own biases. These often rely upon individual imaging tools, or predictors, such as the $Gini-M_{20}$ methodology, or the asymmetry of the galaxy light in imaging. One approach to overcome these biases is to utilize simulations of merging galaxies to better understand the shortcomings of individual tools and to characterize the observability timescales of these methods. For example, \citet{Lotz2008,Lotz2010a,Lotz2010b} use simulations of merging galaxies to measure the length of time that a major merger is observable by the $Gini-M_{20}$ and asymmetry metric. They find $0.3 - 0.5$ Gyr observability timescales,  meaning that merging galaxies are visible as mergers using these techniques for only a short time during the $\sim$few Gyr duration of the merger. Another strategy is to combine the predictors to create a single classification tool that dramatically lengthens the observability timescale by capitalizing on the strengths of the individual methods  (e.g., \citealt{Goulding2018,Snyder2019}). 

In \citet{Nevin2019} (henceforth N19), we pursue both of these approaches and utilize \texttt{GADGET-3/SUNRISE} simulated galaxies to build a merger identification technique. This technique combines seven imaging predictors to create one more accurate and precise classifier that incorporates the strength of all of these predictors, lengthening the observability timescale to $>$2 Gyr. Using this approach to simulate merging galaxies, we achieve high temporal resolution (relative to cosmological simulations), which enables us to construct a more complete picture of the different stages of a merger. The suite of simulated mergers also provides a known sample of merging and nonmerging galaxies from which we can understand the limitations of the identification technique before it is applied.

Recent years have witnessed an increase in the quantity and quality of integral field spectroscopy (IFS) data sets. With these advancements, kinematic predictors provide a promising addition to imaging predictors in the merger identification toolkit. Kinematic predictors are able to directly probe the dynamical histories of galaxies by tracing baryonic \textit{and} dark matter (\citealt{Glazebrook2013}). Disturbances in the stellar kinematics are dynamically long-lived and can identify a merger long after the imaging signatures have faded. For instance, morphological disturbances like tidal tails can fade on a $\sim$500 Myr timescale following final coalescence and are faint compared to the light of the galaxy (e.g., \citealt{Hung2014,Wen2016}) whereas kinematic disturbance in the stars of a galaxy can persist for longer (up to $\sim$Gyr after final coalescence; \citealt{Hung2016}).

Kinematic predictors may additionally clear up ambiguities in imaging. For instance, some clumpy star-forming galaxies appear to be mergers in imaging due to their disturbed morphologies (\citealt{Miralles-Caballero2011,Petty2014}), yet galaxies with clumpy morphologies can actually be nonmerging spiral galaxies with clumps of star formation in their centers or in their spiral arms (\citealt{Alonso-Herrero2006,Haan2011}). Kinematics have shown promise as an additional tool to determine if a star-forming galaxy is disk-like (e.g., \citealt{White2017}). 

This type of clumpy star-forming galaxy is even more abundant at intermediate and high redshifts, where a higher fraction of galaxies are expected to be actively merging, yet many isolated (non-merging) galaxies are also inherently clumpy (e.g., \citealt{Guo2015}). In addition to their clumpy and rapidly evolving morphologies, high redshift galaxies also have distinct kinematic features such as high velocity dispersions regardless of whether they are actively merging or isolated (e.g., \citealt{Law2012a,Law2012b}). The decreasing spatial resolution and surface brightness dimming of high redshift galaxies also confound the identification of mergers. Since high redshift galaxies present a host of additional complications, in this work we focus on local galaxies in order to develop the groundwork for a method that could eventually be extended to the more distant universe.

Like every other merger identification tool, kinematic predictors also have their own set of ambiguities and limitations. For instance, in gas-rich mergers, disks are able to survive the merger and these recently-merged galaxies can masquerade as isolated disk galaxies (e.g., \citealt{Robertson2006}). \citet{Hung2015} find that relying upon kinematics alone to classify a sample of ULIRGs identifies many merging galaxies as isolated disks and would provide a false-negative merger identification for up to 50\% of ULIRGs. Additionally, the identification technique depends strongly on the merger stage and the choice of kinematic predictor. Other work confirms that some mergers with highly disturbed visual morphology exhibit a distinct lack of disturbance in the stellar kinematics (\citealt{Bellocchi2013,Hung2016}). It is therefore important to probe the kinematics of merging galaxies using simulations in order to understand the biases and limitations of these tools before applying them to real galaxies.

There is currently a wealth of work dedicated to the imaging approach to identifying merging galaxies from large surveys. While there are many detailed case studies of the kinematics of individual local mergers (e.g., \citealt{Dasyra2006,Piqueras-Lopez2012}), there is a lack of detailed statistical-sized kinematic studies of local mergers. Recent years have brought a revolution in more and more capable IFS surveys, creating opportunities to identify merging galaxies using kinematic signatures. Surveys such as ATLAS-3D (\citealt{Cappellari2011}), CALIFA (\citealt{Sanchez2012}), SAMI (\citealt{Croom2012}), MaNGA (\citealt{Bundy2015}), and HECTOR (\citealt{Bryant2016}) offer a promising avenue to study the spatially-resolved spectral properties of an astounding number of galaxies. Here, we focus on the nearing-completion Mapping Nearby Galaxies at Apache Point Observatory (MaNGA) survey. MaNGA is an IFS survey of $>$10,000 local galaxies (with a median redshift $z\sim0.03$) with a spectral resolution of $R\sim2000$ and a spatial sampling of $1-2$ kpc (\citealt{Bundy2015}). One of MaNGA's secondary scientific goals is to help disentangle the evolutionary pathways of galaxies and to focus on incorporating simulations of merging galaxies with observations. It is thus uniquely well-suited for this project, where the goal is to create a merger classification technique from the kinematics of simulated galaxies, which we will then apply to the kinematics of the $>$10,000 galaxies in the MaNGA survey in order to identify mergers in future work.

This paper is organized as follows: In \S \ref{methods4} we review the \texttt{GADGET-3/SUNRISE} simulations from N19, describe the process of creating mock stellar kinematic maps from the synthetic spectra of the galaxy merger simulations, introduce the kinematic predictors, and review the linear discriminant analysis (LDA) technique used in N19 and in this work. In \S \ref{results4} we describe the results of the LDA classification, including the coefficients of the LDA, the observability timescales, and the accuracy and precision of the method. In \S \ref{discuss4} we describe the LDA coefficients in the context of previous work on mergers, how the classification changes with mass ratio, and examine the performance of the kinematic classification technique in the context of other tools and statistical methods. We present our conclusions in \S \ref{conclusions4}. In this work we focus on creating the kinematic classification from simulated galaxies. In future work we plan to apply the classification to galaxies in the MaNGA survey. A cosmology with $\Omega_m = 0.3$, $\Omega_\Lambda = 0.7$, and $h = 0.7$ is assumed throughout.

\section{Methods}
\label{methods4}

In order to construct a merger identification framework from the kinematics of simulated galaxies, we follow a detailed procedure to best mimic observations from the MaNGA survey. We introduce the galaxy merger simulations in \S \ref{simdeets} and describe the process for preparing mock kinematic maps from the simulated galaxies in \S \ref{mimic}. Finally, we introduce the kinematic predictors that we utilize in the kinematic classification in \S \ref{kinclass}. 

We dedicate several appendices to discussions that are informative but ancillary to the goals of this paper. We make the deliberate choice to extract the stellar kinematics from the \texttt{SUNRISE} spectra, as opposed to relying directly on particle velocities. We discuss the implications of this choice and compare the extracted stellar velocity and velocity dispersion maps to the inherent velocity of the simulation particles in \S \ref{scatter}. In \S \ref{scatter} we also discuss the effects of dust on the simulated observations. In \S \ref{noise} we include more details about adding noise to the mock spectra. In \S \ref{AGNprobs} we address AGN contamination and how we extract stellar kinematics from galaxies that host AGN.

\subsection{\texttt{GADGET-3/SUNRISE} Overview}
\label{simdeets}
As in N19, we utilize \texttt{GADGET-3/SUNRISE} simulations of merging galaxies. \texttt{GADGET-3} (\citealt{Springel2003,Springel2005}) is a smoothed particle hydrodynamics (SPH) and N-body code that models processes such as radiative heating, radiative cooling, star formation, supernova feedback, and the multi-phase interstellar medium (ISM) using sub-resolution prescriptions. \texttt{GADGET-3} also includes SMBH accretion as well as AGN feedback (this is achieved by coupling 5\% of the accreted luminosity to the gas as thermal energy). \texttt{GADGET} has been used for many different astrophysical applications, including wide use in studies of merging galaxies (e.g., \citealt{DiMatteo2005,cox06,Hopkins2006,Robertson2006,Hopkins2008,Blecha2011,Blecha2013b,Hopkins2013a,Hopkins2013b}).

We present the five galaxy merger simulations and the matched isolated simulations in Table \ref{simulations}. The framework for these simulations is established in \citet{Blecha2018}, and the simulations themselves are presented in N19.  Three of the simulations are major mergers (where the mass ratio, $q$, of the progenitors is greater than $q = 0.25$, or 1:4, \citealt{Rodriguez-Gomez2015,Nevin2019}) and two of the simulations are minor mergers. The major mergers have mass ratios of 1:2, 1:3, and 1:3. We define the gas fraction of these simulations as  $f_{\mathrm{gas}} = M_{\mathrm{gas,disk}}/(M_{\mathrm{gas,disk}} + M_{\mathrm{*,disk}})$. The 1:2 and 1:3 mass ratio major mergers have a relatively high gas fraction of 0.3 and one of the 1:3 mass ratio major mergers has a relatively low gas fraction of 0.1. 

We verify that the different gas fractions of the simulations (0.1 and 0.3) cover the full range in gas fractions of the MaNGA galaxies. The mean gas fraction in MaNGA is defined by \citet{Barrera-Ballesteros2018} as: $$\mu_{\mathrm{gas}} = \frac{\sigma_{\mathrm{gas}}}{\sigma_{\mathrm{gas}}+\sigma_{\mathrm{*}}}$$ where $\sigma_{\mathrm{gas}}$ is the gas mass density and $\sigma_*$ is the stellar mass density. The gas mass and stellar mass densities are derived from the Balmer decrement and stellar template fitting, respectively.

\citet{Barrera-Ballesteros2018} find that the mean gas fraction for the MaNGA sample ranges from 0.16 to 0.32. A $f_{\rm{gas}}$ value of 0.1 is therefore below the mean for MaNGA galaxies and thus relatively gas poor, while $f_{\rm{gas}} = 0.3$ is at the top of the range of mean values for the sample and relatively gas rich.

These simulations are named for their mass ratio and gas fraction; for instance, the gas rich 1:2 mass ratio major merger is q0.5\_fg0.3, the gas rich 1:3 mass ratio major merger is q0.333\_fg0.3, and the gas poor 1:3 mass ratio major merger is q0.333\_fg0.1. All of the major merger progenitors have a bulge-to-total (B/T) mass ratio of 0, meaning that they are a pure disk initially. Both of the progenitor galaxies of the minor mergers have a B/T ratio of 0.2 and both are gas rich. These simulations are q0.2\_fg0.3\_BT0.2, which is the 1:5 mass ratio minor merger, and q0.1\_fg0.3\_BT0.2, the 1:10 mass ratio minor merger. 

We build the matched sample of isolated galaxy simulations for each merger simulation from two sources. First, we use a stand-alone sample of isolated galaxies that are matched for mass and gas fraction to each of the simulations. Some simulations have more than one matched isolated galaxy, but for the case where there is only one isolated galaxy, it is matched to the mass of the larger merging galaxy from the corresponding merger simulation. Second, we define snapshots of each simulated merger that fall before first pericentric passage or $>0.5$ Gyr after final coalescence as isolated galaxies. We refer to the isolated galaxies that are from the snapshots before first pericentric passage as `pre-merger' isolated galaxies and the snapshots that happen $>0.5$ Gyr after final coalescence as `post-merger' isolated galaxies. This distinction is useful because the properties of these two populations differ.

\begin{table*}
    \centering
    \begin{tabular}{c|cccc}
         Simulation& Mass Ratio& Gas Fraction &Stellar Mass of Primary &  Matched Isolated Galaxies  \\
         
        & &  &[10$^{10}$ M$_{\odot}$] &  \\
         
         \hline
         q0.5\_fg0.3&1:2 & 0.3 & 3.9 & m0.5\_fg0.3, m1\_fg0.3\\
         q0.333\_fg0.3 & 1:3 & 0.3 & 3.9& m1\_fg0.3 \\
         q0.333\_fg0.1 & 1:3 & 0.1 & 4.7& m0.333\_fg0.1, m1\_fg0.1\\
         q0.2\_fg0.3\_BT0.2 & 1:5 & 0.3 & 4.2  &m1\_fg0.3\_BT0.2 \\
         q0.1\_fg0.3\_BT0.2 & 1:10 & 0.3 & 4.2 & m1\_fg0.3\_BT0.2 \\
    \end{tabular}
    \caption{Key simulation parameters and matched isolated galaxies. The simulations are named for the mass ratio, gas fraction and bulge-to-total mass ratio of the merging galaxies. For instance, q0.5\_fg0.3 is a 1:2 mass ratio merger where each progenitor galaxy has a gas fraction of 0.3 and an initial B/T ratio of 0. The stellar mass of the primary (more massive) galaxy is $3.9\times10^{10}$ M$_{\odot}$. The matched isolated galaxies are mass-matched to the merging galaxies and are named for which merging galaxy they are matched to (i.e., m0.5\_fg0.3 is matched to the smaller of the two merging galaxies in the q0.5\_fg0.3 merger). }
    \label{simulations}
\end{table*}

We couple \texttt{GADGET-3} with \texttt{SUNRISE} in order to directly compare the simulated galaxies with
observations. \texttt{SUNRISE} is a 3D polychromatic Monte-Carlo
dust radiative transfer (RT) code (\citealt{Jonsson2006,Jonsson2010}) that is used to model a wide range of isolated and merging galaxies (e.g., \citealt{Narayanan2010,hayward11,Blecha2013a,Snyder2013,Hayward2014}). The full details of the \texttt{SUNRISE} prescription are presented in \citet{Blecha2013b,Blecha2018} and N19. Briefly, \texttt{SUNRISE} performs Monte Carlo radiative transfer on a 3D adaptively-refined grid to compute the emission from stars, HII regions, and AGN. \texttt{SUNRISE} uses the STARBURST99 stellar population synthesis models (\citealt{Leitherer1999}) to calculate the age- and metallicity-dependent spectral energy distributions for each star particle.  The treatment for dust includes dust self-absorption and thermal re-emission as well as polycyclic aromatic hydrocarbon (PAH) absorption and emission. We additionally include kinematic (doppler) effects, which requires very high spectral resolution. Ultimately, \texttt{SUNRISE} calculates the emergent, attenuated resolved UV-to-IR spectra ($3300-6990$\AA, $\Delta \lambda = 0.3$\AA) for seven isotropically positioned viewing angles.

We utilize the datacube of \texttt{SUNRISE} optical synthetic spectra from the seven isotropically positioned viewpoints from each merger snapshot to produce the mock datacubes. In N19, a `snapshot' is the \texttt{SUNRISE} image; in this work, we use the term `snapshot' to refer to the full datacube from a specific point in time. These snapshots occur at 50-100 Myr intervals during each merger, and we refer to them as early-stage, late-stage, and post-coalescence stage snapshots. We define these stages using the $r-$band images from N19. The early-stage mergers occur after first pericentric passage and have (view-point) average stellar bulge separations $\Delta$x $\ge$ 10 kpc, late-stage mergers have separations 1 kpc $<$ $\Delta$x $<$ 10 kpc, and post-coalescence mergers are no longer resolvable with separations $\Delta$x $\leq$ 1 kpc until 0.5 Gyr after final coalescence. With a 50-100 Myr cadence for snapshots, we have 5-10 snapshots for each of these stages. In total, there are $\sim$20 snapshots per simulation and seven viewpoints per snapshot, which amounts to 100-200 observations per merger simulation.

We further discuss the importance of running RT and incorporating dust attenuation and scattering on the merger snapshots in Appendix \ref{scatter}; briefly, the stellar kinematic maps are affected by both the presence of dust and dust scattering. The implication is that for this type of kinematic analysis, it is important to use velocities derived directly from the RT product (synthetic spectra) as opposed to the original SPH particle velocities.

\subsection{Preparing Mock MaNGA Kinematic Maps}
\label{mimic}

\begin{figure}
    
    \includegraphics[width=0.47\textwidth]{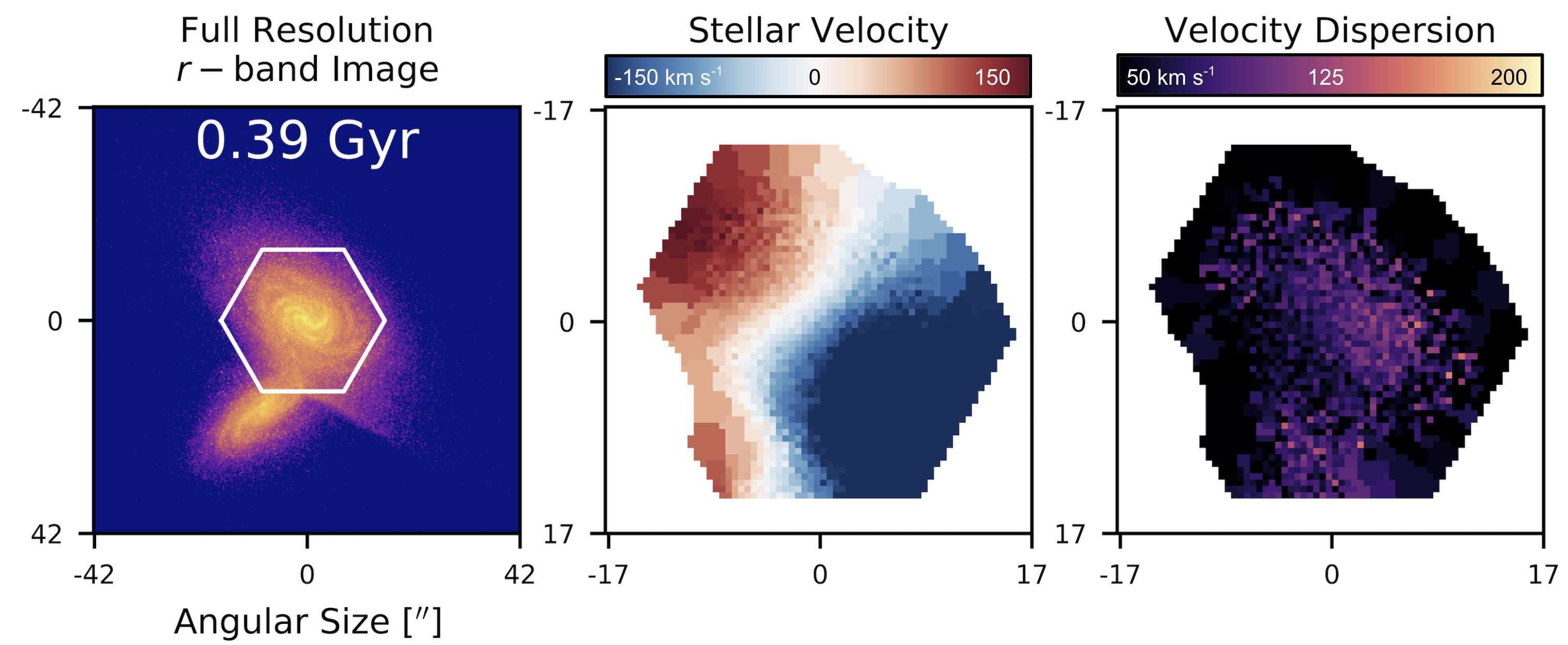}
    
   \includegraphics[scale=0.66, trim=1.1cm 2.7cm 1.2cm 3.0cm, clip]{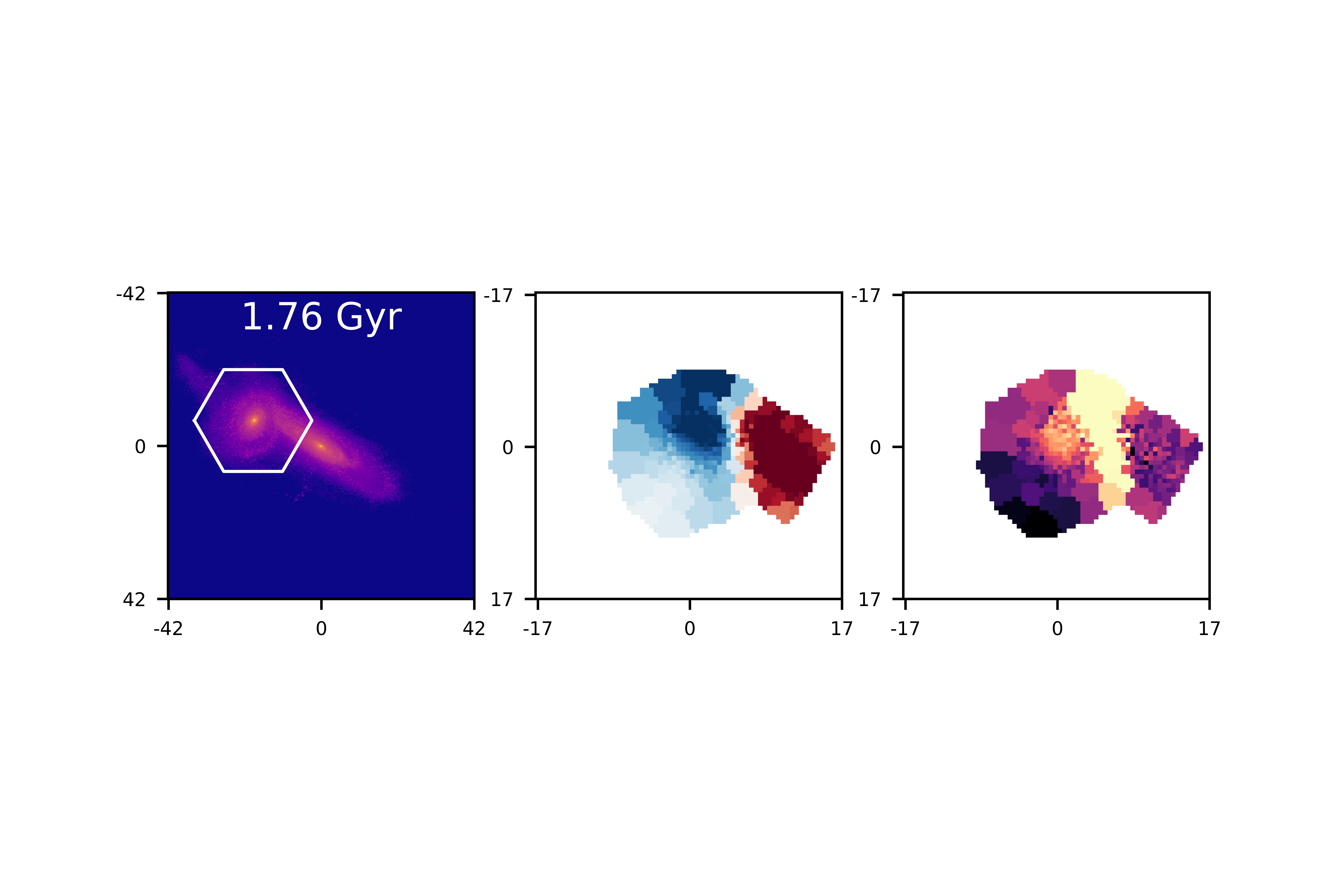}

    \includegraphics[scale=0.66, trim=1.1cm 2.7cm 1.2cm 3.0cm, clip]{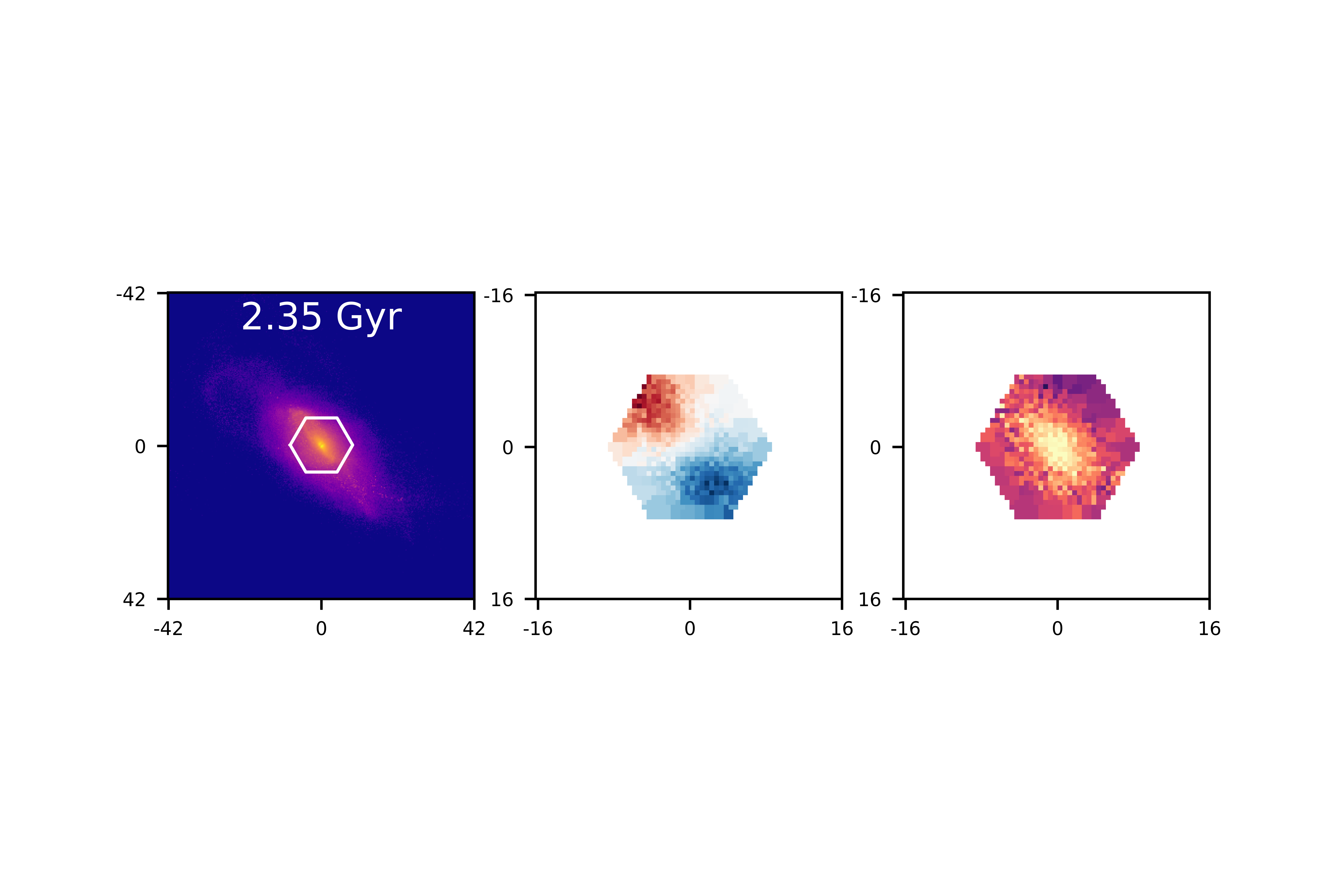}
    
    \caption{Snapshots of images (left column), stellar velocity maps (middle column), and stellar velocity dispersion maps (right column) from different epochs of the q0.5\_fg0.3 simulation. The $r-$band image is the log-scaled full-resolution simulation prior to the mock-up process in order to show all of the features of the merger. The colorbar for the middle and right columns is in km s$^{-1}$. The spatial position for all panels is in arcsec and the stellar velocity and stellar velocity dispersion columns have the same spatial coverage. We include a snapshot that is an early-stage merger (first row), a late-stage merger (second row), and a post-coalescence merger (third row). The stellar kinematics change over the course of the merger. For instance, the stellar velocity map is distorted due to the superposition of two merging galaxies, while the velocity dispersion map undergoes a global enhancement with time.}
    \label{mosaic_fg3_m12}
\end{figure}

To produce stellar kinematics for our sample of simulated galaxies, we use the specifications of MaNGA to create a datacube of spectra and then we mimic the MaNGA Data Analysis Pipeline (DAP) to extract stellar kinematics to use in our kinematic classification. Examples of finalized `MaNGA-ized' stellar velocity and stellar velocity dispersion maps are presented in Figure \ref{mosaic_fg3_m12}. In this section, we describe how we mimic the specifications of MaNGA. This involves reducing the spatial and spectral resolution of the simulations to create a MaNGA-ized datacube, placing an appropriately sized fiber bundle over each galaxy, and fitting each spaxel with \texttt{ppxf} (a penalized pixel-fitting method from \citealt{Cappellari2004,Cappellari2017}) to obtain the velocity and velocity dispersion of the stars at each spatial position.

SDSS-IV/MaNGA is an IFS survey that targets a sample of $>$10,000 nearby galaxies that are selected to span a wide range of environments and stellar masses (\citealt{Gunn2006,Smee2013,Bundy2015,Drory2015,Law2015,Blanton2017}). Spectra are obtained using the BOSS spectrograph on the 2.5m telescope at Apache Point Observatory (\citealt{Gunn2006}), which has a spectral resolution of $R\sim2000$. Fibers are bundled into integral field units (IFUs); MaNGA has five different fiber bundles, equipped with 19, 37, 61, 91, and 127 fibers (the largest fiber bundle is known as the `frankenbundle'); each individual fiber has a $2\farcs0$ diameter with $2\farcs5$ spacing between fibers (\citealt{Smee2013,Drory2015,Yan2016,Wake2017}). These fiber bundles range from $12\farcs5$ to $32\farcs5$ in diameter. 

MaNGA has a median point spread function (PSF) of $2\farcs5$, which roughly corresponds to a spatial resolution of 1-2 kpc. The primary sample of galaxies (which is 2/3 of the full sample) has coverage out to 1.5 times the $r-$band effective radius ($R_e$) and the secondary sample has coverage to 2.5 $R_e$ (\citealt{Yan2016,Wake2017}). The redshift range of the MaNGA survey is $0.01 \leq z \leq 0.15$.

The most recent internal MaNGA product launch (MPL-9) includes 8000 unique galaxies, observed and reduced by the Data Reduction Pipeline (DRP; \citealt{Law2016}). The publicly available version is released as DR-15, which includes 4621 unique galaxies (\citealt{Aguado2019}). The derived properties (including the stellar kinematics) are produced by the Data Analysis Pipeline (DAP; \citealt{Westfall2019,Belfiore2019}) in the format of a single datacube per galaxy (\citealt{Yan2016calib}). The MaNGA team also creates and maintains Marvin, which is a useful tool for visualizing MaNGA data (\citealt{Marvin}).

To create the mock datacubes, we begin with the \texttt{SUNRISE} synthetic spectra, which we extract at the median redshift of the MaNGA survey ($z = 0.03$). We also use the \texttt{SUNRISE} SDSS $r-$ and $g-$band images to construct the mock datacubes, since they are essential for certain steps of the process. We follow this procedure (which mirrors the MaNGA DAP whenever possible):

\begin{enumerate}
    \item Convolve the datacube with the $2\farcs5$ MaNGA PSF.
    
    Here we use a $2\farcs5$ Gaussian kernel, which is a good approximation of the effective PSF of the MaNGA datacubes. A model of the effective PSF is 
automatically computed for each datacube as part of the MaNGA DRP, which convolves a simulated point source with the fiber footprint of a given set of observations, incorporating as-observed details of the seeing, transparency, differential atmospheric refraction, dithering, and other instrumental
 effects  (\citealt{Law2016}). We briefly investigate the difference between using our simplified Gaussian kernel and the effective PSF model provided by the DRP, and find that there are small differences in the maps when the reconstructed PSF is used, in particular a slight increase in the spread of values in the velocity dispersion maps. This is to be expected, given that the reconstructed PSF does not have a perfectly Gaussian shape. However, the differences in the kinematic maps are minimal and do not cause significant differences in the classification. We define what it means for the classification to be `significantly different' in \S \ref{viewpoint}. Therefore, we conclude that using the $2\farcs5$ Gaussian PSF is adequate for this work.
 
    \item Rebin to match the spatial ($0\farcs5$ spaxels) and spectral sampling ($R \sim 2000$) of MaNGA. The spectral sampling varies as a function of wavelength.
   \item Use a mock $g-$band image that is rebinned to the $0\farcs5$ spatial scale to mask all spaxels in the datacube that fall below a $g-$band S/N cutoff value of 1.
   
   Follow the procedure from N19 to convolve, rebin, and introduce noise characteristic of SDSS imaging to the mock $g-$band images to match the $0\farcs5$ spatial binning of the MaNGA cubes. Then find the average $g-$band S/N per spaxel and mask all spaxels that fall below a S/N cutoff value of 1. This procedure directly follows the MaNGA DAP (\citealt{Westfall2019}), which masks all spaxels using the same $g-$band S/N cutoff.
   
    \item Use the MaNGA procedure to select which sized fiber bundle to use for each mock datacube and mask the spaxels that are external to this hexagonal footprint. 
    
    We use \texttt{statmorph} (\citealt{Rodriguez-Gomez2019}) to measure the effective radius of the mock $r-$band images from N19. We then determine the smallest fiber bundle needed to cover each galaxy to 1.5 R$_e$ (this is how MaNGA's primary sample is defined). We select the smallest fiber bundle if the total angular extent of the galaxy (2$\times$1.5 R$_e$) is smaller than $12\farcs5$ and the largest fiber bundle if the angular size exceeds $32\farcs5$.

     \item We introduce noise to each spaxel to produce a datacube with noise and a sqrt(variance) datacube (from here on, `error datacube'). 
     
     We first produce a typical noise spectrum that demonstrates how the noise trends with wavelength for MaNGA observations. We then normalize this noise spectrum using the $g-$band S/N value for each spaxel. The end result is a sqrt(variance), or error spectrum, which we use to introduce random noise to each spaxel in the datacube. The noisy spectra and the accompanying error spectra are the inputs to \texttt{ppxf}. More details of this process can be found in Appendix \ref{noise}.  
     
     To verify that the S/N of the simulated spectra are representative of the MaNGA sample, we use the peak $g-$band S/N as a comparison statistic, which is the maximum value of the $g-$band S/N (per pixel) from a single galaxy observation. The peak $g-$band S/N ratio of a sample of MaNGA galaxies that span the full range of sizes, surface brightnesses, and stellar masses of the MaNGA sample ranges from 10 - 60, with a median of 25. The same statistic for the simulation suite ranges from 10 - 100, with a median of 30. In \S \ref{limitssnz}, we experiment with changing the S/N of simulated spectra, and investigate how this affects the classification.

    Since the MaNGA datacubes oversample the effective PSF, they also contain significance covariance in the errors between adjacent spaxels such that the S/N ratio of binned spectra does not increase as $\sqrt{N}$.  This covariance is irrelevant for the fitting of individual spectra, but we account for it in our Voronoi binning by following the analytic approximation given by \citet{Law2016}, as we discuss below.
     
      \item  After completing the masking steps, we further exclude regions that are background dominated.
    
    At this stage, we notice that the datacubes have `patchy' outskirts, or regions of low S/N data that are surrounded by masked regions. The MaNGA datacubes do not have this feature; instead, they exclude regions that can be characterized as `background dominated'. This patchiness does not affect the results of the classification, instead we choose to correct it for cosmetic purposes. To do this, we mask spaxels where the $g-$band signal is less than 3$\sigma$ above the background value, where $\sigma$ is the standard deviation of the noise given above. This produces the desired effect, where the mask has a sharper cutoff, matching the appearance of the MaNGA cubes.  
    
    \item Rebin spatially using a Voronoi binning scheme with $g-$band S/N of 10 (\citealt{Cappellari2003}).
    
    We create spatial bins that have a $g-$band S/N of 10, reproducing the procedure described in \citet{Westfall2019}. When a Voronoi bin contains more than one spaxel, the new spectrum is the masked average of all constituent spectra while the error spectrum for that bin is determined by co-adding the error spectra. It is important to account for covariance between neighboring spaxels in our Voronoi bin calculation. In order to avoid the computational cost of calculating the covariance matrix for all spaxels, we instead use the correction from \citet{Law2016}.
    The correction is an analytic function of the number of spaxels in a bin ($N_{\mathrm{bins}}$): $$n_{\rm{measured}}/n_{\rm{no\ covar}} = 1 + 1.62 \times \mathrm{log} (N_{\mathrm{bins}})$$
where $n_{\rm{measured}}$ is the corrected noise level after the correction is applied to the co-added error where covariance is not considered ($n_{\rm{no\ covar}}$) and $N_{\mathrm{bins}}$ is the number of spaxels in a bin.

\end{enumerate}

The final step of the creation of mock kinematic maps is to pass the Voronoi binned spectra through \texttt{ppxf} (\citealt{Cappellari2004,Cappellari2017}). \texttt{ppxf} is a penalized pixel-fitting method which assumes that a galaxy spectrum is a combination of stellar templates that are convolved with the line-of-sight velocity distribution (LOSVD).

To run \texttt{ppxf}, we follow these steps from the DAP:
\begin{itemize}
    \item Normalize the flux data so that the mean over all templates is unity. 
    \item  Mask the spectra to match the wavelength range of the \texttt{MILES-HC} library (3600-7400 \AA).
    \item Mask the emission lines using the DAP module StellarContinuumBitMask().
    \item Use the 42 template \texttt{MILES-HC} spectral library to globally fit each datacube. 
    
   The templates are first convolved to the spectral resolution of MaNGA.\footnote{This is a departure from the DAP. However, as noted in \citet{Westfall2019}, there is no mathematical difference between our approach and later subtracting the difference in resolution in quadrature from the \texttt{ppxf} result.}
    \item We use the `\texttt{NZT}', or non-zero template iteration mode to fit all bins with \texttt{ppxf}. 
    
    In this mode (which is also used in the DAP), we first fit the masked average of all spectra in the datacube and use this global fit to isolate the subset of templates allocated non-zero weights. This template subset is then used to individually fit each bin. 
    
    \item Each fit iteration of \texttt{ppxf} uses an additive eight-order Legendre polynomial and a Gaussian line of sight velocity dispersion (LOSVD) with two moments. As in the DAP, due to limited spectral resolution, we do not solve for the higher order moments $h_3$ and $h_4$ (\citealt{Westfall2019}).
   
\end{itemize}

The final product of our MaNGA-izing procedure is the first two moments of the LOSVD, or a stellar velocity map and a stellar dispersion map, both with associated error maps from the fit to the stellar continuum.

\subsection{Preparing Kinematic Predictors}
\label{kinclass}

Here we define and describe the predictors extracted from the stellar kinematic maps. The goal is to create a set of kinematic predictors that adequately describe the different types of merger-induced kinematics in the velocity and velocity dispersion maps. 

To develop this kinematic identification tool, we use the stellar kinematics instead of the warm ionized gas kinematics (henceforth, `gas kinematics'). The stellar and gas kinematics trace different physical regions and processes in the merging galaxies. We select the stellar kinematics because they directly trace the assembly history of a galaxy's stellar population. On the other hand, the gas kinematics can be subject to a number of non-gravitational forces. The stellar kinematics and the gas kinematics diverge in the presence of shocks, inflows, and/or outflows, all of which are processes that are not limited to merging galaxies. An analysis built on gas kinematics is a compelling direction for future work but is beyond the scope of this paper (i.e., see \citealt{Khim2020}).\footnote{Gas kinematics are not available for many MaNGA galaxies (since many are non-starforming), but are easier to obtain than stellar kinematics for many high redshift galaxies and could be more compelling direction to pursue in this context.}

The kinematic predictors are based on previous work to identify merging galaxies from the stellar kinematics of observed and simulated galaxies. All of these predictors are sensitive to different orientations, merger stages, mass ratios, and/or gas fractions of merging galaxies. Our goal is to combine them into one LDA classification to best identify a variety of different types and epochs of merging galaxies. In total, we extract the following predictors (which are all introduced in Table \ref{tab:predictors}): $A$, $A_2$, $\Delta$PA, \vasym, \sasym, resid, \lambdare, $\epsilon$,  $\Delta x_V$, $\Delta x_\sigma$, \muv, \mus, \sigv, \sigs, \hthreev, \hthrees, \hfourv, and \hfours. We include a brief definition for all predictors in Table \ref{tab:predictors} but focus the remainder of this section on the kinematic predictors that were selected by the random forest term selection technique described in \S \ref{RFR}: $A_2$, $\Delta$PA, resid, \lambdare, \muv, \mus, \sigv, \sigs, \hthreev, \hthrees, \hfourv, and \hfours. These terms are the most informative for identifying the merging galaxies and we discuss them throughout the rest of the paper. We further describe the kinematic predictors that were not selected in Appendix \ref{predcont}. 

\renewcommand{\arraystretch}{1.5}
\begin{table*}

    \centering
    \begin{tabular}{l|l|l}
         Predictor Name&Description & Derivation  \\ 
         \hline
         
        $A$ &The weighted asymmetry in the position angle,  &
        $A = \frac{\sum_i \delta\hat{\theta}_i}{2 N_{i,j}}w_{i,j}$\\  
        & which is calculated from the Radon profile & where $\hat{\theta}$ is the best fit kinematic position angle\\ 
        \hline
        
        \textcolor{purple}{$A_2$} & The error-weighted asymmetry in the position angle &
        $A_2 = \sum_i \frac{\delta\hat{\theta}_i}{\sigma_{\delta\hat{\theta},i}}$ \\
        
        \hline
         
        \textcolor{purple}{$\Delta$PA} & The difference between the global kinematic and & $\Delta \rm{PA} = |\rm{PA}_{\rm{kin}} - \rm{PA}_{\rm{img}}|$\\
        & photometric position angles, which are measured &   \\
        &from \texttt{kinemetry} and the $g-$band image&   \\
       
        \hline
        
        \sasym & Describes the degree of smoothness of the  &   $\sigma_{\mathrm{asym}} = < {\frac{\sum_{n=1}^5 k_{\mathrm{n}, v}/5}{A_{0,v}}} >_r$ \\
        & velocity dispersion map & which is the sum of the higher \\
        & &order coefficients from \texttt{kinemetry} \\

        \hline
        
        \vasym  &The deviation of the velocity dispersion map & $\rm{v}_{\mathrm{asym}} = < {\frac{\sum_{n=2}^5 k_{\mathrm{n}, v}/4}{B_{1,v}}} >_r$\\
        &from ordered rotation & same as above but excluding the $k_{1,v}$ term\\

        \hline
        
        \textcolor{purple}{resid} & The residual between the best fit \texttt{kinemetry} & $\mathrm{resid} = \frac{\sum_{i,j}^N |V_{*} - V_{\rm{model}}|}{N}$   \\
        &  model  and the velocity map& \\

        \hline

        \textcolor{purple}{\lambdare} & The approximate spin parameter & \lambdare $= \frac{\sum_{n=1}^N F_n R_n |V_n|}{\sum_{n=1}^N F_n R_n \sqrt{V_n^2+\sigma_n^2}}$\\

        \hline
        
       $\epsilon$ & Galaxy ellipticity &Measured using \texttt{statmorph} from the $r-$band imaging   \\
        \hline

        $\Delta x_V$  &  The spatial distance between the center of the& The imaging center is measured from the $r-$band image and \\
        & velocity map and the imaging center in kpc &  the kinematic center is from the Radon Transform\\

        \hline
        $\Delta x_{\sigma}$ & Same as above, but for the velocity dispersion map & The center of the velocity dispersion map is determined \\
        & & using a low pass filter\\
        
        \hline

        \textcolor{purple}{\muv\ and \mus} & The mean of the distribution of the & The distributions for each map are created\\
         &  velocity and velocity dispersion maps &  by collecting the values from all spaxels \\

        \hline
        
        \textcolor{purple}{\sigv\ and \sigs} &The variance of the distributions&  \\
        
        \hline
        
        \textcolor{purple}{\hthreev\ and \hthrees} & The skewness of the distributions &  \\
        
        \hline
        \textcolor{purple}{\hfourv\ and \hfours} & The kurtosis of the distributions & \\

    \end{tabular}
    \caption{Synthesis of all of the kinematic predictors measured in this paper. We highlight the predictors that are selected as important in \textcolor{purple}{purple}. We include a brief description and derivation for each predictor. For more details, see \S \ref{kinclass} for the predictors that are selected as important and Appendix \ref{predcont} for the predictors that are not used in the classification. }
    \label{tab:predictors}
\end{table*}

\renewcommand{\arraystretch}{1}

To define the asymmetry in the kinematic position angle ($A_2$), we utilize the Radon Transform from \citet{Stark2018}. We transform the velocity maps into circular coordinates ($\rho$,$\theta$) where $\rho$ is the distance from the spaxel to the center of the velocity map, which is the kinematic center (defined below), and $\theta$ is the angle between the positive x-axis and the line segment from the kinematic center to the spaxel. The angle $\theta$ ranges from 0 to 180 in the CCW direction. Positive values of $\rho$ are regions of the velocity map above the x-axis and negative values of $\rho$ are below the positive x-axis. 

The Radon Transform is defined as:

\begin{equation}
    R(\rho,\theta) = \int_0^L v(x,y) dl
    \label{eq:radon}
\end{equation}
where the velocity is summed along line integrals that are centered on the point ($\rho$, $\theta$) and perpendicular to the kinematic center of the galaxy. The Radon Transform is a 2D array that is calculated at all values of $\rho$ and $\theta$. 

\begin{figure*}
    \centering
    \raisebox{-0.5\height}{\includegraphics[scale=0.43, trim=0cm 2cm 0cm 2cm, clip]{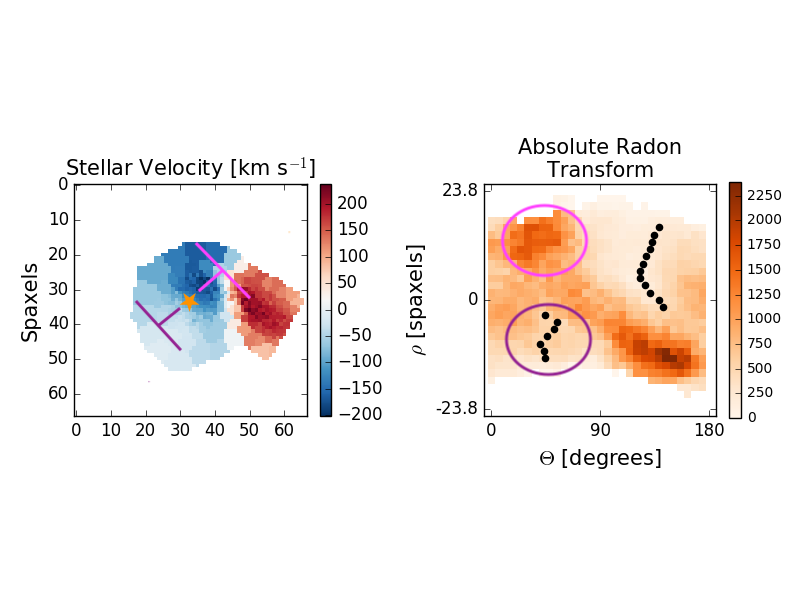}}
    \raisebox{-0.5\height}{\includegraphics[scale=0.43]{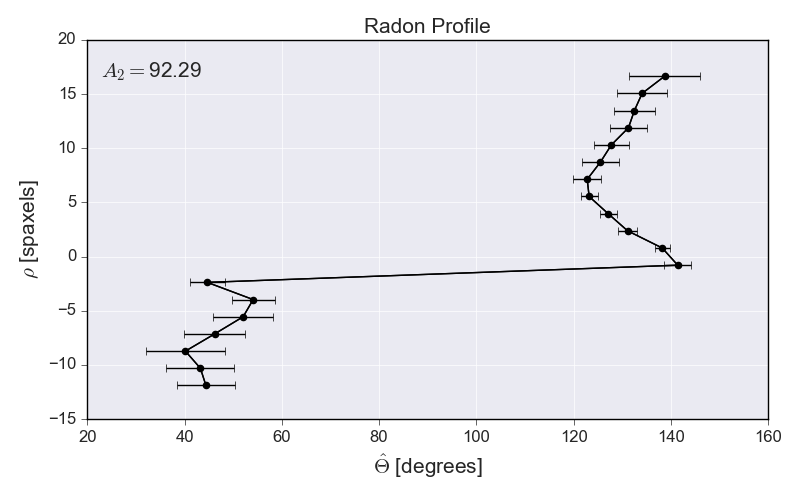}}
    \caption{Stellar velocity map (left), bounded Absolute Radon Transform (middle), and Radon profile (right) for a snapshot during the late stages of the q0.5\_fg0.3 merger. In this case, the primary galaxy is \textbf{blueshifted} at the center of the velocity map (systemic velocity is $\sim $-100 km s$^{-1}$ and the secondary galaxy approaches from the right and is redshifted relative to the primary galaxy. To compute the Radon Transform, the velocity field is transformed into $\theta$ and $\rho$ coordinates, where $\theta \subseteq \ $[0,180] and $\rho \subseteq \ $ [-$\infty$,+$\infty$], where $\theta$ is measured CCW from the top of the map. The bounded Absolute Radon Transform is then calculated by creating line integrals over a grid of ($\rho$,$\theta$) positions, where the line is perpendicular to the kinematic center of the map. It is `bounded' because the line integral is limited to the length $R_e$. In the left panel, the kinematic center is a yellow star and the magenta and purple line segments demonstrate the calculation of the Absolute Radon Transform at $\theta \sim 45$ for positive and negative $\rho$ values, respectively. The magenta and purple regions in the middle panel have large and small values, respectively, which demonstrates that the value of the Absolute Radon Transform is smaller in the regions where the spaxel velocities vary less along the line integral. We find the minima (shown in lighter yellow) of $R_{AB}$ at each value of $\rho$, to measure the Radon profile (right), which is used to calculate the error-weighted asymmetry in the kinematic position angle, $A_2$.}
    \label{radon}
\end{figure*}

We then calculate the bounded Absolute Radon Transform, $R_{AB}$, which is integrated over a distance $R_e$ and is the absolute value of the difference between the velocity at each point and the mean value along the line segment. 

We present the bounded Absolute Radon Transform and the Radon profile in Figure \ref{radon}. The Radon profile is computed by determining the minimum value of $\theta$ ($\hat{\theta}$, where the hat operator denotes an estimated value) for each value of $\rho$ from the bounded Absolute Radon Transform. The value of $\hat{\theta}$ traces the direction of maximal rotation in the stellar velocity maps at each radial position. 

We follow the procedure from \citet{Stark2018} to determine the galaxy's kinematic center, which we describe in more detail in Appendix \ref{predcont}.

We quantify the asymmetry of the Radon profiles using the kinematic predictor $A_2$, from \citet{Stark2018}:
\begin{equation}
    A_2 = \sum_i \frac{\delta\hat{\theta}}{\sigma_{\delta\hat{\theta},i}}
    \label{eq:A2}
\end{equation}
where $\delta \hat{\theta}$ is the absolute magnitude of the difference between $\theta_i$ on one side of the Radon profile to the other (same $\rho$, different sign), $\sigma_{\delta \hat{\theta}}$ is the uncertainty on $\delta\hat{\theta}$, and the expression is summed over the $i$ values of $\hat{\theta}$.

The $A_2$ predictor incorporates the absolute magnitude of the difference between the measured kinematic PA on one side of the galaxy to the other. We therefore expect that $A_2$ will be enhanced for merging galaxies, since mergers can cause warps in the stars in a galaxy (e.g., \citealt{Shapiro2008}).

We use \texttt{kinemetry} to measure both $\Delta$PA and resid from the LOSVD (\citealt{Krajnovic2006}). Functionally, \texttt{kinemetry} measures the kinematic asymmetry from the line of sight velocity maps by dividing them into a set of nested elliptical rings. The best fit model at each radius is determined using a ring defined by the kinematic PA and the flattening factor $q_f$ = 1-$e$, where $e$ is the ellipticity of the ring in the plane of the sky. These models use a decomposition of the moment maps into harmonic Fourier coefficients in polar coordinates. For instance, a velocity map, $K(r, \psi)$ can be expanded into a finite number of harmonic frequencies:
\begin{equation}
    K(r, \psi) = A_0(r) + \sum^N_{n=1} A_n(r)\ \mathrm{sin} \ (n\psi) + B_n(r)\ \mathrm{cos} \ (n \psi)
    \label{eq:harmonics}
\end{equation}
where $r$ is the semimajor axis of the ellipse, $\psi$ is the azimuthal angle, $A_0(r)$ is the systemic velocity, $N$ is the number of ellipses fit, and $A_n$ and $B_n$ are the coefficients of the harmonic expansion. The best-fitting ellipses are obtained by minimizing $\chi^2$ for the linear combination of the $A_n$ and $B_n$ coefficients. 

An ideal rotating disk can be described using only the $B_1$ term, which represents the cosine term for the circular velocity of a galaxy's rotating disk:
\begin{equation}
V(r,\psi) = V_c(r)\  \mathrm{sin} \ i \  \mathrm{cos} \ \psi
\label{eq:vel}
\end{equation}
where $r$ is the radius in the plane of the galaxy, $\psi$ is the azimuthal angle, $V_c(r)$ is the circular velocity, and $i$ is the inclination of the galaxy disk.

To determine the best fit Fourier coefficients, we run \texttt{kinemetry} multiple times. We first allow the best fit kinematic PA and value of $q_f$ to vary for each radius. We define the kinematic position angle (PA$_{\mathrm{kin}}$) to be the median value of the best fit kinematic PAs. We then allow the value of $q_f$ to vary and determine the median value. After determining the global values for kinematic PA and $q_f$, we do a final run to determine the values of the higher order kinematic moments and therefore the best fit disk model. We then compare PA$_{\mathrm{kin}}$ to the imaging major axis (PA$_{\mathrm{img}}$, which is measured using \texttt{statmorph} from the $r-$band imaging) to create the predictor $\Delta$PA. Since $\Delta$PA traces the recent global misalignments of stars, it should be elevated for the merging galaxies that have misaligned stellar disks.  

We use the global kinematic position angle from \texttt{kinemetry} to measure $\Delta$PA instead of the median of the kinematic position angles from the Radon Transform. The main motivation for this choice is that \texttt{kinemetry} uses an adaptive binning scheme; at each step outwards, the ellipses are larger, which gives less weight to the kinematic confusion at the outskirts of the galaxy. The Radon Transform, however, is equally sampled in $\rho$ (see Figure \ref{radon}), so it can be more influenced by the measurement of the kinematic PA at the outskirts of the galaxy. In most cases, the two measurements agree within error, but in cases where the kinematic maps are disturbed, the global kinematic PA from \texttt{kinemetry} more closely matches our by-eye assessment of the kinematic PA.

\begin{figure*}
    \centering
    \includegraphics[scale=0.8, trim=0cm 5cm 0cm 5cm, clip]{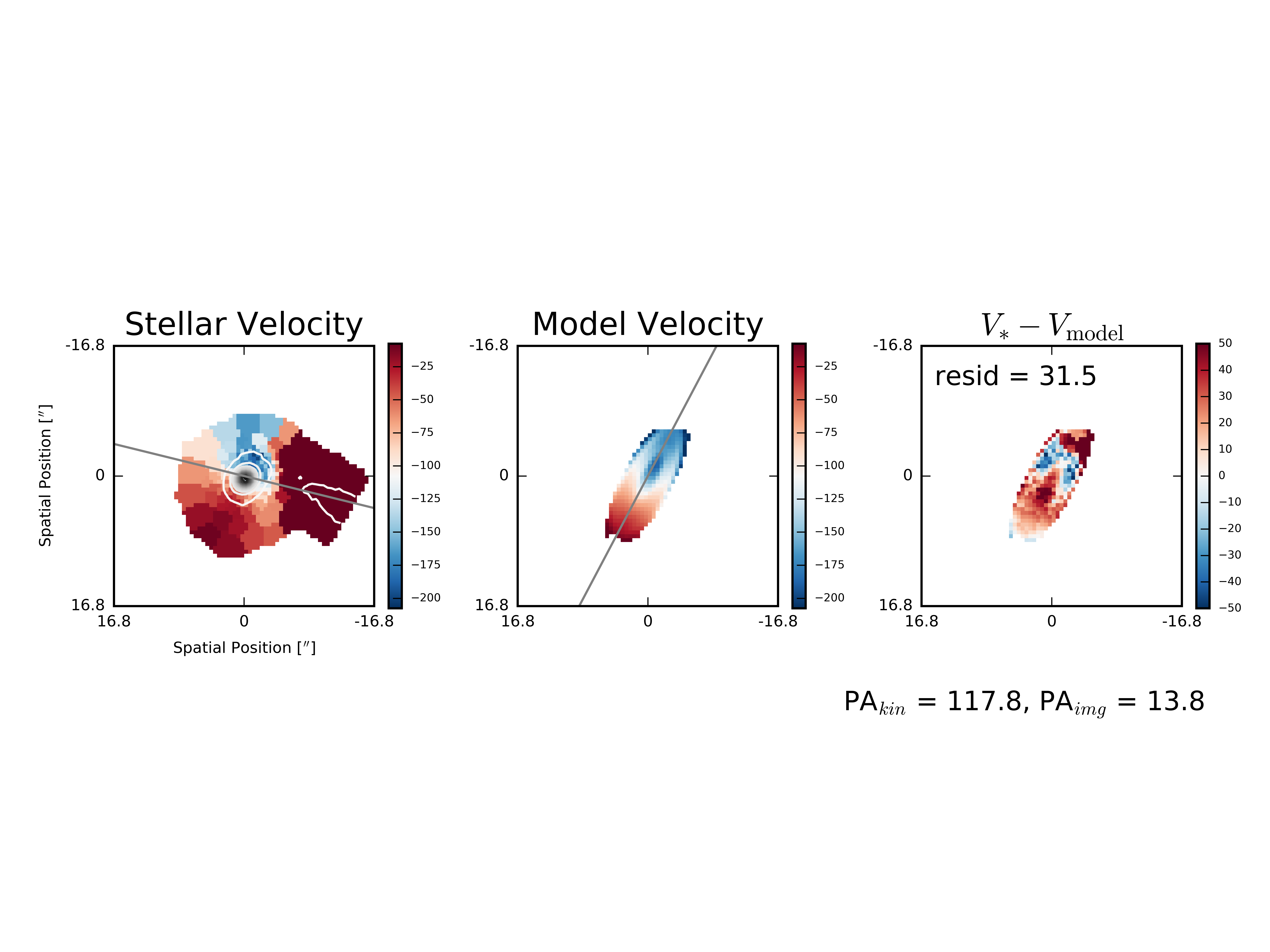}
    
    \caption{An example \texttt{kinemetry} fit to a snapshot of the q0.5\_fg0.3 merger simulation with observed stellar velocity map (left), best fit \texttt{kinemetry} model (middle), and the model velocity subtracted from the stellar velocity map (right). Note that this is the same snapshot shown in Figure \ref{radon}. The color bars show the velocity in km s$^{-1}$. In the left panel we overplot the contours from the $r-$band imaging and the imaging position angle. The kinematic position angle (from \texttt{kinemetry}) is the straight line in the middle panel. We utilize the normalized residuals as a predictor (right), which we refer to as `resid', which is the sum over all spaxels of the absolute value of the difference of the stellar velocity and the model velocity, normalized by the number of spaxels in the model.}
    \label{kinexample}
\end{figure*}

We also extract `resid', or the \texttt{kinemetry} residuals between the best fit rotating disk and the velocity map. This predictor is defined as:

\begin{equation}
\mathrm{resid} = \frac{\sum_{i,j}^N |V_{*} - V_{\rm{model}}(r,\Psi)|}{N}
\label{eq:resid}
\end{equation}
where $V_{*}$ is the observed velocity map, $V_{\rm{model}}(r,\Psi)$ the circular velocity model from \texttt{kinemetry}, and $N$ is the number of spaxels fit. We include this normalization factor in order to penalize the fits that converge to a very inclined galaxy. For these galaxies, the fit is attempting to avoid fitting disordered kinematics in the exterior regions of the galaxy by fitting a smaller region. We show an example of a simulated galaxy snapshot from the q0.5\_fg0.3 simulation fit with \texttt{kinemetry} and its velocity residuals in Figure \ref{kinexample}.

We measure $\lambda_{R_e}$, the approximate spin parameter, from the stellar velocity and velocity dispersion maps, which is defined by \citet{Emsellem2007}:
\begin{equation}
\lambda_{R_e} =  \frac{\sum_{n=1}^N F_n R_n |V_n|}{\sum_{n=1}^N F_n R_n \sqrt{V_n^2+\sigma_n^2}}
\label{eq:lambdare}
\end{equation}
where $F_n$ is the ($r-$band) flux of a spaxel, $R_n$ is the distance from the kinematic center, $V_n$ is the stellar velocity, and $\sigma_n$ is the stellar velocity dispersion. We measure $\lambda_{R_e}$ to the $r-$band effective radius. Since the fiber bundles are designed to provide coverage of each galaxy to 1.5$R_e$, if a secondary nuclei falls towards the outside edge of the hexagonal FOV, it is excluded from the measurement of \lambdare. This effect is more relevant for the minor mergers, where the secondary component covers a smaller effective area of the hexagonal FOV.

We measure the ellipticity of a galaxy, $\epsilon$, from the $r-$band photometry using \texttt{statmorph}. It is distinct from the ellipticity parameter used by \texttt{kinemetry} to fit rotation curves. We do not use $\epsilon$ as a kinematic predictor. Instead, we use it to construct the \lambdare-$\epsilon$ diagnostic diagram in \S \ref{discuss:useless}, where the division between fast and slow rotators is defined by \citet{Cappellari2016}:

\begin{equation}
\lambda_{R_e} = 0.08 + \epsilon/4
\label{eq:slowfast}
\end{equation}
where slow rotators fall below this line.

On the \lambdare-$\epsilon$ diagram, \lambdare\ is the more predictive of the two axes; it decreases dramatically for the `slow-rotating' population of galaxies, which are dynamically disordered and dispersion-dominated. We predict that \lambdare\ will decrease for merging galaxies since mergers are kinematically disordered and can contribute to bulge-growth, which is associated with enhanced velocity dispersion.

\begin{figure}
    \centering
    \includegraphics[scale=0.4, trim=2cm 0cm 0cm 0cm]{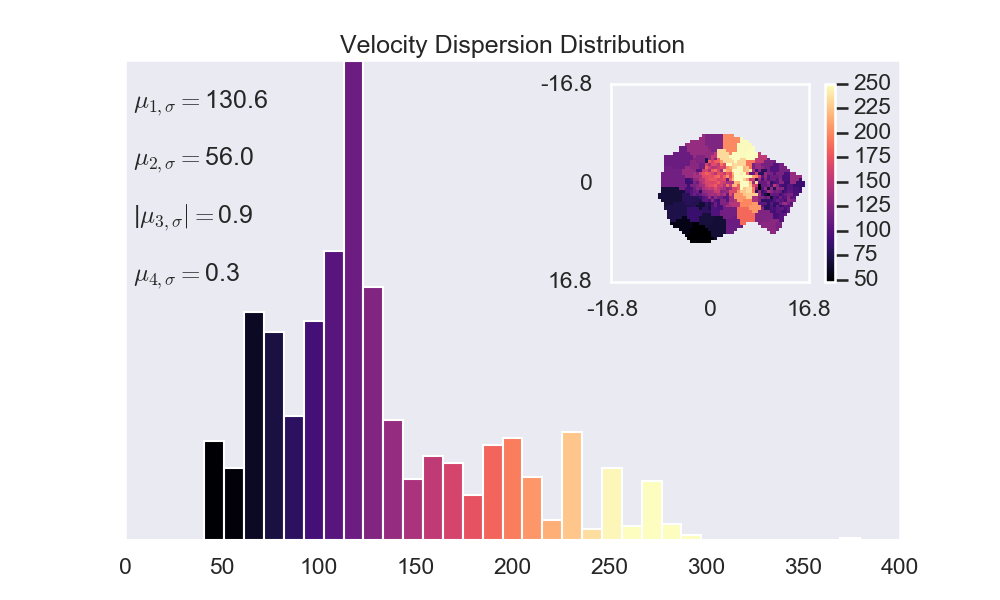}
    \caption{Distribution of the velocity dispersion values (in km s$^{-1}$) taken from each spaxel in the velocity dispersion map (inset, velocity dispersion bar is in km s$^{-1}$ and the spatial axis is in arcsec). This snapshot is also showcased in Figures \ref{radon} and \ref{kinexample}. We also include the measured values of the mean (\mus), dispersion (\sigs), skew (\hthrees), and kurtosis (\hfours) of this distribution. A distribution with a larger skew is asymmetric about the mean. A distribution with a positive kurtosis has a high degree of peakedness relative to a normal distribution.  }
    \label{fig:distribution}
\end{figure}

In addition to kinematic predictors that were utilized in previous work, we define a new set of predictors based on the distributions of values in the velocity and velocity dispersion maps. These predictors include \muv/\mus, \sigv/\sigs, \hthreev/\hthrees, and \hfourv/\hfours, which are the standardized moments of the stellar velocity/velocity dispersion maps. These predictors are similar to the formulation from \citet{Sweet2020}, which calculates the moments of PDF($s$), where $s$ is the normalized specific angular momentum.

To determine the values of these predictors, we measure the four standardized moments of the distribution; mean ($\mu_1$), variance (standard deviation; $\mu_2$), skewness ($\mu_3$), and kurtosis ($\mu_4$). This produces eight different predictors (four each from the velocity and velocity dispersion distributions). These quantities are different from the higher order moments $h_3$ and $h_4$, which are typically measured by \texttt{ppxf}. We show an example of these predictors measured from a velocity dispersion map in Figure \ref{fig:distribution}.

We expect to see an offset in the mean velocity (\muv) from systemic for the merging systems and an enhanced mean velocity dispersion (\mus). The spread in the velocity distribution (\sigv) and the dispersion of the velocity dispersion distribution (\sigs) could identify superpositions of dynamically distinct stellar components. This could include a secondary merging galaxy or features like a stellar bulge.

The higher order moments could be useful for identifying subtler features of mergers, beyond bulk shifts in \muv, for example. The skewness of a distribution is sensitive to the tails; we take the absolute value to treat positive and negative skew identically. A skewed velocity or velocity dispersion distribution (\hthreev\ and \hthrees) could have a faint secondary source in the field of view, where the distribution is actually a combination of two galaxy rotation curves. Kurtosis measures how peaked a distribution is relative to normal; a flatter distribution has a negative kurtosis and a more peaked distribution has a higher peak. A smoothly rotating velocity dispersion field has a normally shaped distribution whereas a disturbed field may have a negative (flatter) kurtosis (\hfours). On the other hand,  post-coalescence mergers with recent bulge growth could have a positive kurtosis in the velocity dispersion distribution.

To summarize, we extract the following kinematic predictors: $A$, $A_2$, $\Delta$PA, \vasym, \sasym, resid, \lambdare, $\epsilon$,  $\Delta x_V$, $\Delta x_\sigma$, \muv, \mus, \sigv, \sigs, \hthreev, \hthrees, \hfourv, and \hfours. We then use the techniques described in the following sections (\S \ref{LDA}, \S \ref{outliers}, and \S \ref{RFR}) to select the most informative of these predictors. We ultimately use the following predictors in the LDA classification: $A_2$, $\Delta$PA, resid, \lambdare, \muv, \mus, \sigv, \sigs, \hthreev, \hthrees, \hfourv, and \hfours.

\section{Results}
\label{results4}
After creating mock MaNGA datacubes from the five simulations of merging galaxies (and matched isolated galaxies), we extract the kinematic predictors introduced in \S \ref{kinclass}. We then prepare the input data, select the predictors that are most informative, and create and assess the classification itself.

In \S \ref{LDA}, we describe the LDA technique. We then provide an overview of our process for preparing the data and examining it in the context of the assumptions made by the LDA in \S \ref{outliers}. Prior to running the LDA classification, we perform an initial term selection using a random forest regressor, which we describe in \S \ref{RFR}. We present the classification results in \S \ref{LDAresults} and measure performance statistics in \S \ref{accuracy4}. We present the LDA observability time in \S \ref{analyzeobservability}. Then, in \S \ref{fails} we explore some failure modes of the classification. Finally, we analyze how the classification changes with redshift and decreasing signal-to-noise (S/N) in \S \ref{limitssnz}. More details of the classification are discussed in the appendices, where we analyze possible biases of the classification in \S \ref{fair}.

\subsection{Linear Discriminant Analysis}
\label{LDA}

The classification in this work relies upon an LDA technique that separates nonmerging galaxies from merging galaxies based upon a combination of the input predictors (for a review of LDA, see \citealt{James2013}). This approach was first presented in N19 for imaging predictors; here, we use this approach for kinematic predictors. 

LDA is one of many statistical learning tools that perform classification tasks. Using pre-defined features (predictors) as inputs, LDA solves for the hyperplane in multi-dimensional predictor space that maximizes the separation between different classes of objects (i.e., mergers and nonmergers). The solution is a linear combination of the input predictors; the classification is therefore relatively easy to interpret because its complexity is low. Recent work has employed other techniques to identify merging galaxies, such as random forest regressors (e.g., \citealt{Snyder2019,Goulding2018}) and convolution neural networks (CNNs; e.g., \citealt{Bottrell2019}). These techniques have various advantages and disadvantages based upon the dataset at hand and the goals of the work. Since we aim to optimize the interpretability of the method, we select LDA over an approach like a CNN. CNNs might increase the number of correct classifications, but they achieve this using complex non-linear features, which are not easily interpreted.

In this work, we have made several important changes to the technique from N19. We first recap the relevant details from the LDA in N19, and then we discuss the changes. Relevant details of the LDA technique from N19 include:
\begin{itemize}
    \item All predictors are linearly standardized prior to the LDA technique, meaning that predictors with large numerical values (such as $A_2$) do not have an outsized effect on the analysis.
    
    \item We utilize priors on the relative fraction of merging and nonmerging galaxies in nature versus in the simulations. This accounts for the fact that we have more merging galaxy snapshots (relative to nonmerging snapshots) for each simulation. We use the same priors from N19; $f_{\mathrm{merg}} = 0.1$ for the major mergers and $f_{\rm{merg}} = 0.3$ for the minor mergers. These priors are based on the fraction of nonmerging and merging galaxies from observation and simulations (e.g., \citealt{Rodriguez-Gomez2015, Lotz2011,Conselice2009,Lopez-Sanjuan2009,Shi2009}).
    \item We include interaction terms to explore correlations between predictors. 
   
     \item In order to select which coefficients are necessary for the classification, we use a forward stepwise selection technique, which orders and includes only the most important terms. This technique adds additional terms to LD1 only if they improve the F1 statistic, which is defined in \S \ref{accuracy4}. It also protects against the unnecessary addition of terms by finding the minimum number of terms that produce an F1 statistic that is consistent with the maximum (within 1$\sigma$ errors, where $\sigma$ is the standard deviation on the F1 statistic measured from each $k-$fold cross-validation set). 
     
     \item We use $k$-fold cross-validation to obtain 1$\sigma$ errors on the predictor coefficients. At each step of the forward stepwise selection process, we divide the sample into $k$ subsets. We then train the LDA on the first $k-1$ subsamples and test on the remaining subsample, which is the `cross-validation' set. We repeat these steps $k$ times for all combinations of subsamples and the variation in predictor coefficient values from the cross-validation subsamples is the 1$\sigma$ error.

\end{itemize}

For complete details, including the full mathematical formulation for LDA, see N19.

We make several changes to the technique motivated by the additional challenges of the kinematic data:
\begin{itemize}
    \item Due to the number of predictor terms in this work, instead of including all of the kinematic predictors in the final classification, we first utilize the RFR technique as a selection technique to eliminate predictors that are uninformative from the analysis (see \S \ref{RFR}).
   
    \item We adjust the model optimization statistic. In N19, we minimized the number of misclassifications in order to both select the predictors and determine their coefficients. Here, we utilize the F1 statistic defined in \S \ref{accuracy4} instead; it does a better job of balancing the number of false negatives and false positives in each classification. 
   
    \item We also adjust the $k-$fold cross-validation, Instead of using $k=10$, we find that a value of $k=5$ improves the performance of the LDA by creating a training set that is 80\% of the sample and a cross-validation set that is 20\% of the sample (as opposed to the 90\%/10\% divide in N19).

    We then train the LDA on nine of the subsamples, and test on the tenth sample. We repeat this procedure
ten times, and the mean number of misclassifications all ten
test samples allows us to decide which set of input predictors to select
\end{itemize}

We use the LDA both as a term selector and to determine the coefficients and standard errors for each selected predictor. In order to directly compare the imaging classification to the kinematic classification, we utilize the same snapshots from all simulations and we rerun the imaging analysis using all the same methods as the kinematic classification.

\subsection{Data Preparation and LDA Assumptions}
\label{outliers}

Prior to term selection and classification, we examine the distributions of predictor values. We screen for outliers and examine the data in the context of the assumptions made by the LDA. The goal is to gain an understanding of the properties of the data prior to classification.

First, we remove outliers by transforming the distribution of each predictor into log space. We define data points that fall more than 5$\sigma$ above or below the mean of the distribution for each predictor as outliers. The combination of the log transformation and 5$\sigma$ cutoff allow us to identify outliers that are caused by errors in the creation of the mock maps and not simply related to very disturbed kinematics. There are $\sim4$ outliers per simulation, amounting to 4 out of 100 or 200 datapoints.

Second, we check the input data for significant violations of the LDA assumptions. LDA operates under the assumptions that the predictors are normal distributed (multivariate normality), the covariance among the merging and nonmerging classes is equal (homoscedasticity), and the predictors are not strongly correlated with one another (multicollinearity).

Here we test these three assumptions by closely examining the data. We carry out the same statistical tests from N19 to test for normality, homoscedasticity, and multicollinearity. We find that the data violates all three assumptions. Additionally, we plan to introduce interaction terms into the LDA classification, which further increases the multicollinearity. 

As discussed in N19, LDA has been shown to be robust to the violations of multivariate normality and homoscedasticity (\citealt{Duda2001,TaoLi2006}). To ensure that the LDA technique in this work is robust to these violations, we directly compare the LDA results to those of a logistic regression. The logistic regression and LDA produce similar results both in terms of the relative importance of the predictors for each simulation and in the performance of the method. This is an indication that the LDA is converging even though it nominally violates several assumptions.

\subsection{Random Forest Regressor Term Selection}
\label{RFR}

In N19, we used a forward stepwise selection technique within the LDA to select informative predictors. Here, motivated by both an increase in the number of initial terms\footnote{This is partially due to a dearth of historically utilized kinematic predictors, so we initially introduce many more terms to determine which are informative.} and a decrease in the predictive power of these terms, we modify the term selection procedure. We introduce a random forest regressor (RFR) into the methodology to select a subset of predictors for each simulation, which will then be presented to the LDA classifier.

An RFR (\citealt{Ho1995}) is an ensemble learning technique that aggregates the result of many individual decision trees run in parallel. We specifically utilize the \texttt{scikit-learn} implementation of RFR (\citealt{scikit-learn}). In an RFR, the number of features that can be used to split at each node of the decision tree is limited to a percentage of the total number of features, ensuring that the ensemble model does not rely too heavily on any one feature. This means that the RFR is able to combine all potentially predictive variables in a fair way. It is also able to incorporate non-linear features to capture some higher-order interaction terms. In practice, we find that the RFR is an efficient method to initially identify the useful features in the dataset from the extensive list of kinematic predictors.\footnote{We do not use it as the main classification technique because the features can be highly nonlinear and more opaque to interpretation. Additionally, this technique is designed to directly complement the LDA technique in N19 for comparison's sake.}

In order to select the informative terms from the RFR, we include an additional predictor. This predictor is assigned a random number for each galaxy snapshot and therefore shows no significant difference between the nonmerging and merging galaxies. We use this technique to eliminate all of the terms that have a feature importance less than the random term for all simulations. In this step we eliminate the \vasym, \sigasym, $A$, $\Delta x_V$, and $\Delta x_{\sigma}$ predictors. Then, for each individual simulation, we additionally eliminate predictors that have an importance less than the random value prior to initiating the LDA classification. The terms eliminated in this step vary from simulation to simulation.

\subsection{Classification Results}
\label{LDAresults}

After using the RFR term selection to narrow the number of kinematic terms down to 11 ($\Delta$PA, resid, \lambdare, $A_2$, \muv, \sigv, \sigs, \hthreev, \hthrees, \hfourv, and \hfours), we run the LDA classification for each simulation individually. We also combine the three major mergers into a combined major merger classification and the two minor mergers into a combined minor merger classification. We run the LDA with interaction terms; the result is a linear combination of selected predictors and coefficients which is unique for each simulation. We present the term coefficients and standard errors for the four most important terms and the intercept term in Table \ref{table:LDAall}. Finally, we briefly discuss the main results of the LDA classification for each simulation, which we will examine in more detail in \S \ref{discuss4}.

\begin{table*}

  \begin{center}
    \caption{The final LD1 predictor coefficients ($\hat{\vec{w}}$) with 1$\sigma$ confidence intervals after term selection for the first four most important terms and the intersect ($\hat{w_0}$) for all simulations. }
    \label{table:LDAall}
     \begin{tabular}{c|cccc|c}
      
      & \multicolumn{4}{c}{\Large{$\hat{\vec{w}}$}} & \Large{$\hat{w_0}$} \\
      \hline
      Simulation & 1 & 2  & 3 & 4 &     \\

\hline

All Major &-6.76 $\pm$ 0.45   \lambdare  &
4.99 $\pm$ 0.6   \hthrees  &
4.54 $\pm$ 0.36   \mus*\lambdare  &
-4.44 $\pm$ 0.51   \mus*\hthrees  & -1.21 $\pm$ 0.07\\

All Minor &-4.99 $\pm$ 0.74   \sigs  &
-4.97 $\pm$ 0.59   \sigs*\hfourv  &
3.47 $\pm$ 0.62   \hfourv*\hfours  &
2.44 $\pm$ 0.38   \hfourv  &
-0.76 $\pm$ 0.04 \\

q0.5\_fg0.3 &-7.15 $\pm$ 0.78   \mus*\hthrees  &
-6.7 $\pm$ 0.63   \mus*\sigs  &
6.65 $\pm$ 0.53   \sigs  &
5.75 $\pm$ 0.2   \mus  &-2.57 $\pm$ 0.05   \\

q0.333\_fg0.3 & 8.27 $\pm$ 0.35   \mus  &
-7.84 $\pm$ 0.71   \mus*\sigs  &
5.92 $\pm$ 0.52   \sigs  &
5.21 $\pm$ 0.73   \hthrees  &
-0.77 $\pm$ 0.18 \\

q0.333\_fg0.1 & -7.78 $\pm$ 0.91   \mus*\sigs  &
7.09 $\pm$ 0.59   \sigs  &5.97 $\pm$ 0.61   \mus  &-- &-0.26 $\pm$ 0.28 \\

q0.2\_fg0.3\_BT0.2 &  -6.51 $\pm$ 1.09   \mus  &
-6.2 $\pm$ 0.93   \sigs*\lambdare  &
-5.75 $\pm$ 1.65   $A_2$  &
5.5 $\pm$ 0.67   \mus*\lambdare  &-0.79 $\pm$ 0.05\\

q0.1\_fg0.3\_BT0.2 &25.06 $\pm$ 5.11   \mus*\hfourv  &
16.02 $\pm$ 3.19   \mus  &
-12.88 $\pm$ 2.51   \hfourv  &
6.8 $\pm$ 0.99   \hfours  &-1.06 $\pm$ 0.07 \\
    \end{tabular}
  \end{center}
     
\end{table*}

\begin{figure*}

    \centering
    \includegraphics[scale=0.3]{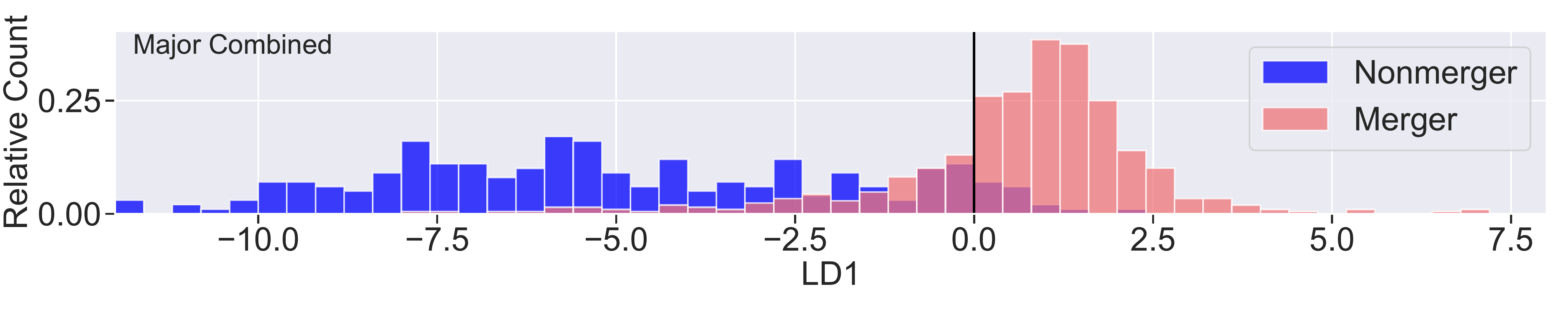}
    \includegraphics[scale=0.3]{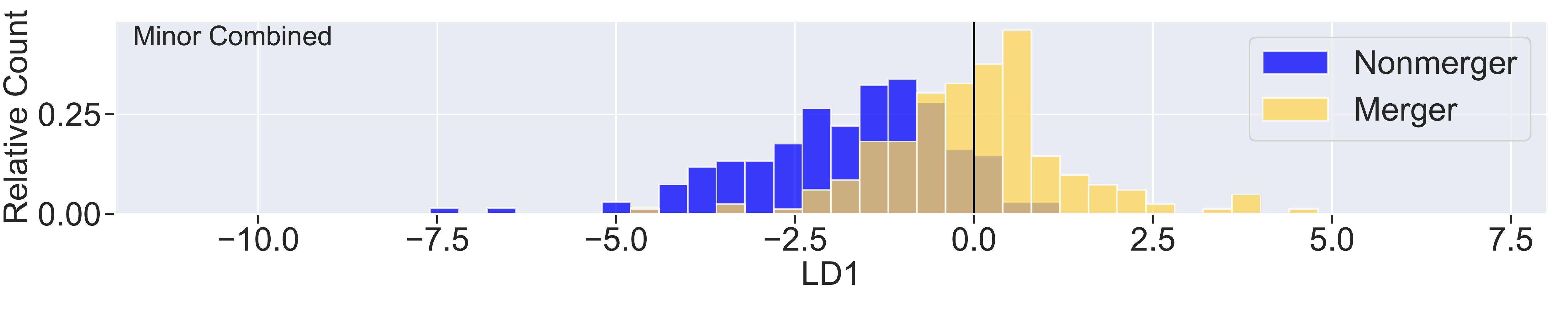}
    
    \caption{Histograms of LD1 for the populations of merging and nonmerging galaxies for the combined major merger (top) simulation and the combined minor merger (bottom) simulation. The blue nonmerging samples include both the stand-alone isolated galaxies and the pre- and post-merger isolated galaxies. The nonmerging galaxies in the top and bottom plots span different ranges in LD1 because they are composed of different samples of nonmerging galaxies and because the selected linear combination of predictors is different for the major and minor merger combined simulations. The vertical black line is the decision boundary; it is the midway point between the mean of the nonmerger and merger populations. If the LD1 value of a galaxy falls above this line, the galaxy is more likely to be a merger.}
    \label{histograms_major}
\end{figure*}

We first introduce the mechanics of the classification. LD1, which is the first linear discriminant axis, is formed from the linear combination of coefficients multiplied by the standardized predictors and an intercept term: $$\mathrm{LD1} =  C*X + B$$
where $C$ is the matrix of coefficients, $X$ is the standardized values of the selected predictors, and $B$ is the intercept term. 

LD1 is the hyperplane that best separates the populations of merging and nonmerging galaxies for each simulation. We use the result from the major merger classification as an illustrative example of how to interpret the LDA results. The LD1 for the major merger combined run (truncated after seven terms) is:

\newcommand\numberthis{\addtocounter{equation}{1}\tag{\theequation}}

\begin{align*}
\mathrm{LD1}_{\mathrm{all\ major}} & = -6.8  \lambda_{R_e}
+ 5.0   |\mu_{3,\sigma}|
+ 4.5    \mu_{1,\sigma}*\lambda_{R_e}  \\
& \ \ \ \ - 4.4   \mu_{1,\sigma}*|\mu_{3,\sigma}| -1.0 \mu_{1,\sigma}*\mathrm{resid}  \\
& \ \ \ \ +1.7 \mu_{1,V}*\lambda_{R_e} +1.7 \mu_{4,\sigma}*\mu_{4,V}  \\
& \ \ \ \ + ... -1.2 \numberthis \label{eq:LD1interaction}
\end{align*}

where LD1 is a linear combination of all selected terms, which are composed of a coefficient (positive or negative) followed by the standardized value of a predictor. The last term is the intercept term.

The higher the value of LD1, the more likely the galaxy is to be classified as merging. We calculate the values of LD1 for all of the galaxies from the major and minor combined simulations in Figure \ref{histograms_major}. The horizontal line at an LD1 value of zero is the decision boundary that corresponds to a $p_{\mathrm{merg}}$ value of 0.5; all galaxies with an LD1 value greater than zero would be classified as merging using a threshold value of 0.5\footnote{This decision boundary can be moved either before the creation of the LDA or after to be more or less tolerant of false negatives and false positives.}. We find that the classification is better able to separate the merging and nonmerging classes for the major merger simulations and this is reflected in Figure \ref{histograms_major}.

There are important nuances to the interpretation of the selected predictors and their coefficients in Equation \eqref{eq:LD1interaction} because the interaction terms complicate the analysis. For instance, in Equation \eqref{eq:LD1interaction}, the first selected predictor is \lambdare, which has a negative coefficient. Ignoring the rest of the equation, this means that if the \lambdare\ value is large, then the probability that a given galaxy is merging will decrease. However, there are other \lambdare\ terms in the equation that are coupled with other predictors in interaction terms. This means that tweaking the value of \lambdare\ will not linearly change the value of LD1. While the interaction terms complicate the analysis, they are an integral part of the classification. Many of the most important terms for LD1 in Table \ref{table:LDAall} are interaction terms, and including them significantly improves the performance of the LDA. As we discuss in more detail in \S \ref{discuss:interaction}, these terms are able to capture the non-monotonic movement of mergers through predictor parameter space. 

While it may be difficult to untangle many of the contributing terms, we can use Table \ref{tab:predictors} to determine which predictors are most prevalent and therefore informative for each simulation. For instance, the \sigs\ and \mus\ predictors are selected as either primary or interaction terms for all simulations. They are therefore universally useful kinematic predictors (for a full discussion of why these terms are important see \S \ref{discuss:musigma}). The selected predictors from the q0.333\_fg0.1 simulation are similar to the q0.333\_fg0.3 simulation. These two simulations are matched for mass ratio but not for gas fraction. However, the difference between a gas fraction of 0.1 and 0.3 is insignificant so we hesitate to make any conclusions about the impact of gas fraction on the stellar kinematics. On the other hand, the minor merger simulations differ from the major merger simulations in the selected predictors. We find that \lambdare\ and \hthrees\ are important for the major mergers while the minor mergers rely more on some of the higher order terms like \hfourv\ and \hfours. We explore the implications of these findings for the physical nature of the kinematics of mergers in the discussion (\S \ref{discuss4}).

\subsection{Performance Statistics and Hyperparameter Tuning}
\label{accuracy4}

\begin{figure}
    \centering
    \includegraphics[scale=0.45, trim = 0cm 0cm 2cm 0cm, clip]{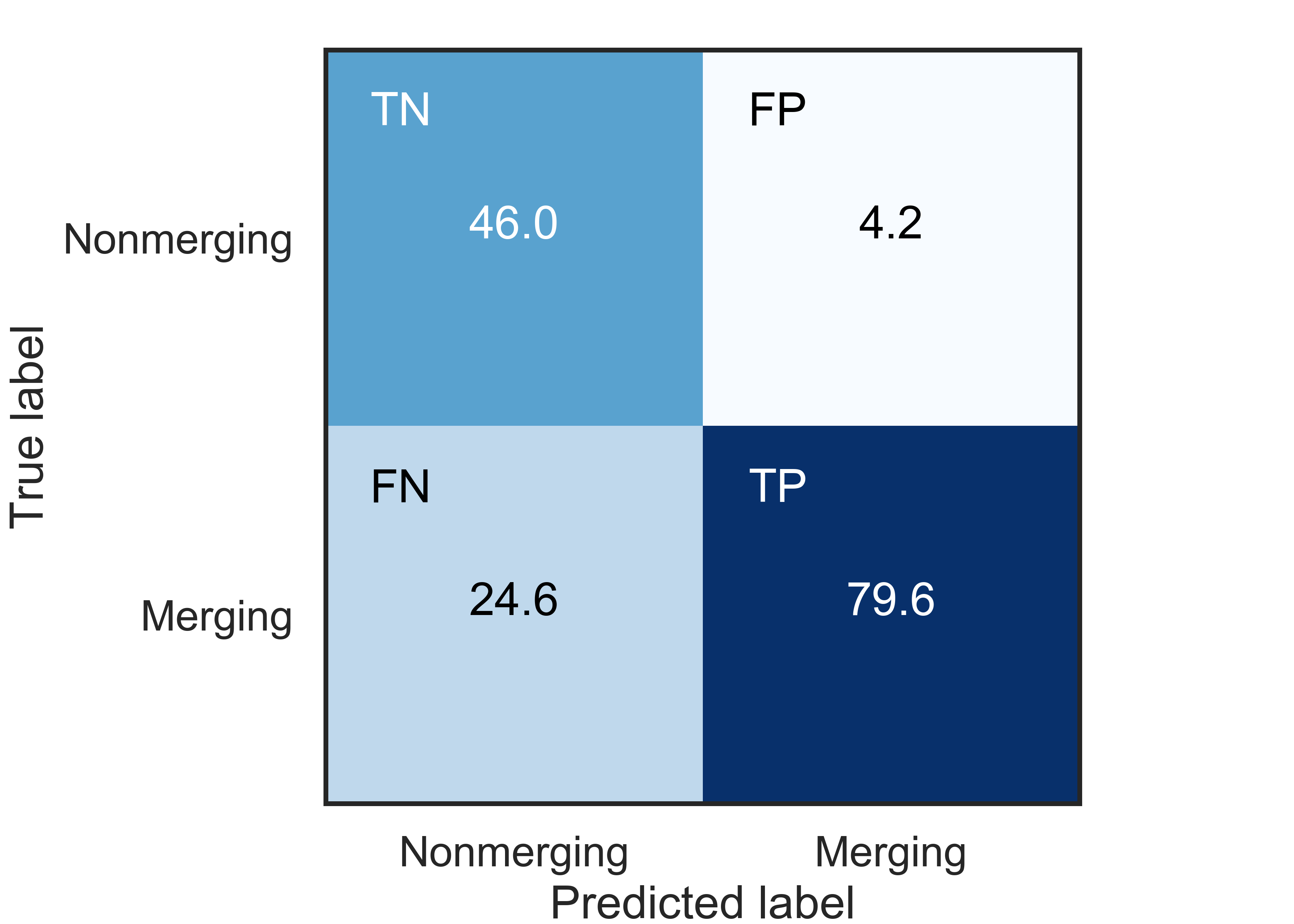}
   \includegraphics[scale=0.45, trim = 0cm 0cm 2cm 0cm, clip]{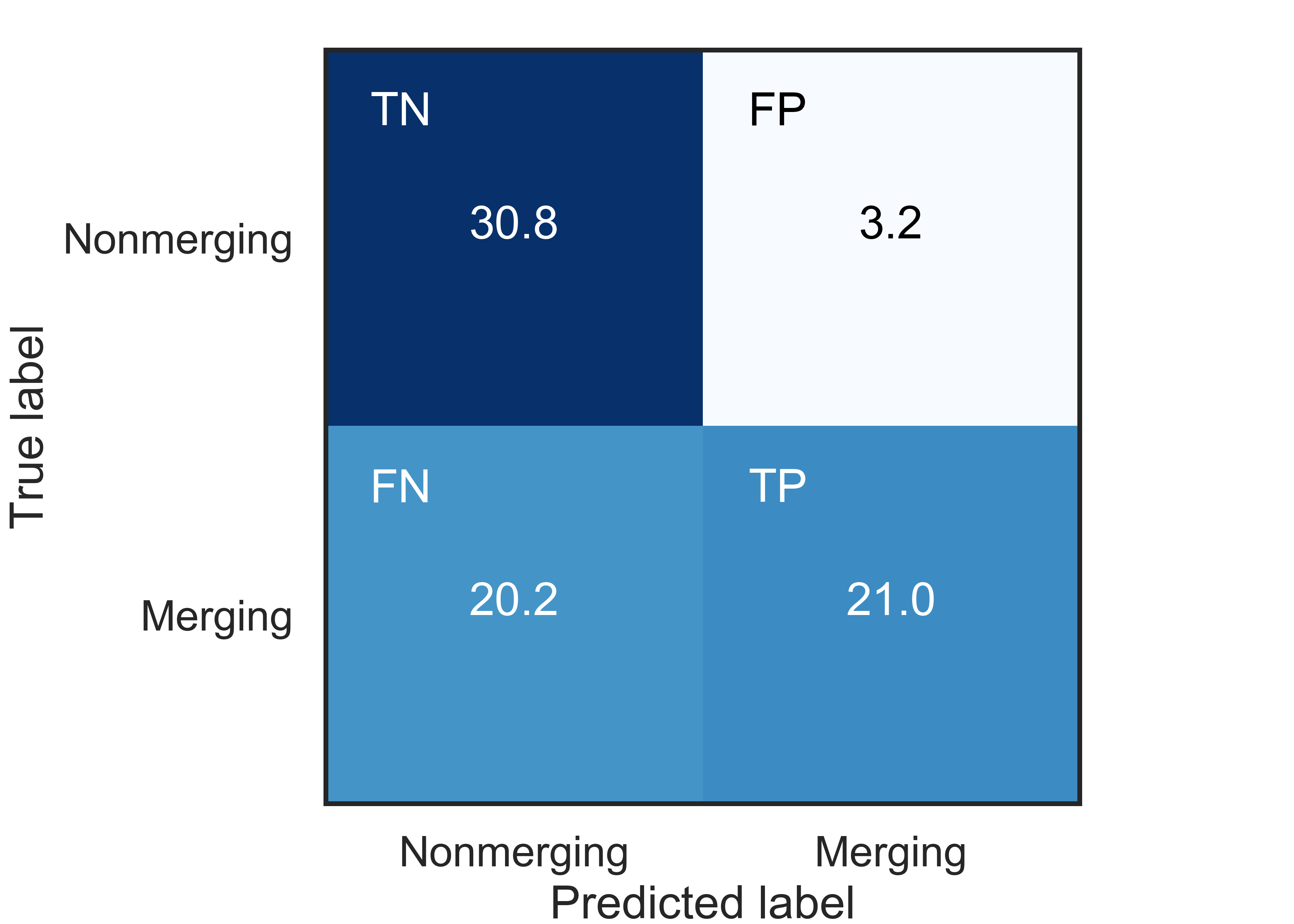}
    \caption{Confusion matrices with the number of true negatives (upper left quadrant), false positives (upper right), false negatives (lower left), and true positives (lower right) for the major merger (top) and minor merger (bottom) combined simulations. These matrices show the mean number of galaxy snapshots in each category from the five ($k=5$) different CV samples. }
    \label{confusion4}
\end{figure}

Here, we define and measure the accuracy, precision, recall, and F1 statistic of the simulations (for a review, see \citealt{Fawcett2006}). We present these results using a confusion matrix for the major and minor combined simulations in Figure \ref{confusion4}, which shows the relative fraction of known mergers and nonmergers in the cross-validation samples that are classified by the LDA as merging and nonmerging. These quantities are derived by taking the mean of the performance statistics measured on each of the cross-validation samples. We quantify the accuracy, precision, recall, and F1 score for all simulations in Table \ref{tableaccuracy}.

\begin{table}
    \centering
    \begin{tabular}{c|cccc}
        Simulation & Accuracy & Precision & Recall & F1 \\
        \hline
        All Major & 0.81 & 0.95 & 0.76 & 0.84 \\
        All Minor & 0.69& 0.87&0.51&0.64\\
        
         q0.5\_fg0.3&0.81&0.92 &0.60&0.73\\
         q0.333\_fg0.3 & 0.80&0.90&0.79 &0.84\\
         q0.333\_fg0.1 &0.80 & 0.93& 0.80&0.86\\
         q0.2\_fg0.3\_BT0.2 & 0.73& 0.83 & 0.62 & 0.71\\
         q0.1\_fg0.3\_BT0.2 &0.81 & 0.83 & 0.69 & 0.75\\
    \end{tabular}
    \caption{Accuracy, precision, recall, and F1 score for all LDA runs. We define these statistics in Equations \ref{eq:accuracy}, \ref{eq:precision}, \ref{eq:recall}, and \ref{eq:F1}, respectively. The recall value is much lower than precision in all cases because there is a much higher fraction of false negatives, or mergers that are missed by the method, yet there is a low value of contaminants, or false positives. The performance statistics of the major merger classifications are $\sim$10\% higher than the minor merger classifications.}
    \label{tableaccuracy}
\end{table}

The accuracy for a given simulation is defined as the number of correct classifications of mergers as mergers (true positives) and the number of correct classifications of nonmergers as nonmergers (true negatives) divided by the number of total classifications:

\begin{equation}
A = \frac{TP+TN}{TP+TN+FP+FN}
\label{eq:accuracy}
\end{equation}
where $FP$ is the number of false positives, or nonmerging galaxies that are classified as mergers, and $FN$ is the number of false negatives, or mergers that are classified as nonmerging. A classifier has a higher accuracy when it is able to increase the number of true classifications relative to false classifications.

Precision is defined as the number of true positive classifications over the total number of positive classifications:
\begin{equation}
P = \frac{TP}{TP+FP}
\label{eq:precision}
\end{equation}
A precise classifier maximizes the fraction of true positive classifications relative to false positives. Precision is also known as the `positive predictive value'. In this work, we seek to eliminate false positives from the sample, or nonmerging galaxies that are incorrectly classified as mergers.

Recall is defined as the number of true positive classifications over the total number of known mergers:

\begin{equation}
R = \frac{TP}{TP+FN}
\label{eq:recall}
\end{equation}
A classifier with high recall is also known as `complete' because it correctly identifies the majority of mergers as such.

Finally, we measure the F1 score or the F1 statistic, which is the harmonic mean of the recall and precision:

\begin{equation}
F1 = \frac{2 P*R}{P+R}
\label{eq:F1}
\end{equation}

F1 ranges in value from 0 to 1 and is strongly penalized if either precision ($P$) or recall ($R$) is small. We maximize the F1 statistic within the LDA during cross-validation in order to select the predictor terms that we use in the classification.

Figure \ref{confusion4} presents the number of true negatives, false positives, false negatives, and true positives (left to right, top to bottom) for the combined major and minor merger simulations. It also quantifies the accuracy, precision, recall, and F1 score. The major merger classification performs better, with an accuracy/precision/recall values of 0.81/0.95/0.76 while the minor mergers have values 0.69/0.87/0.51. The imbalance between precision and recall is due to the priors utilized in the classification (we use the priors from N19 where $f_{\mathrm{merg}} = 0.1$ for the major mergers and 0.3 for the minor mergers). We have designed the classification with these strong priors so that when it is applied to galaxy surveys (where there are many less mergers), the classifier will be more balanced. As a result, the classifier produces more false negatives than false positives when tested on the training set.

We experiment with adjusting the performance statistics of the classification, which is also known as `hyperparameter tuning'. It is possible to increase the number of false positives while decreasing the number of false negatives by either adjusting the decision boundary (i.e., the threshold of $p_{\mathrm{merg}}$) or by changing the priors. This could be a direction to pursue in future work if we find that we are no longer tolerant of false negatives or if we wish to adjust the priors on $f_{\rm{merg}}$. As a test, we adjust the priors so that $f_{\mathrm{merg}} = 0.5$ and find that it produces a similar classification, lower precision, and higher recall for the major merger classifications and results in slightly different selected predictors and higher performance statistics for the minor mergers. While the classification with adjusted priors performs better on the training and cross-validation datasets, we find that it is not a fair representation of the fraction of merging galaxies in nature, so we present the original classification with $f_{\rm{merg}} = 0.1,0.3$ in this work.

Overall, we find that the kinematic classifications generally score lower on all performance statistics than the imaging classifications. This is true for all viewpoints (see \S \ref{viewpoint} for a discussion of the effect of viewpoint on the kinematic classification). For instance, the accuracy/precision/recall/F1 value for the combined major merger run with the imaging predictors are 0.88/0.98/0.84/0.90. For the combined minor merger run with imaging predictors, the values are 0.80/0.89/0.72/0.80. Another result is that the kinematic minor merger classifications generally score $\sim$10\% lower on all statistics than the major mergers. We discuss the implications of the performance of the kinematic predictors in comparison to the imaging predictors and the performance of the major versus minor merger classifications and in \S \ref{discussaccuracy}.

\subsection{Observability Timescale}
\label{analyzeobservability}

\begin{table*}
    \centering
    \begin{tabular}{c|ccc}
      Simulation & LDA Observability Time [Gyr] & Total Merger Time [Gyr] & Observability Fraction\\

        \hline
         q0.5\_fg0.3&0.9 & 2.2  & 0.4\\ 
         q0.333\_fg0.3 &2.2 & 2.6  &0.8\\
         q0.333\_fg0.1 & 2.4 & 2.8  &0.9\\
         q0.2\_fg0.3\_BT0.2 &3.0 & 3.5  &0.9  \\
         q0.1\_fg0.3\_BT0.2 &6.6 & 9.2  &0.7\\
    \end{tabular}
    \caption{The duration of the merger, LDA observability time, total merger time, and observability fraction (LDA observability time/total merger time) for each simulation. }
    \label{time}
\end{table*}

\begin{figure*}
    \centering
    \includegraphics[scale=0.21, trim=6cm 0cm 0cm 0cm]{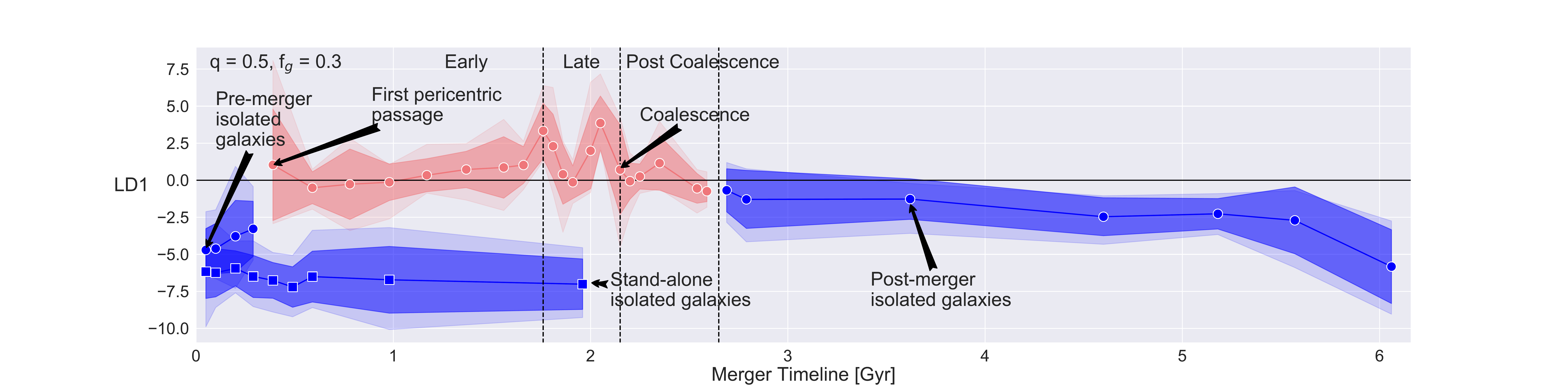}
    \includegraphics[scale=0.21, trim=6cm 0cm 0cm 0cm]{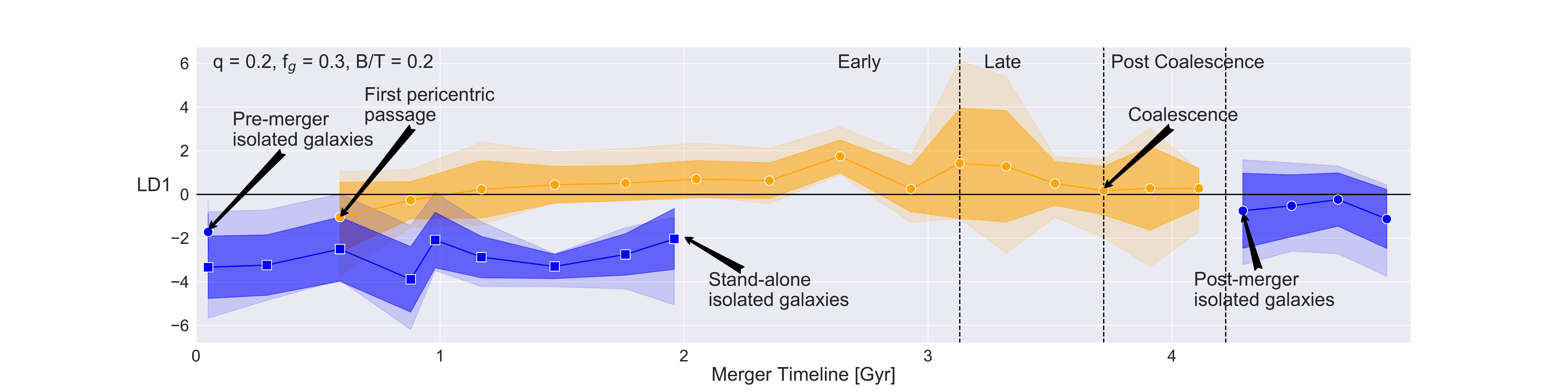}
    \caption{LD1 sensitivity with time for the q0.5\_fg0.3 (top) and q0.2\_fg0.3\_BT0.2 simulations (bottom). These two simulations are chosen because they are representative of the major and minor merger simulations, respectively. The points are the viewpoint-averaged value of LD1 for each snapshot in time along with the shaded 1$\sigma$ confidence intervals (darker shade) based on the scatter of the LD1 values for each snapshot. We also include the full range of values for each snapshot (lighter shade). We divide each plot into the early, late and post-coalescence stages of the merger. The blue lines and shaded 1$\sigma$ confidence intervals are associated with the isolated galaxies for each simulation. This includes the pre- and post-merger isolated galaxies (circles) and the stand-alone isolated galaxies (squares). The horizontal black line is the decision boundary, which marks the divide between the merging and nonmerging galaxies, or $p_{\mathrm{merg}} = 0.5$. This figure demonstrates that the major mergers have little to no overlap with the isolated galaxies, which produces a more accurate and complete classification (see \S \ref{accuracy4}). The LD1 sensitivity plots for all of the simulations will be available in an interactive figure.}
    \label{mountain4}
\end{figure*}
The LDA observability timescale is defined as the sum of all consecutive snapshots where the viewpoint-averaged LD1 value for a given snapshot is greater than zero.  We present the observability timescales for all of the simulations in Table \ref{time} along with the total merger duration for each simulation and the fraction observability, or fraction of the duration of the merger that it is observable by the LDA technique. We exclude the combined major and minor mergers from this table since they are built from mergers that progress at different rates.

All of the simulations have a relatively long timescale of observability ($2-6$ Gyr). The exception is the q0.5\_fg0.3, where the observability timescale is 0.9 Gyr due to a decline in the LD1 values for a handful of snapshots in the late stage of the merger. We present a visualization of how the mean values of LD1 change throughout the lifetime of each merger in Figure \ref{mountain4}. Here and throughout the remainder of this paper, we show the q0.5\_fg0.3 and q0.2\_fg0.3\_BT0.2 simulations as examples of major and minor mergers, respectively. This shows the viewpoint-averaged value of LD1 for each snapshot as well as the 1$\sigma$ confidence interval on this value and the total range. We also plot the decision boundary for each simulation, which falls at an LD1 value of zero (horizontal line). The minor mergers do not fall significantly above this line; even though the viewpoint-averaged LD1 values for the q0.2\_fg0.3\_BT0.2 simulation fall above the decision boundary, they overlap the decision boundary to 1$\sigma$ confidence at almost all points in time. This means that not all viewpoints are significantly above this boundary. On the other hand, the major merger simulations are significantly above this boundary for the majority of their duration. While this is not seen in Figure \ref{mountain4}, the q0.5\_fg0.3 simulation is an outlier when it comes to the LDA observability time. 

Figure \ref{mountain4} also demonstrates how sensitive the LDA classification is to the merger stage. For instance, there are a number of false negatives from the early stage of the mergers when the galaxies are more disk-like. Another key finding is that the sensitivity of the technique decays slowly during the post-coalescence and post-merger stages. We discuss the implications of the different observability timescales and the variations with time in more depth in \S \ref{discussobservability}, \ref{discuss:end}, and \ref{discuss:decoupled}.

\subsection{Where and why does the LDA fail?}
\label{fails}

\begin{figure*}
\centering
\includegraphics[scale=0.28]{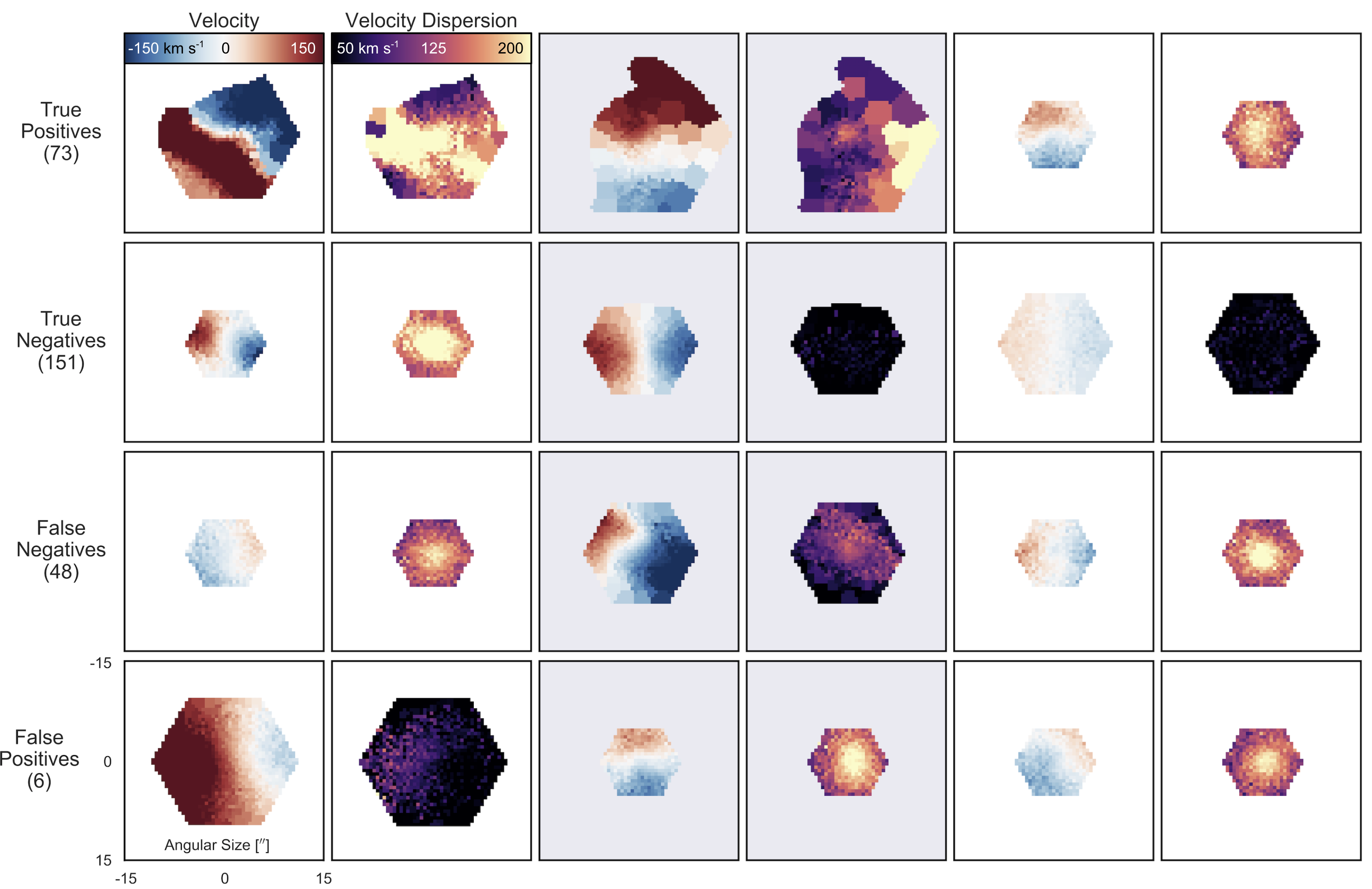}

\caption{Correct and incorrect classifications from the cross-validation sets for the q0.5\_fg0.3 simulation, which is representative of the major merger simulations. The correct classifications include true positives (first row) and true negatives (second row) and the incorrect classifications include false negatives (third row) and false positives (fourth row). We include the number of (non-repeated) galaxies in each category and three examples per row of galaxies from the cross-validation sample. The velocity and velocity dispersion maps for each example galaxy cover two consecutive panels, which is shown with alternating white and grey backgrounds. }
\label{fig:fp_fn_maj}
\end{figure*}

\begin{figure*}
\centering
\includegraphics[scale=0.28]{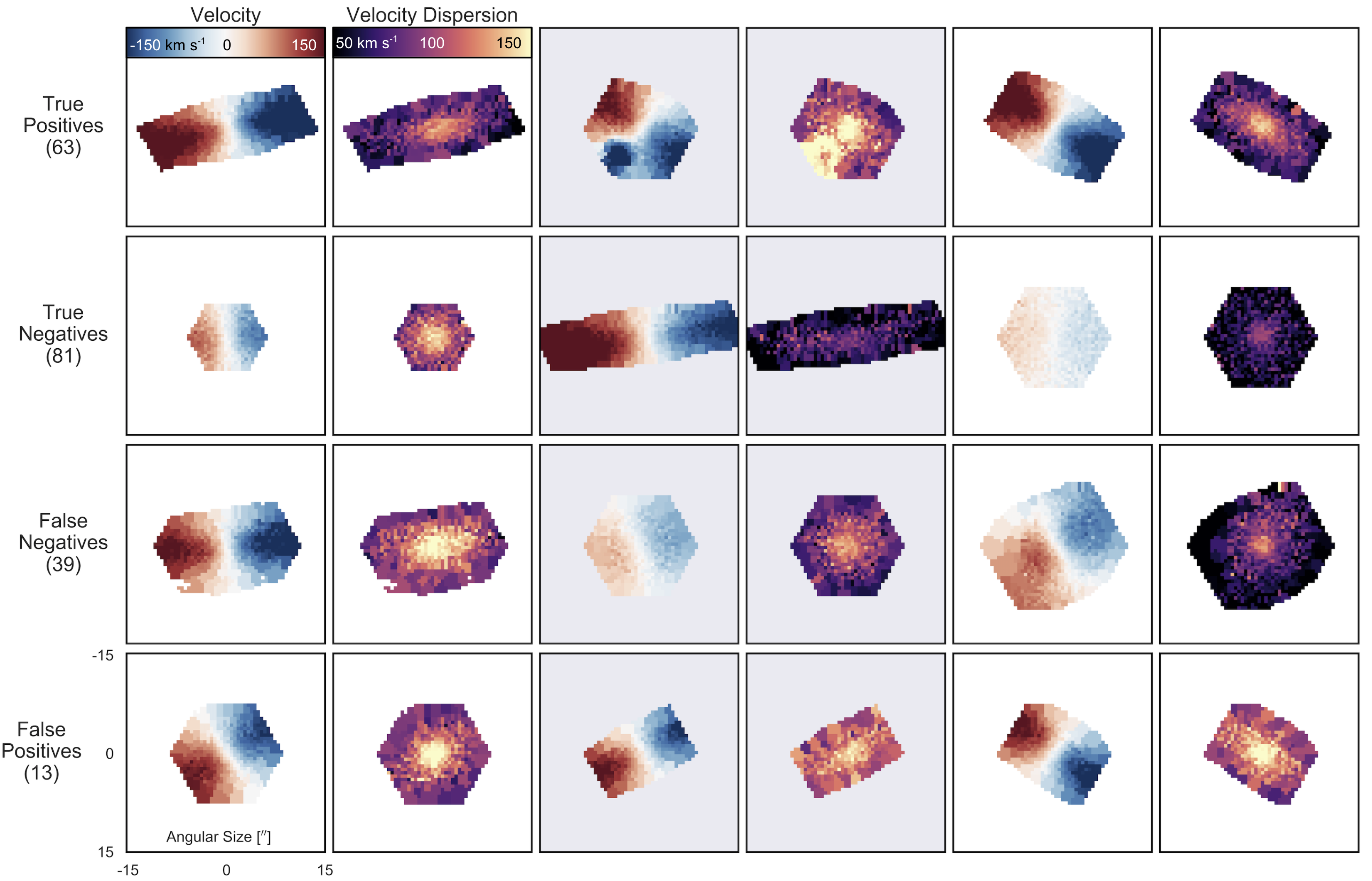}

\caption{Same as Figure \ref{fig:fp_fn_maj} but for the q0.2\_fg0.3\_BT0.2 simulation, which is representative of the minor merger simulations.}
\label{fig:fp_fn_min}
\end{figure*}

Here we summarize the factors that are most likely to lead to false classifications (false positives and false negatives) for the different simulations. Our goal is to identify the primary failure modes of the classification and assess if it is making reasonable choices. In other words, we should be concerned if our \textit{by-eye} classification disagrees with the majority of the false classifications.

We present a visual version of a confusion matrix for the q0.5\_fg0.3 and q0.2\_fg0.3\_BT0.2 classifications in Figures \ref{fig:fp_fn_maj} and \ref{fig:fp_fn_min}, respectively. These simulations are representative of the results from the major and minor mergers, respectively. We generate example velocity and velocity dispersion maps for each classification category (in rows, top to bottom: TP, TN, FN, FP) by combining the results of each iteration of the $k-$fold cross-validation and then randomly selecting example snapshots from each category.

In Figure \ref{fig:fp_fn_maj}, after a by-eye examination, it makes sense that many of the false negatives and false positives are incorrectly classified in the q0.5\_fg0.3 simulation. The false negatives (third row) are orderly rotating with relatively low velocity dispersions. These look like the pre-merger isolated population shown in the true negatives row. The majority of the false positives (fourth row) are post-merger snapshots that have kinematic disturbances. 

The incorrect classifications for the major mergers can mostly be attributed to two factors. First, the false negatives are due to a lack of disturbed features, meaning that it is difficult to correctly classify many of these snapshots as mergers. Despite these limitations, the classification does correctly identify the majority of early stage snapshots as mergers, meaning that it is out-performing the by-eye assessment in many cases. Second, kinematic disturbances that are induced by the merger persist into the post-merger stages, producing a number of false positives. These kinematic features are very similar to the features in the post-coalescence stages so it makes sense that these are commonly classified as false positives. As we discuss in more detail in \S \ref{discuss:end}, our definition of the `end' of the merger (the dividing line between post-coalescence and post-merger stages) is somewhat arbitrarily defined, and results in a number of classifications from these two stages.

In Figure \ref{fig:fp_fn_min} we find that it is more challenging to correctly classify the nonmerging and merging galaxies in the q0.2\_fg0.3\_BT0.2 simulation using a by-eye assessment. For instance, the false negatives are very visually similar to the true negatives in their kinematic features. The same is true for the false positives, which are similar to the example true positives. The exception is a number of obvious merger snapshots (we show an example of one such snapshot in the upper middle pair of panels) where the kinematics are dramatically affected. However, these disturbances are short-lived, so the majority of the merging snapshots appear like the example in the upper right corner of the diagram. This diagram illustrates the crux of the problem for the minor mergers. While the LDA is able to pick up on a number of subtle features (such as stellar bulge enhancements), it ultimately struggles with a number of limitations related to the lack of identifiability of all of the stages of the merger as such. These challenges contribute to a minor merger classification that has lower performance statistics than the major merger classification.

Overall, the LDA is not misclassifying obvious (by-eye) mergers or nonmergers. The lack of identifiability of mergers/nonmergers given their kinematic maps is therefore the largest challenge for this technique. Other work highlights this same challenge with kinematic predictors; \citet{Hung2016} find that a significant fraction of merging galaxy kinematics remain indistinguishable by-eye relative to the nonmerging kinematics. This indicates something fundamental about galaxy kinematics: namely, that we are not missing obvious features and instead that merging galaxies are often indistinguishable from nonmergers. 

\subsection{The role of viewing angle in the classification}
\label{viewpoint}
It is well known that many kinematic predictors, such as \lambdare, are correlated with galaxy inclination (e.g., \citealt{Cappellari2007,Emsellem2011,Harborne2019}). In this section we examine how the viewing angle, which is a proxy for inclination, affects the kinematic predictors and ultimately, the LDA. 

As described in \S \ref{simdeets}, there are seven isotropically distributed viewpoints (0-6) at each snapshot. Critically, the inclinations are not an exact match between the stand-alone isolated galaxies and the merging galaxies. For instance, viewpoints 0 and 4 are the most face-on viewpoints for the merging galaxies, but viewpoints 4, 5, and 6 are the most face-on for the stand-alone isolated galaxies.

\begin{figure*}
    \centering
    \includegraphics[scale=0.55]{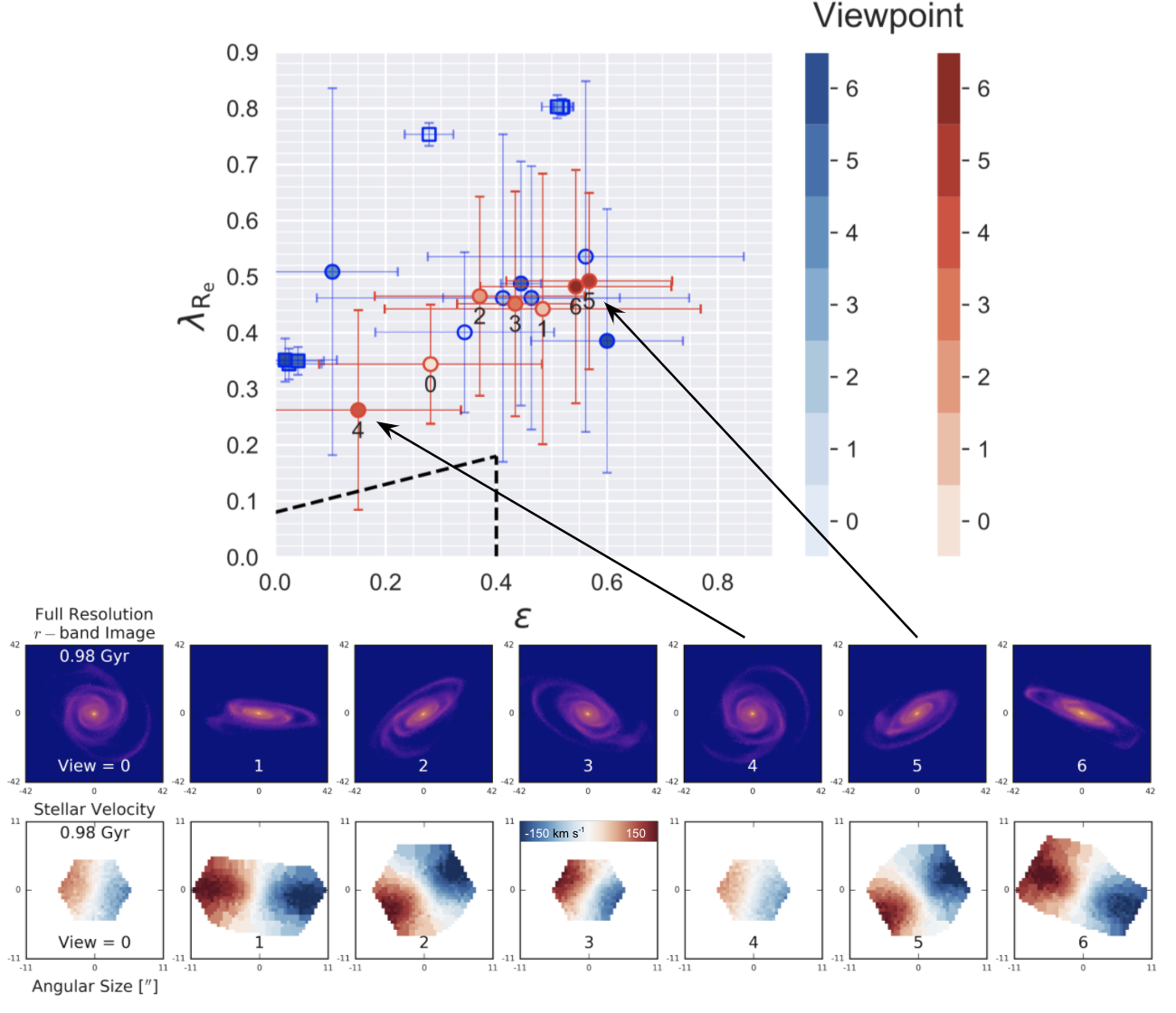}
    \caption{Distribution of the mean values of \lambdare\ and $\epsilon$ as a function of viewpoint (top) and the full resolution $r-$band images (middle) and stellar velocity maps (bottom) for all of the different viewpoints from a snapshot in time for the q0.5\_fg0.3 simulation. The more face-on viewpoints (i.e., 4), tend towards lower values of \lambdare\ and $\epsilon$, while the more edge-on viewpoints (i.e., 5) tend to have a larger \lambdare\ value. We also include error bars to demonstrate the standard deviation of the spread at each viewpoint from all of the different moments in time of this simulation. While there is a relationship between inclination and \lambdare, the trend is borderline significant.  \textbf{This is consistent with the trend of varying \lambdare\ and $\epsilon$ values with viewing angle from \citet{Emsellem2011}}. }
    \label{fig:inc_lambdar}
\end{figure*}

We first explore how inclination affects the \lambdare\ predictor in Figure \ref{fig:inc_lambdar}, where  \lambdare\ increases as the galaxy inclination increases. For instance, viewing angles 0 and 4 are the most face-on and they also have the lowest values of \lambdare. When the 1$\sigma$ errorbars are taken into consideration, the difference in \lambdare\ values is marginally significant. \textbf{This is fully consistent with the results from  \citet{Emsellem2011}, who predict that the measured \lambdare\ and $\epsilon$ values of an axisymmetric rotating oblate spheroid vary with viewing angle (see Figure 3 of \citealt{Emsellem2011}).} These errorbars are the standard deviation in \lambdare\ values from all of the different moments in time of the merger. We observe a larger difference in the \lambdare\ values as a function of time (\S \ref{approxspin}) than as a function of viewpoint.   
\begin{figure*}
    \centering
    \includegraphics[scale=0.42]{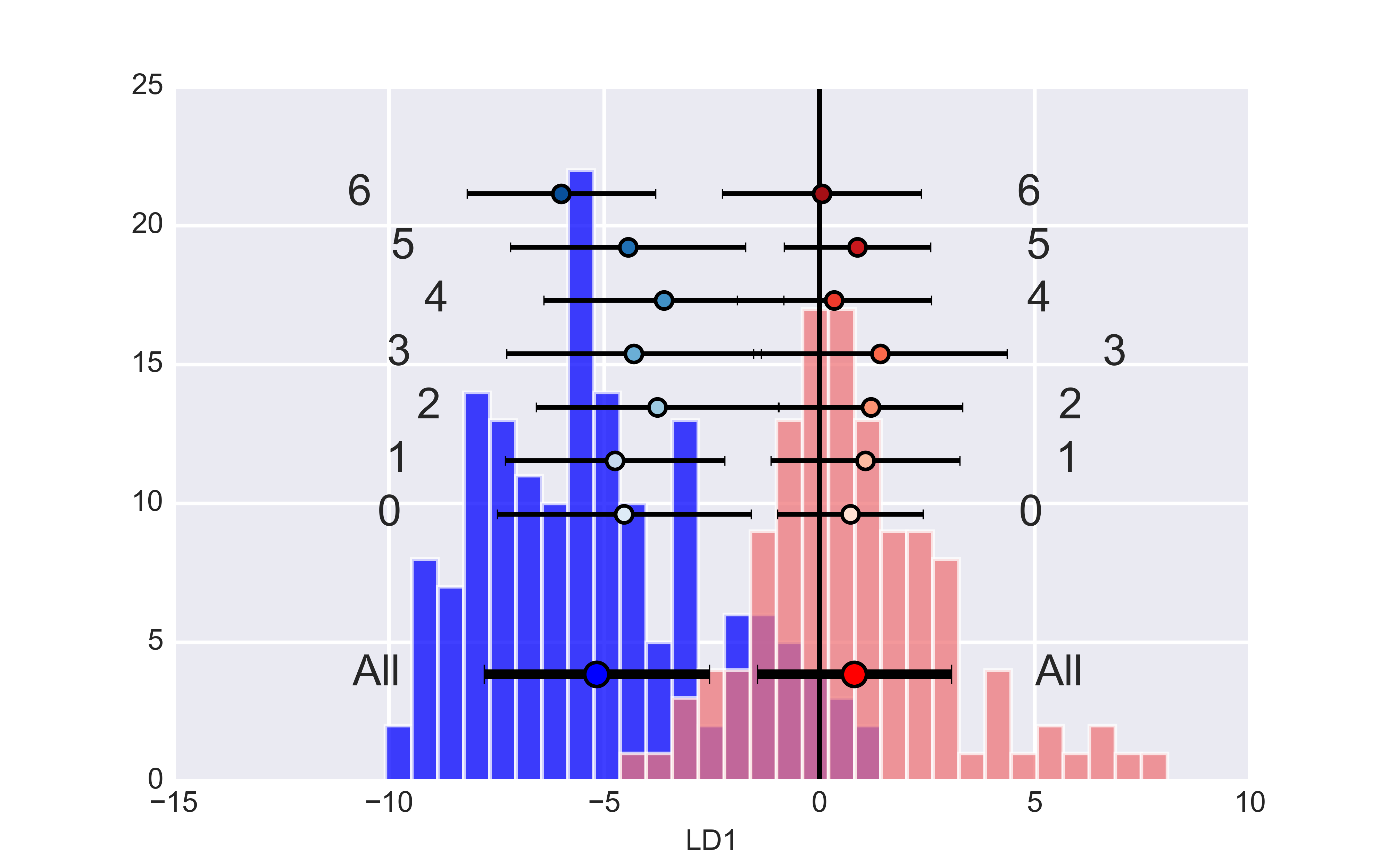}
    \includegraphics[scale=0.42]{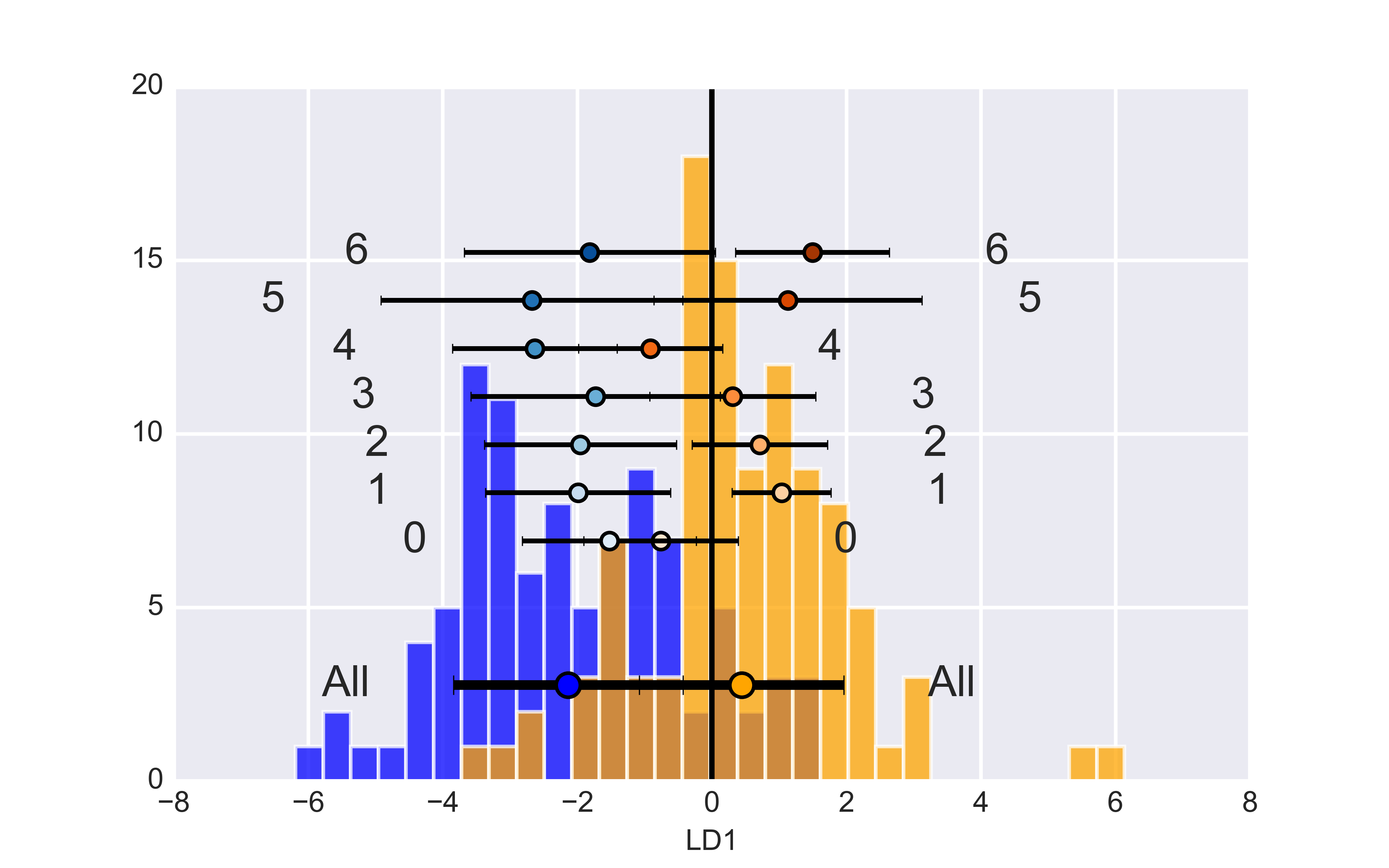}
    \caption{As in Figure \ref{histograms_major}, histograms of the LD1 value for the q0.5\_fg0.3 simulation (left) and the q0.2\_fg0.3\_BT0.2 simulation (right). We also overplot the time-averaged LD1 values for each viewpoint and errorbars to demonstrate the 1$\sigma$ variation amongst these values for all moments in time. There is less variation as a function of viewpoint than as a function of time (shown in Figure \ref{mountain4}).} 
    \label{fig:stackedhist}
\end{figure*}

We next investigate how the LDA classification changes as a function of viewpoint. To visualize this, we plot the distribution of LD1 values in Figure \ref{fig:stackedhist}. We include the histograms of the LD1 value for the nonmerging (blue) and merging (red and orange) galaxies from both the q0.5\_fg0.3 simulation (left panel) and the q0.2\_fg0.3\_BT0.2 simulation (right panel). We then overplot the mean and standard deviation of the LD1 values for all snapshots of each specific viewpoint. Focusing on the mean LD1 values for the merging sample, we can determine if the LDA is varying as a function of viewpoint.  

Focusing first on the left panel of this figure, which is for the q0.5\_fg0.3 simulation, the means for both the nonmerging and merging galaxies are not significantly different as a function of viewing angle. In fact, the mean LD1 values are more similar than the variation we observe in LD1 as a function of time in Figure \ref{mountain4}. The implication is that the major merger LDAs are fairly robust to viewing angle. For the q0.2\_fg0.3\_BT0.2 simulation, we observe slightly more variation in the LD1 distribution as a function of viewpoint, and the most face-on viewpoints (0 and 4) have lower LD1 values, which result in more false negative detections at these viewpoints. 

To further quantify if the LDA is changing as a function of viewpoint, we iteratively drop the merging galaxies at each viewpoint from the analysis and rerun the classification for the q0.5\_fg0.3 and q0.2\_fg0.3\_BT0.2 simulations. If the classification changes, this could indicate that a given viewing angle and/or inclination is significantly more or less accurate than the other viewpoints, which would point to inclination itself being the primary driver of this difference.

From here on, we determine that the LDA is `significantly different' from the fiducial run if either of the following criteria are met: first, the majority of the selected predictors in the top four selected terms must change or second, the performance statistics in Table \ref{tableaccuracy} must change by more than 10\% on average. This quantification of a significantly different classification applies to this section, where we explore the role of different viewing angles, and also to \S \ref{limitssnz} where we experiment with changes in the data reduction (i.e., changing the S/N or redshift of the simulated galaxies).

When we rerun the LDA classification for the q0.5\_fg0.3 simulation iteratively without each viewpoint, the LDAs are not significantly different than the fiducial run. This confirms our findings in Figure \ref{fig:stackedhist}. For the q0.2\_fg0.3\_BT0.2 classification, we find that when viewpoints 2, 5, and 6 are absent, the classification is significantly different with lower performance statistics and different selected predictors. Our interpretation is that the minor mergers are best identified when the secondary nuclei is within the field of view, which happens most often in viewpoints 2, 5, and 6. Therefore, the significant changes to the classification as a function of viewpoint both in Figure \ref{fig:stackedhist} and in the rerun of the LDA without these viewpoints can be attributed to the chance positioning of the secondary galaxy as a function of viewpoint. This means that inclination-related effects on the intrinsic kinematic properties of the primary galaxy are not primarily responsible for the differences in the q0.2\_fg0.3\_BT0.2 LDA as a function of viewpoint. As a final note, in Appendix \ref{fair} in our discussion of between-class biases, we introduce the inclination itself as a predictor in the LDA. We ultimately determine that $\epsilon$, which we use as a proxy for inclination, is not an important predictor. This further supports the finding that changes in the kinematic predictors purely due to inclination effects are not biasing the LDA classification itself.

To conclude, we have determined that while the kinematic predictors themselves can vary as a function of viewing angle and/or galaxy inclination, the LDA classification is only sensitive to viewpoint in the sense of the relative positioning of the secondary nucleus relative to the line of sight vector.

\subsection{Limitations of the technique in $z$ and S/N}
\label{limitssnz}
\begin{figure}
     \centering
     \includegraphics[scale=0.47, trim = 4cm 0.75cm 1cm 1.25cm, clip]{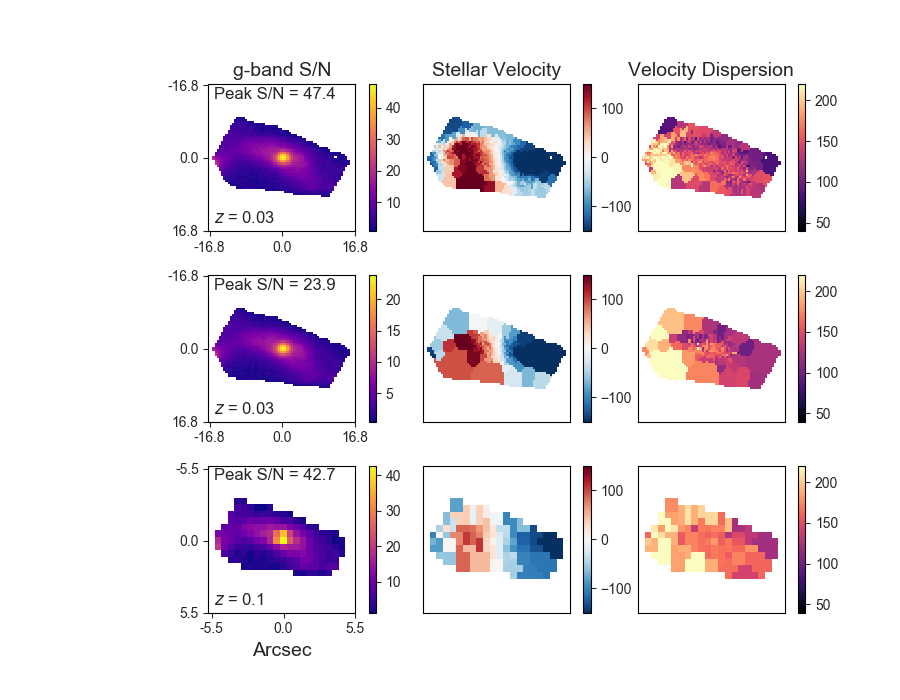}
     \caption{The $g-$band S/N (left), stellar velocity (middle), and velocity dispersion (right) maps from a snapshot of the q0.5\_fg0.3 simulation. We have decreased the S/N by a factor of two (second row) and redshifted the galaxy to $z = 0.1$ (third row) to demonstrate the point at which the classification begins to change. The classification has a higher failure rate when the S/N is decreased by a factor of two, mostly due to the sparsity of the Voronoi bins. Additionally, when the spaxel size is increased to mock a galaxy that is redshifted to $z=0.1$, the classification begins to change, as this is the point at which the larger-scale kinematic features are distorted by the large spaxel size.}
     \label{fig:comparison}
 \end{figure}

As with the imaging identification technique in N19, the kinematic technique is sensitive to both S/N and resolution, meaning that as the S/N decreases in the spectra and/or as the redshift of the galaxy increases, the technique will undergo significant changes. We test the sensitivity by decreasing the S/N and by moving the mock galaxies to higher redshift.

To test how sensitive the classification is to decreased S/N, we decrease the average S/N of q0.5\_fg0.3 simulation by factors of 1.5 and 2. In Figure \ref{fig:comparison} we compare a snapshot with S/N that has been decreased by a factor of two to the same snapshot from the fiducial run. When the S/N is decreased by a factor of 2, the classification is significantly different. While many of the predictors stay the same, the performance statistics decrease overall and there is an increase in the number of false negatives during the early stages of the merger. When the S/N is decreased, the Voronoi bins increase in size in the exterior regions of the galaxy. This obscures the large-scale kinematic features, which lowers the performance of the classification. We predict that MaNGA galaxies with low S/N may therefore be more likely to be misclassified.

There are a couple of approaches we plan to explore when classifying MaNGA galaxies with different S/N ratios. One option is to implement a S/N cut when we apply the fiducial classification to the MaNGA galaxies. However, MaNGA galaxies with lower S/N need not be excluded from classification. Instead, another option is to use the classification and completeness correction from the lower S/N ratio LDA run to classify these galaxies separately. We predict that we should be able to classify the majority of the MaNGA sample since the fiducial (non-decreased) S/N of the simulation suite is representative of the MaNGA sample. 

\begin{figure}
     \centering
     \includegraphics[scale=0.63, trim=5.5cm 8cm 1.4cm 2cm, clip]{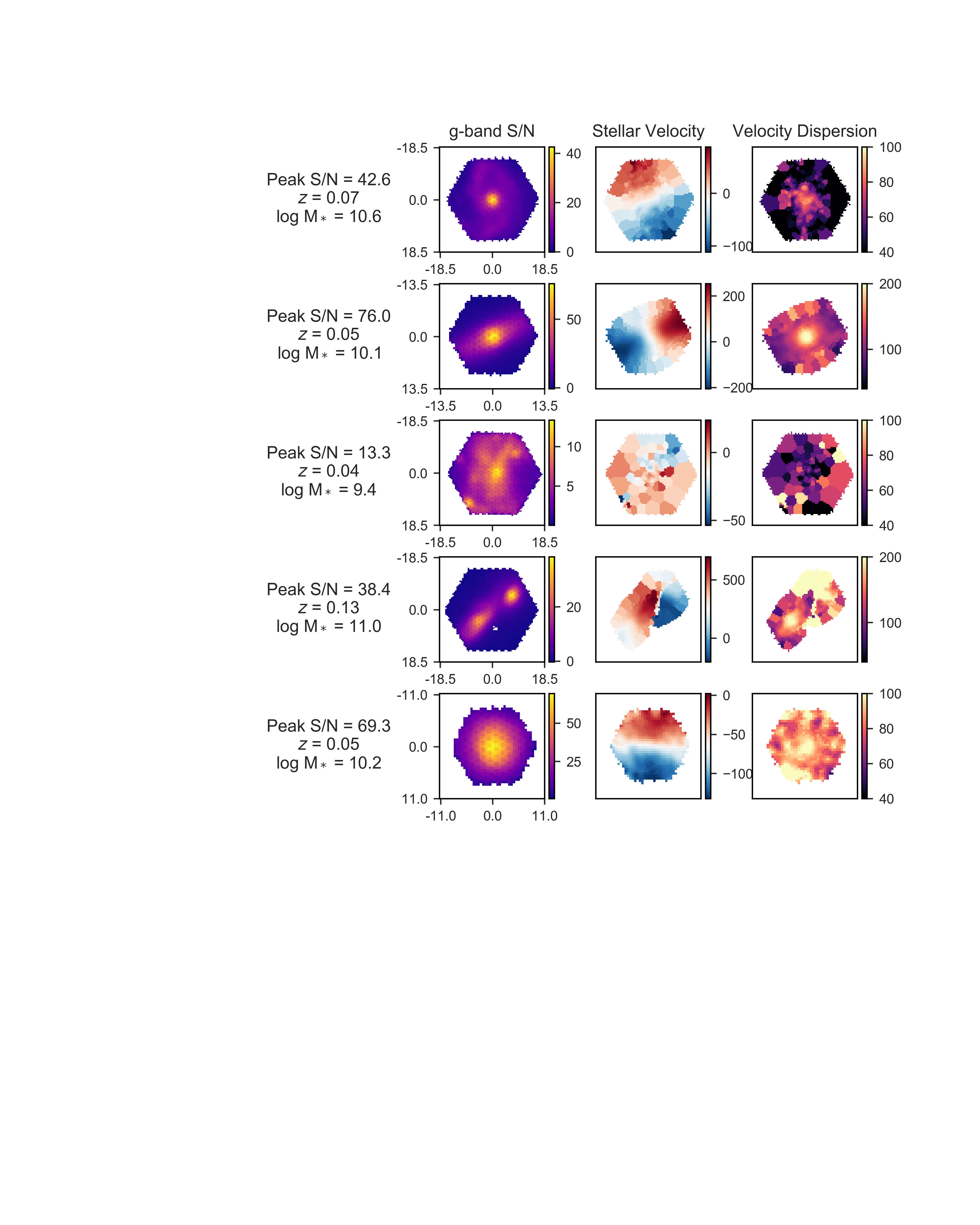}
     \caption{Same as Figure \ref{fig:comparison} but for a sample of MaNGA galaxies that span a range in surface brightness, redshift, and stellar mass. At low S/N (i.e., first and third rows) the velocity maps have large Voronoi bins and some kinematic predictors will be difficult to measure. As discussed in the text, the higher redshift galaxies (first and fourth row) in MaNGA also tend to be larger and more massive.}
     \label{fig:manga}
 \end{figure}

For comparison's sake, in Figure \ref{fig:manga} we present a sample of MaNGA galxies that span a range in surface brightness, redshift, and stellar mass. The approximate stellar mass is from the NSA catalog and is estimated using the kcorrect code (\citealt{Blanton2007}). The simulated galaxies, which have stellar masses log M$_* \sim$ 10.6, are intermediate mass galaxies relative to the full MaNGA sample. 

We also experiment with increasing the S/N by a factor of 2. The classification undergoes minimal changes, with a slight increase in the performance statistics. When the S/N is higher, we predict that the current classification will be better able to determine if a galaxy is merging. Since the selected predictors do not significantly change, we can conclude that the fiducial run is identifying all relevant kinematic features. 

The mock galaxies are placed at a redshift of $z$ = 0.03, which is the median redshift of galaxies observed by MaNGA. In order to understand the limitations of the identification over the full range of redshifts for the MaNGA survey (0.01 < $z$ < 0.15), we experiment with increasing the redshift of the mock galaxies. To do this, we increase the spaxel size from $0\farcs5$ to $1\farcs0$ and $1\farcs5$ and we increase the PSF size to $5\farcs0$ and $7\farcs5$. This mimics the effects of moving the simulated galaxies to a redshift of $z=0.07$ and $z=0.1$, respectively. When we artificially redshift the galaxies we do not introduce cosmological dimming, meaning that the galaxies have the same S/N as the sample at $z = 0.03$. This is because we want to understand the effects of the apparent size of galaxies independently of S/N effects. 
 
The classification does not change significantly when the galaxies are placed at a redshift of $z=0.07$. At $z=0.1$, the classification is significantly different; the number of misclassifications increases and the selected terms are different. However, these results are based on a galaxy-galaxy merger where each galaxy has a stellar mass on order $10^{10}M_{\odot}$. While it is a valid conclusion that the technique will struggle on an intermediate mass galaxy at $z=0.1$, the MaNGA sample does not tend to include this type of galaxy. Instead, MaNGA is designed to maintain roughly uniform coverage in log $M_*$ and radial coverage, meaning that higher mass galaxies ($> 10^{11} M_{\odot}$), which are more luminous and have larger angular sizes (i.e., the fourth row of Figure \ref{fig:manga}), are observed primarily at higher redshift, somewhat alleviating this concern (\citealt{Bundy2015,Wake2017}).

\subsection{Limitations of the technique in stellar mass and B/T}
\label{results:mass}
The simulation suite of merging galaxies is limited in stellar mass and B/T ratio. The mergers can all be characterized as intermediate-mass disk-dominated galaxies that span a range of $3.9 - 4.7 \times 10^{10}$ M$_{\odot}$ in stellar mass and 0 - 0.2 in initial B/T ratio. These limitations are especially important given that many of the leading kinematic predictors (\lambdare, \muv, \mus, \sigv, and \sigs) are related to the intrinsic kinematic properties of galaxies. For instance, \mus\ is a proxy for stellar mass, so we are skeptical if this classification can be reliably applied to galaxies that differ in properties, i.e., bulge-dominated elliptical galaxies. 

One possible approach to circumvent these concerns is to remove these predictors from the classification. We rerun the LDA for all simulations without the \lambdare, \muv, \mus, \sigv, and \sigs\ predictors and find that the performance significantly decreases. Specifically, the accuracy, recall, and F1 score of the major mergers decrease by $20-50$\%. This is unsurprising given that the leading predictors presented in Table \ref{tab:predictors} include all of the predictors that are tied to the intrinsic kinematic properties of galaxies (\lambdare, \muv, \mus, \sigv, \sigs). Interestingly, when the intrinsic kinematic predictors are excluded, the performance of the minor merger simulations is not significantly affected. In fact, the performance of the major merger simulations is comparable or worse than that of the minor merger simulations. This highlights that the major mergers undergo a more dramatic global transformation during the merging process, which is reflected in the intrinsic kinematic properties of the remnant galaxy.

Since removing these predictors significantly decreases the performance of the classification, we choose to include all predictors in this work and to attach the following caveat to this paper: Since this analysis focuses on kinematic predictors that are sensitive to intrinsic galaxy properties, we advise against applying this classification to all galaxy types in MaNGA. In \S \ref{discuss:extrapolate}, we discuss possible strategies for carefully applying the classification to MaNGA galaxies.

\section{Discussion}
\label{discuss4}
In the discussion portion of this paper, we consider the implications of the kinematic LDA classifications for merging galaxies. We focus on the individual LD1 coefficients in \S \ref{discuss:useless} where we examine why some of the kinematic predictors that have been useful in the past are not informative in this technique. We then examine the most important kinematic predictors in \S \ref{discusssign}. In \S \ref{discussmass} we explore the impact of mass ratio on the stellar kinematics of mergers. We consider the physical meaning of the interaction terms and their importance to the classification in \S \ref{discuss:interaction}. We examine the observability timescale of the kinematic LDA technique and how the observability of a merger varies with time in \S \ref{discussobservability}. We specifically focus on the definition of the `end' of a merger in \S \ref{discuss:end} and the kinematics of the merger remnants in \S \ref{discuss:decoupled}. In \S \ref{discussaccuracy} we compare the performance of the imaging classifications to the kinematic classifications. Finally, we end with a note on applying this technique to MaNGA IFS observations in \S \ref{discuss:extrapolate}.

\subsection{Why are some traditionally utilized kinematic predictors not useful in this classification?}
\label{discuss:useless}
Some of the kinematic predictors that are traditionally utilized to identify merging galaxies are uninformative in this analysis. In this case, `uninformative' means that the predictor is discarded during the RFR term selection steps or that it has a small relative coefficient value in the LDA. The uninformative predictors are $\Delta$PA, \vasym, \sigasym,  $\Delta x_{V}$, and $\Delta x_{\sigma}$.  

\subsubsection{The misalignment between the kinematic PA and imaging PA ($\Delta$PA) is most sensitive to galaxy inclination}

A small fraction of the most dramatic mergers have significantly disturbed stellar kinematic maps. However, this does not translate to a global kinematic PA that is misaligned from the PA in imaging due to two factors. First, many of the warps seen in the stellar kinematic disks are symmetric, which can produce a global kinematic PA that is not misaligned. Second, both the PA from the kinematics and the PA from imaging are not well determined during the most disturbed stages of the merger. This contributes to random deviations around a low $\Delta$PA value.

\subsubsection{The asymmetry in the velocity and velocity dispersion maps (\vasym\ and \sigasym) are only sensitive to the most disturbed times in the major mergers}

 \begin{figure}
    \centering
    \includegraphics[scale=0.5, trim=0cm 0cm 0cm 0cm, clip]{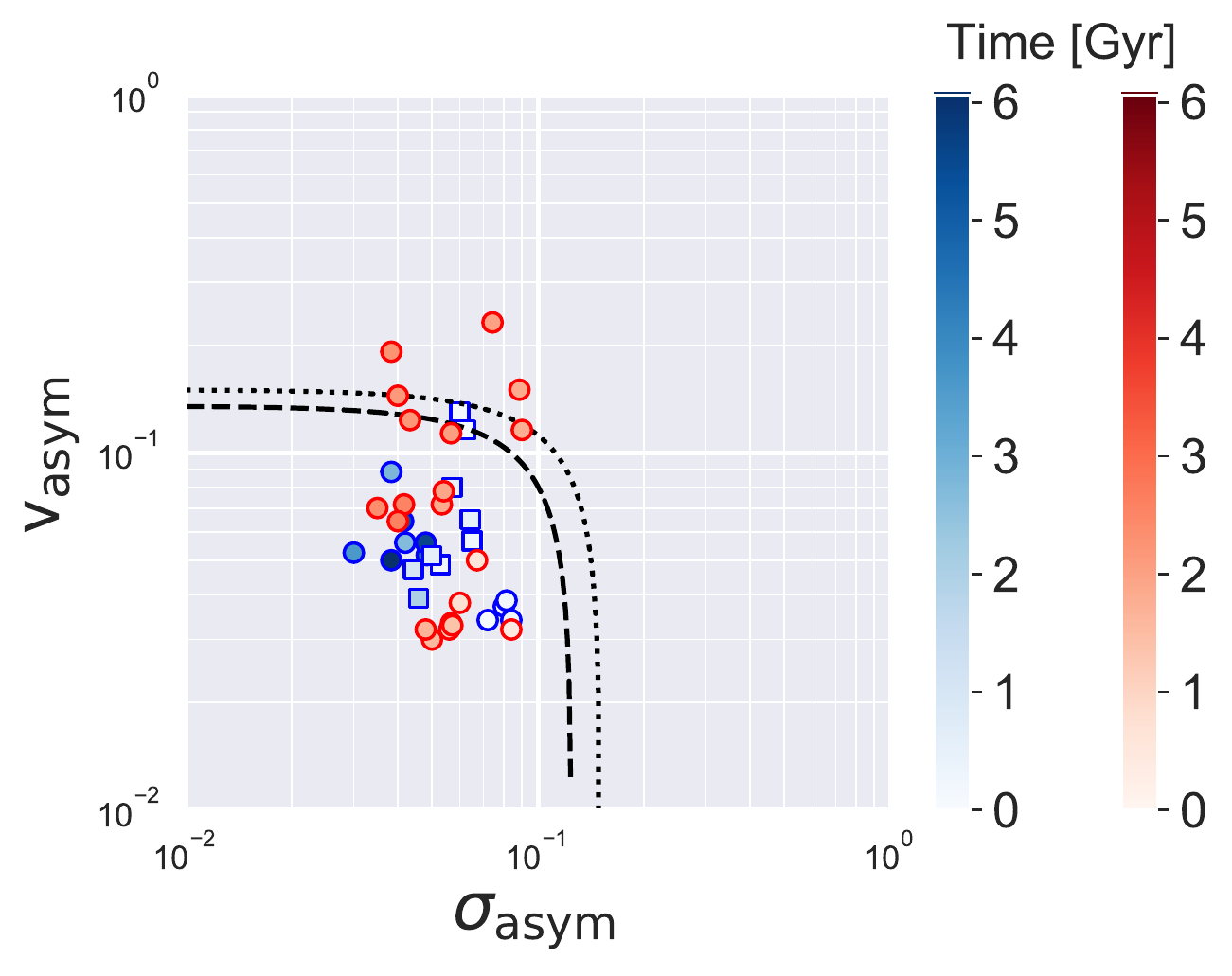}
     \includegraphics[scale=0.5, trim=0cm 0cm 0cm 0cm, clip]{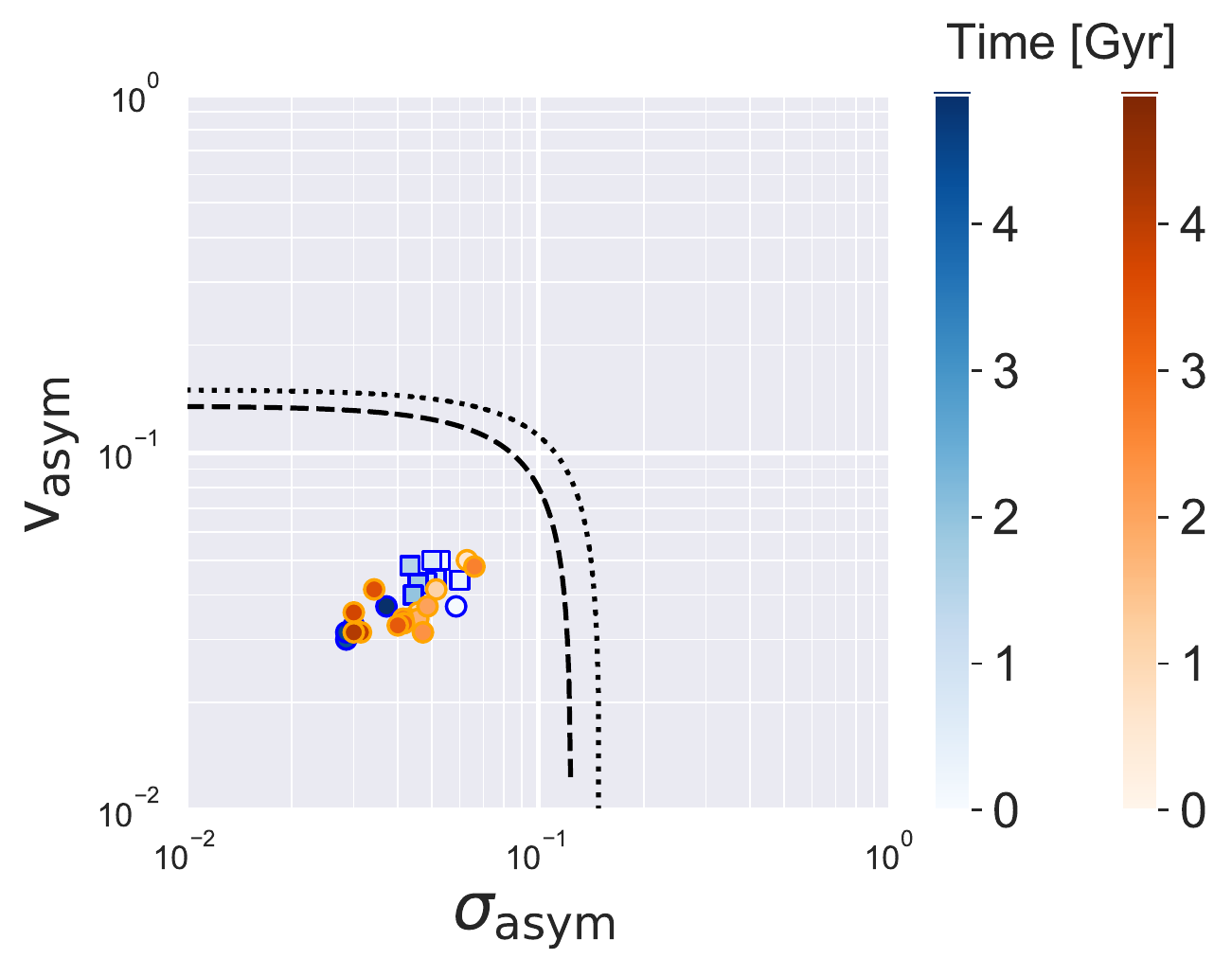}
    
    \caption{Time evolution of the merging (red and orange) and matched nonmerging (blue) galaxies for the q0.5\_fg0.3 (top, red) and q0.2\_fg0.3\_BT0.2 (bottom, orange) on the \vasym-\sigasym\ diagram. The blue squares indicate the matched isolated sample of galaxies, while the blue circles are the pre- and post-mergers. Here, we show the pre-standardized predictor values. The time is zero as the simulations begin and progresses in Gyr. The $K_{\mathrm{asym}} = 0.15$ (dotted) and 0.135 (dashed) threshold lines are included from \citet{Hung2016} and \citet{Bellocchi2012}, respectively, where galaxies above the diagnostic lines are classified as merging.}
    \label{fig:va_sa}
\end{figure}
Previous work with the gas kinematics of simulated and observed mergers finds that merging galaxies have enhanced values of both \vasym\ and \sigasym\  (e.g., \citealt{Shapiro2008,Bellocchi2012,Hung2016,Bloom2017}). These studies define a threshold value in $K_{\mathrm{asym}}$ to identify merging galaxies, where $K_{\mathrm{asym}} = \sqrt{\mathrm{v}_{\mathrm{asym}}^2 + \sigma_{\mathrm{asym}}^2}$. For instance, \citet{Bellocchi2012} study local luminous infrared galaxies (LIRGs) and find a threshold value of $K_{\mathrm{asym}} > 0.135$. \citet{Hung2016} define a $K_{\mathrm{asym}}$ threshold of 0.15 for the galaxies in their work, which is calculated from the velocities of gas particles from simulated \sunrise\ mergers.

The \sigasym\ and \vasym\ predictors are ultimately unimportant in our analysis. We present the viewpoint-averaged values of \sigasym-\vasym\ in Figure \ref{fig:va_sa} for the q0.2\_fg0.3\_BT0.2 and the q0.5\_fg0.3 simulations. We include the $K_{\mathrm{asym}}$ diagnostic lines used to identify mergers from \citet{Bellocchi2012} and \citet{Hung2016}. The top panel of the plot demonstrates that there is minimal time evolution for the predictor values for the minor mergers. The predictor values for the major mergers are only slightly enhanced, falling above the diagnostic line from \citet{Hung2016} for a few snapshots during the late stages of the merger. While Figure \ref{fig:va_sa} presents the predictor values in log space, the standardized predictor values of \sigasym\ and \vasym\ used to construct the classification also have minimal separation between the merging and nonmerging populations. The insensitivity of any of the mergers to enhancement in the values of these predictors results in their exclusion during the RFR selection step. 

Ultimately, the \vasym\ and \sigasym\ predictors are unimportant in this work because \vasym\ is only elevated for specific points in time during the late phases of merging. They are more useful in studies such as \citet{Bellocchi2012}, which focus on LIRGs, which are prototypical gas-rich major mergers, and \citet{Hung2016}, where the simulated galaxies include gas-rich major mergers. Additionally of interest, \citet{Hung2016} find that the mergers in their sample exceed the $K_{\mathrm{asym}}$ value only during the `strong interaction' or late stage of merging.

\subsubsection{The offset between the center of the velocity and velocity dispersion maps and the imaging center ($\Delta x_V$ and $\Delta x_{\sigma}$) are not very sensitive to mergers}

We design the 
$\Delta x_V$ and the $\Delta x_{\sigma}$ statistics to identify galaxies that have offsets in their kinematic centers. We find that these values are most elevated during the late stages of the merger, where there are two visible nuclei. However, since the kinematic maps are disky throughout the merger and not dramatically disturbed at most stages, these two statistics are not noticeably elevated for the duration of the merger and are therefore relatively unimportant. 

Statistics like these have been used in the past for galaxies such as NGC 4473, which is a `double sigma (2$\sigma$)' galaxy, meaning that it has two peaks in its 2D stellar velocity dispersion map that are aligned with the photometric major axis of the galaxy (\citealt{Krajnovic2011}). This type of velocity dispersion map is rare in observations (e.g., \citealt{Krajnovic2011}) and is associated with a co-addition of a counter-rotating stellar disk, produced by retrograde 1:1 mass ratio mergers in simulations (e.g., \citealt{Jesseit2007,Bois2011}). Therefore, these statistics may not be as useful for identifying the more typical types of mergers, which are often more unequal in mass ratio and do not occur under idealized conditions.

\subsection{What can we learn from the most important LDA predictors about the kinematics of stars in mergers?}
\label{discusssign}

Here we examine the most important predictors in the LDA for all simulations and make connections between these predictors and the dynamical evolution of the stars during a merger. We split the discussion by predictor and for brevity we focus only on the leading predictors presented in Table \ref{table:LDAall}. The top predictors include \lambdare, \mus, \sigs, \hthrees, and \hfourv.

\subsubsection{The approximate spin parameter tracks a global `slow-down' in the velocity maps of the major mergers}
\label{approxspin}

The approximate spin parameter (\lambdare) is a key predictor for all merger simulations and is especially important for the major mergers. The angular momentum of the merging galaxies is therefore significantly different than that of the nonmerging population; this effect is more apparent for the major mergers. In this section we examine how \lambdare\ changes with time for the various simulations and how this compares to previous work.

\begin{figure}
    \centering
    \includegraphics[scale=0.5]{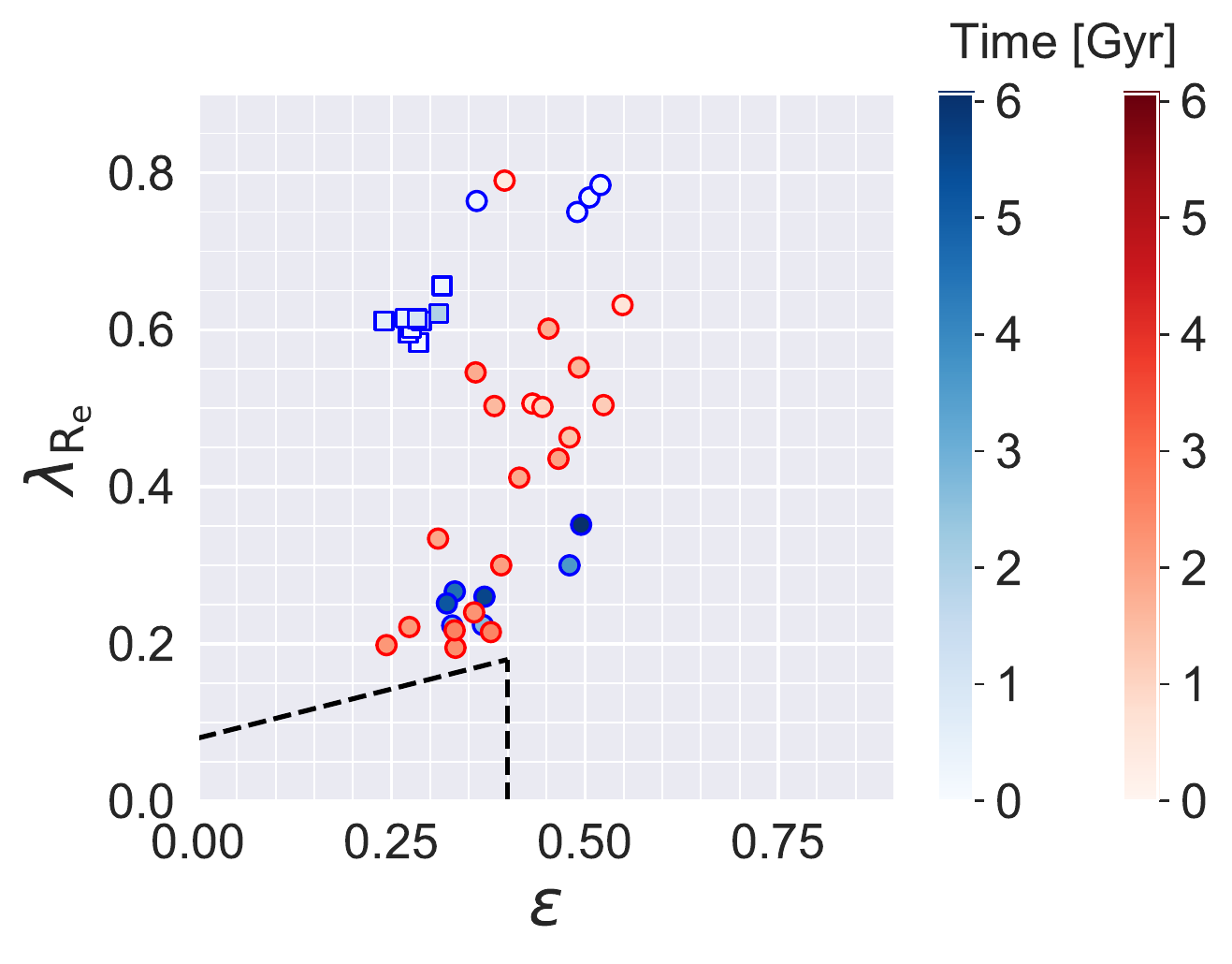}
    \includegraphics[scale=0.5]{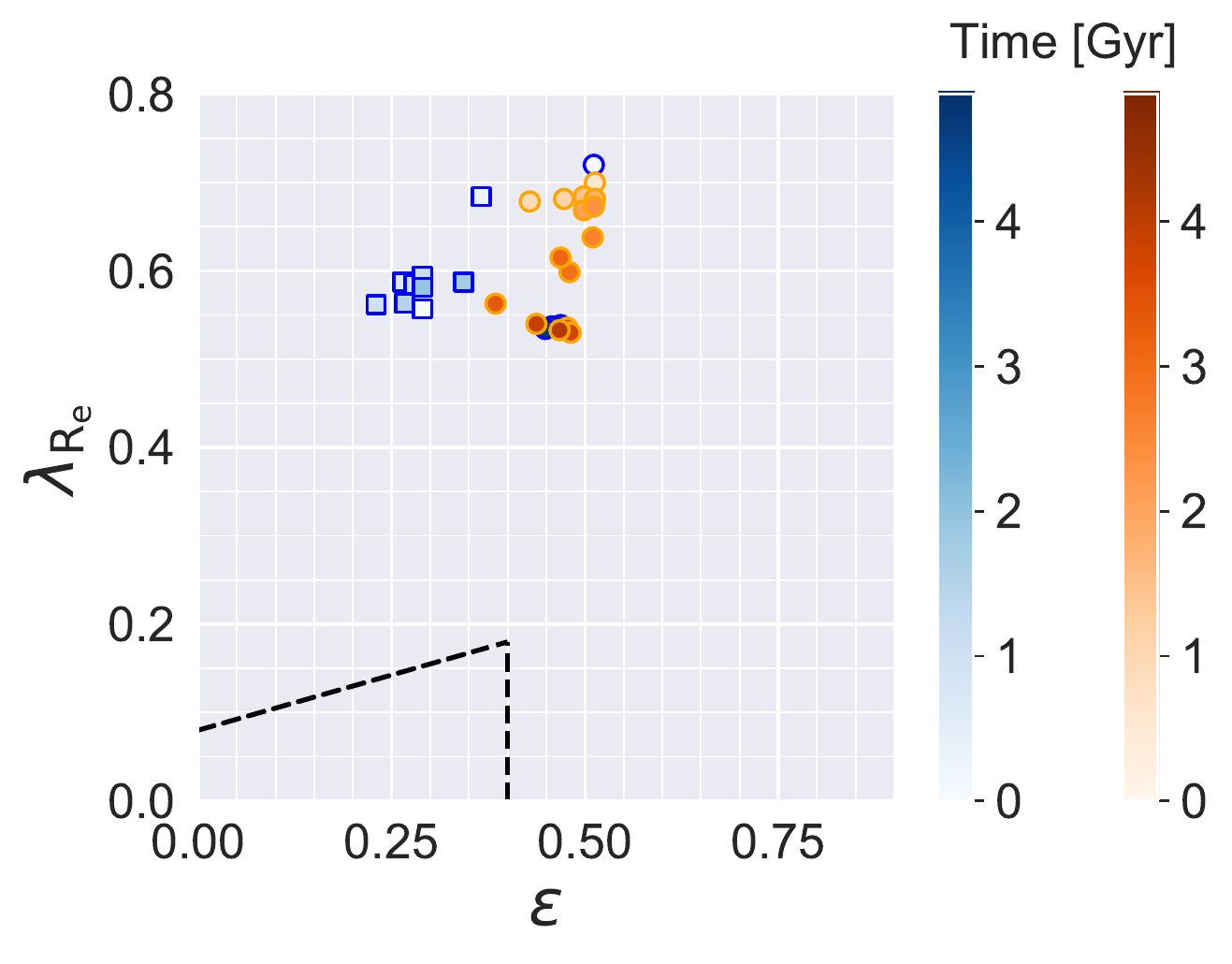}
    
    \caption{Same as Figure \ref{fig:va_sa} but for the time evolution of the merging (red and orange) and nonmerging (blue) galaxies for the q0.5\_fg0.3 (top, red) and q0.2\_fg0.3\_BT0.2 (bottom, orange) on the \lambdare-$\epsilon$ diagram. As the merger progresses (red points), the galaxies evolve towards decreased values of \lambdare, which corresponds to increasing levels of disorder in the kinematic maps. Slow rotators, defined by \citet{Cappellari2016}, fall below the dashed line on these plots.}
    \label{fig:lambdar_epsilon}
\end{figure}

We first directly examine the pre-standardized values of \lambdare\ and $\epsilon$ for the q0.2\_fg0.3\_BT0.2 and q0.5\_fg0.3 simulations in Figure \ref{fig:lambdar_epsilon}. On this diagram, we indicate the `slow-rotator' territory, which is in the lower left corner of this diagram. The \lambdare-$\epsilon$ diagram is often used to kinematically distinguish the slow-rotator population of early-type galaxies from the fast-rotating population. This fast-slow rotator distinction probes the evolutionary histories of galaxies through disk assembly (see \citealt{Cappellari2016} for a review).

Much recent work has focused both on examining the observed populations of fast and slow-rotators and on making predictions for how merging galaxies move through this territory. For instance, \citet{Naab2014} utilize cosmological merger tree simulations to show that major mergers significantly affect the angular momentum content of a galaxy; they can either spin up or spin down the remnant. In our case, all of the simulated galaxies begin with a \lambdare\ value of $\sim0.7$. For the major mergers, the \lambdare\ value dramatically decreases, to the boundary of the slow-rotator region. This confirms that major mergers can dramatically affect the kinematic properties of the remnant, kinematically transforming the galaxy from one that can be described as disk-dominated to one that is still rotating but is dispersion-dominated.

\subsubsection{The mean and dispersion of the velocity dispersion distribution (\mus\ and \sigs) track the growth of a stellar bulge component}
\label{discuss:musigma}

\begin{figure*}
\centering
\includegraphics[scale=0.45]{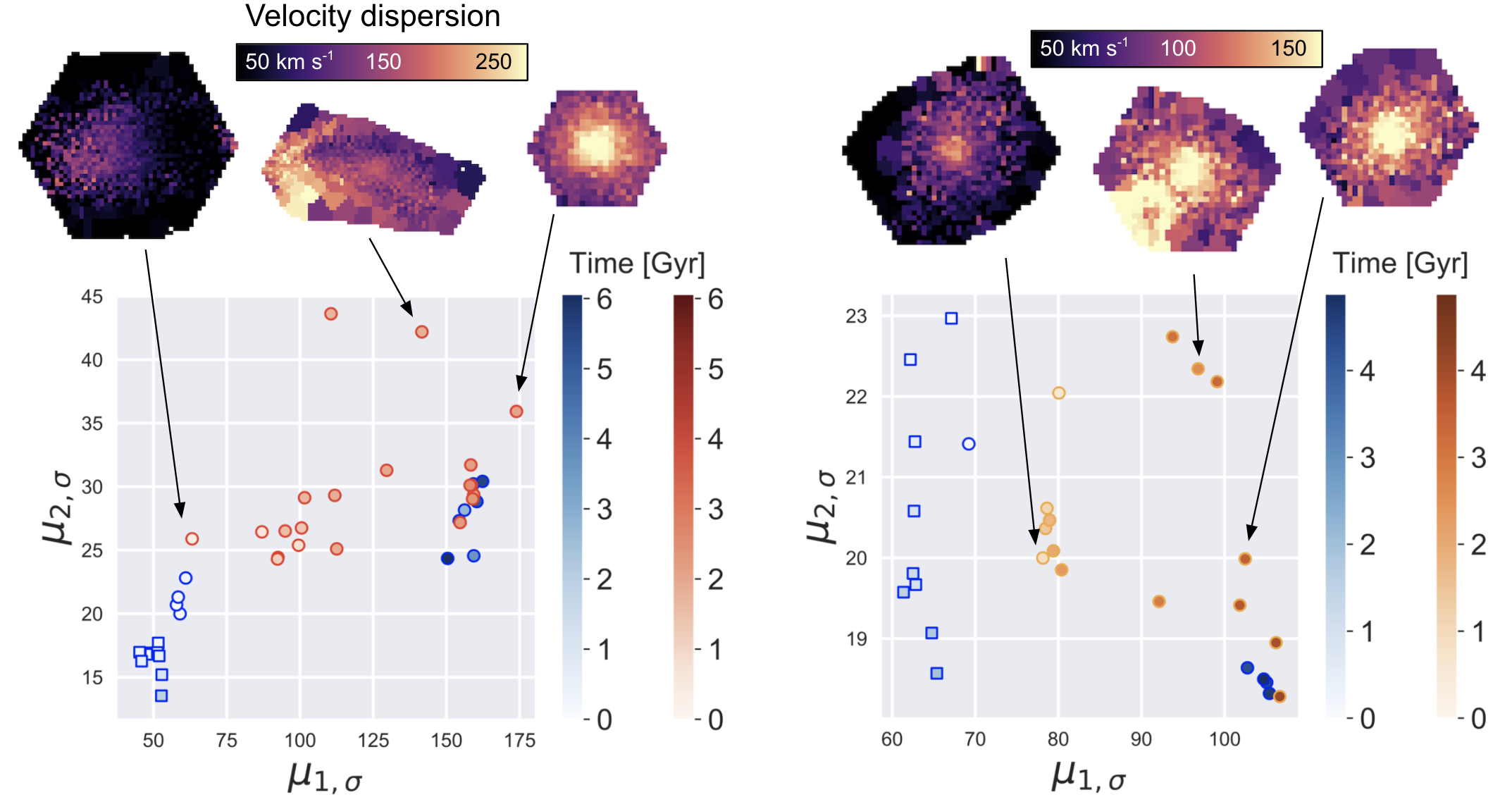}

\caption{Time evolution of the mean values of \mus\ and \sigs\ for the q0.5\_fg0.3 (left plot) and q0.2\_fg0.3\_BT0.2 simulations (right plot). We show the time-evolution of the merging galaxies with the red and orange points and of the nonmerging galaxies with the blue points. The stand-alone isolated galaxies are squares and the pre- and post-merger isolated galaxies are circles. Above each plot, we additionally include representative velocity dispersion maps from key snapshots. We find that the major mergers (left plot) tend to show a consistent evolution in \mus\ and \sigs\ with time; both values increase as the stellar bulge component is built. We include three example velocity dispersion maps (above each plot) that belong to the early (left), late (middle), and post-coalescence (right) stages of the merger. The late and post-coalescence snapshots have elevated values of \mus\ and \sigs; during the late stage the area between the two nuclei has an enhanced velocity dispersion (middle) and during the post-coalescence stage (right), the center of the galaxy has a larger velocity dispersion value. On the other hand, the minor mergers (right plot) show an increase in \mus\ with time but there is not a significant change to \sigs\ with time.  While both types of mergers are contributing to a stellar bulge component, the change to the dispersion maps of the major mergers is more dramatic and global. }
\label{fig:musigma}
\end{figure*}

The overall importance of the \mus\ and \sigs\ predictors reflects the fact that the velocity dispersion map is informative for identifying mergers. We examine how these two predictors evolve with time during a merger in Figure \ref{fig:musigma} for the q0.5\_fg0.3 and q0.2\_fg0.3\_BT0.2 mergers. Here we present the average value for all of the viewpoints of a given snapshot. We also include representative velocity dispersion maps for a handful of informative snapshots.

\textit{For all merger simulations, the \mus\ value increases throughout a merger, tracing the assembly of a stellar bulge component.} This increase is more dramatic for the major mergers, which increase to a \mus\ value of $\sim$200 km s$^{-1}$. Even for the minor mergers, the merger incites growth of the central velocity dispersion with time. This enhancement is still present 0.5 Gyr after coalescence, so the isolated post-coalescence stages are mixed with the merger snapshots along the \mus\ axis in Figure \ref{fig:musigma}. \textit{The signatures of the bulge growth are therefore dynamically long-lived as opposed to imaging features that fade quickly with time following a merger (i.e., in N19 the imaging predictors fade within 0.5 Gyr of final coalescence).} 

The \sigs\ predictor serves different roles in major versus minor mergers, which is reflected in the different evolution of the \sigs\ values with time. An increase of \sigs\ with time for the major mergers traces the presence of two kinematic components by capturing the `bridge' of higher velocity dispersion values between two merging galaxies. This is formed by two overlapping counter-rotating features. Additionally, the post-coalescence major mergers have more significant bulge growth, which is reflected both in the enhancement in \mus\ and in an increase in \sigs, since the entire distribution is broadened in this process. 

The minor mergers show less change in the value of \sigs\ with time. While \sigs\ is still informative (because it increases during the late stages of the merger to track the bridge of higher velocity dispersion), it does not continue to increase into the post-coalescence stages. This could indicate that a smaller fraction of the stars are involved in the buildup of the stellar bulge in the case of the minor mergers. 

\subsubsection{The skewness of the velocity dispersion distribution (\hthrees)  identifies secondary kinematic components}

The skewness in the velocity dispersion distribution, \hthrees, is an important predictor for the major mergers; it is sensitive to secondary kinematic components and disturbances in the velocity dispersion maps. For instance, the stellar bulge region has a higher dispersion, which manifests itself as a small wing on the velocity dispersion distribution (the main contribution to this distribution is the disk rotation component). In this case, the skewness predictor identifies similar features to the \mus\ and \sigs\ predictors. 

The \hthrees\ predictor is additionally important for identifying early-stage mergers. These snapshots tend to have low values of \mus\ and \sigs. Since they have undergone first pericentric passage, they have a slight enhancement in the velocity dispersion map in the area of the primary galaxy that is perturbed by the merger. An example of such a snapshot with this type of velocity dispersion enhancement is the leftmost galaxy in the q0.5\_fg0.3 panel in Figure \ref{fig:musigma}. This galaxy is classified as merging by the classification and the most important predictor leading to this decision is \hthrees. 

\subsubsection{The kurtosis of the velocity distribution (\hfourv) identifies the superposition of two merging galaxies}
The kurtosis of the velocity distribution, \hfourv, is important for both of the individual minor mergers and the combined minor merger classification. This predictor is sensitive to perturbations in the velocity field, specifically cases where there are high velocities in the velocity distribution. When there are extreme velocities in the velocity distribution (due to the superposition of two merging nuclei), the kurtosis becomes more negative due to the flattening of the distribution. The \hfourv\ predictor is significant because it is able to track smaller changes in the velocity distribution, as opposed to global disruptions as in the case of the major mergers. It is a significant predictor for the minor mergers because the velocity dispersion distributions are not dramatically changing during a minor merger. Instead, the LDA must rely upon the extreme velocities caused by the superposition of a secondary nuclei.

\subsection{The classification changes with mass ratio}
\label{discussmass}

Past studies have investigated how the properties of simulated mergers affect the kinematic predictors. \citet{Hung2016} have investigated this for a set of simulated merging galaxies with mass ratios 1:1 and 1:4. They find that the merger signatures in kinematics are most apparent for the 1:1 major merger, where they can be visible for up to twice as long as for the 1:4 major merger. While there are significant differences between the work in this paper and \citet{Hung2016} (i.e., we perform a full RT and create mock IFS maps while \citet{Hung2016} use the \texttt{GADGET-3} particle velocities to create velocity maps), we also find that the classification differs significantly with mass ratio.

The major and minor merger classifications are different in several ways. The minor merger classifications have performance statistics that are $\sim$10-30\% lower. The minor mergers are also more unstable in the terms that are selected, meaning that the coefficient values fluctuate slightly when the classification is re-run. The LDA for the minor mergers is therefore using a couple of key snapshots to create the classification. If these snapshots are excluded from the training set and instead fall in the CV set, then the classification is slightly different. The overall effect is that the minor mergers are less stable and many of the selected terms have similar coefficient values, making it difficult to assess which are the most important. 

Another difference between the major and minor merger classifications is that they are composed of different predictors. While some predictors, such as \mus\ and \sigs\ are important for all classifications, \lambdare\ is more important for the major mergers and \hfourv\ is more important for the minor mergers. As we discuss in \S \ref{discusssign}, both major and minor mergers demonstrate bulge growth, which leads to an enhancement in \mus, but the change is more apparent in the major mergers. The global kinematic properties of the major mergers are more significantly impacted; this includes the \lambdare\ predictor, which traces a global slow-down in the velocity maps of the major mergers. On the other hand, the minor mergers are most sensitive to smaller-scale changes in the kinematic maps, which can be traced by predictors like \hfourv, which track the superposition of the secondary stellar nuclei.

\subsection{The predictors evolve non-linearly with time; the LDA incorporates this behavior with interaction terms}
\label{discuss:interaction}

Many of the kinematic predictors in this analysis evolve with time throughout the merger. In most cases, this evolution is non-monotonic, meaning that merging galaxies evolve back and forth in predictor space as a function of merger time. The LDA technique accounts for this behavior using interaction terms.

An example of an interaction term in action is the $\mu_{\sigma}*\sigma_{\sigma}$ term, which has a negative coefficient for several of the major merger simulations. This means that if the value of \mus\ is relatively large, then \sigs\ must be relatively small for the merger probability to increase. The opposite is also true: if \mus\ is relatively small, then \sigs\ must increase. To be clear, `relatively large' and `relatively small' refer to the standardized values of these predictors, which are measured relative to the distribution of values for the entire merger. So if a term is relatively small, the predictor value will be negative. For instance, if \mus\ is large and \sigs\ is small, then the term becomes: (- coefficient) * (positive) * (negative) = positive value = increase in LD1. 

Consider the $\mu_{\sigma}-\sigma_{\sigma}$ diagram for the q0.5\_fg0.3 merger presented in the left panel of Figure \ref{fig:musigma}. The \mus\ and \sigs\ predictors are correlated, so they occupy the diagonal space of this diagram. This correlation and the interaction term function so that the center of the $\mu_{\sigma}-\sigma_{\sigma}$ diagram is the `merger territory'. This picture is somewhat complicated by the fact that there are other terms in the LDA, but if the $\mu_{\sigma}*\sigma_{\sigma}$ interaction term has a large coefficient, then this interpretation generally applies.

The interaction terms account for the non-monotonic evolution of the predictor values with time. When we create a classifier with only the primary predictors, it is fundamentally different from the LDA with interaction terms. The linear classifier is inaccurate, classifying many of the isolated post-merger snapshots as mergers. When the LDA is forced to generalize to one direction of movement across these diagrams in the monotonic case, it loses key information. 

When the interaction terms are included, the LDA becomes sensitive to the values of the other predictors and is therefore able to create `if-then' cases for the predictor space. For example, \textbf{if} the galaxy is in the later stages of merging (which are characterized by significant bulge growth and a large value of \mus), \textbf{then} to avoid ambiguity with the post-merger isolated galaxies, the classification would like to see a relatively negative standardized value of \sigs\ in order to classify the galaxy as merging. On the other hand, \textbf{if} the galaxy has a relatively small value of \mus, \textbf{then} \sigs\ should be large for the galaxy to be classified as merging. An example of this type of galaxy is the early stages of the merger, where the bulge growth has not yet begun so \mus\ is small. However, the galaxy has undergone its first pericentric passage and is experiencing an enhancement in the velocity dispersion which leads to an increase in \sigs. An example of this type of galaxy is shown in the left panel of Figure \ref{fig:musigma}.

We conclude that it is critical to include the interaction terms in the LDA classification. Not only do they improve the performance of the technique, but they are physically motivated by the non-monotonic evolution in kinematic predictors over the course of a merger lifetime.

\subsection{The observability of a merger varies with time; mergers are missed during the early stages when the kinematics are disk-like}
\label{discussobservability}
In \S \ref{analyzeobservability}, we present the observability timescales of the various simulated mergers. We conclude that the kinematic LDA technique lengthens the observability timescale of the simulated mergers over that measured from the individual predictors. Here we focus specifically on how the observability of a merger changes with the merger stage, and we refer the reader to Figure \ref{mountain4} for a useful visualization of the mean value of LD1 with time in all of the simulated mergers.

When we examine the observability of the mergers in terms of the pre-defined early-, late-, and post-coalescence stages, we notice several differences with stage. First, during the early stages of merging, some simulations show a larger standard deviation in the LD1 values. For all simulations, the value of LD1 tends to fall below the decision boundary for these early stages.  \citet{Hung2015} find that using kinematic predictors to identify mergers results in a significant fraction of false negatives from epochs where the merger is indistinguishable from a rotating disk. We also find that a significant fraction of false negatives occur during the early stages where the rotation is indeed disk-like. 

During the late stage of the merger, the minor mergers do not show much variation from one snapshot to the next; the LD1 values are relatively flat. On the other hand, the major mergers, in particular q0.5\_fg0.3, show variation between the late-stage snapshots. For q0.5\_fg0.3, this is what contributes to the relatively short observability timescale. These changes in LD1 values are significantly greater than the variance due to different viewpoints. \textit{The kinematic features are therefore changing significantly and rapidly with time for some of the major mergers during the late stage.} We also find that most simulations have relatively high LD1 values during the late stage of the merger. The late stage of the merger is therefore characterized by short-lived dramatic kinematic features. This is consistent with \citet{Hung2015}, who find that kinematic tracers of mergers tend to be most informative during the late stage of the merger, which is when the imaging predictors are also most useful.

As the merger progresses into the post-coalescence and post-merger isolated stages, we find that the LD1 value is more stable, which is characteristic of longer-lived kinematic features. The LD1 value does not significantly decline during the post-merger stages. We focus on the kinematics of the post-coalescence and post-merger stages in \S \ref{discuss:end} and \S \ref{discuss:decoupled}.

\subsection{When does a merger end? The kinematic disturbances due to mergers are long-lived}
\label{discuss:end}

In N19, we define the end of the merger as 0.5 Gyr after final coalescence. This cutoff was selected so that the imaging predictors and therefore the value of LD1 decayed smoothly until the end of the merger. The imaging technique is therefore very accurate and precise during the transition from a post-coalescence merger to an isolated post-merger galaxy. In contrast, in this work the kinematic predictors and therefore the LD1 value remain elevated during the period of post-merger isolated snapshots. There are visually apparent warps in the velocity maps and the velocity dispersion maps have elevated values of \mus\ and \sigs\ for the post-merger isolated galaxies. \citet{Hung2016} also find a persistence of kinematic merger signatures up to $\sim$Gyrs following coalescence. We find that the kinematic disturbances fade $2-2.5$ Gyr after coalescence, meaning that in order to improve the classification, we would need to significantly extend the post-coalescence phase.

\textit{Instead of changing the definition of the end of the merger, a more relevant task could be to define the merger more specifically by stage.} For instance, if it is a priority to distinguish post-coalescence from post-merger isolated galaxies (post-coalescence snapshots occur immediately following coalescence until 0.5 Gyr after coalescence and post-merger snapshots occur $>$0.5 Gyr after coalescence), we should rely on the imaging predictors, which do a better job in this specific case. In future work we plan to combine the imaging and the kinematic tools and more directly address this question. One path forward could be to create separate classifications that target different stages of the merger. This could provide a more flexible definition of galaxies that are merging, and allow other users of the technique to target stages of interest in a merger.

\subsection{The kinematics of the post-merger stages track the growth of a stellar bulge and a kinematically decoupled core for the 1:2 mass ratio merger}
\label{discuss:decoupled}

Here we focus on the kinematics of the post-coalescence and post-merger stages. During these stages we observe the build-up of a central component in the stellar velocity dispersion maps. The change is more dramatic for the major mergers and can best be explained as tracing the growth of a stellar bulge component. \citet{Hopkins2009} investigate the effect of mass ratio on the merger remnant and find that the fraction of the primary stellar disk that is relaxed into the bulge is directly proportional to the mass ratio of the merger. This supports the hypothesis that a stellar bulge is built in the post-merger stages, since we would expect major mergers to contribute a larger fraction of stars to the bulge component.

\begin{figure}
    \centering
    \includegraphics[width=0.47\textwidth]{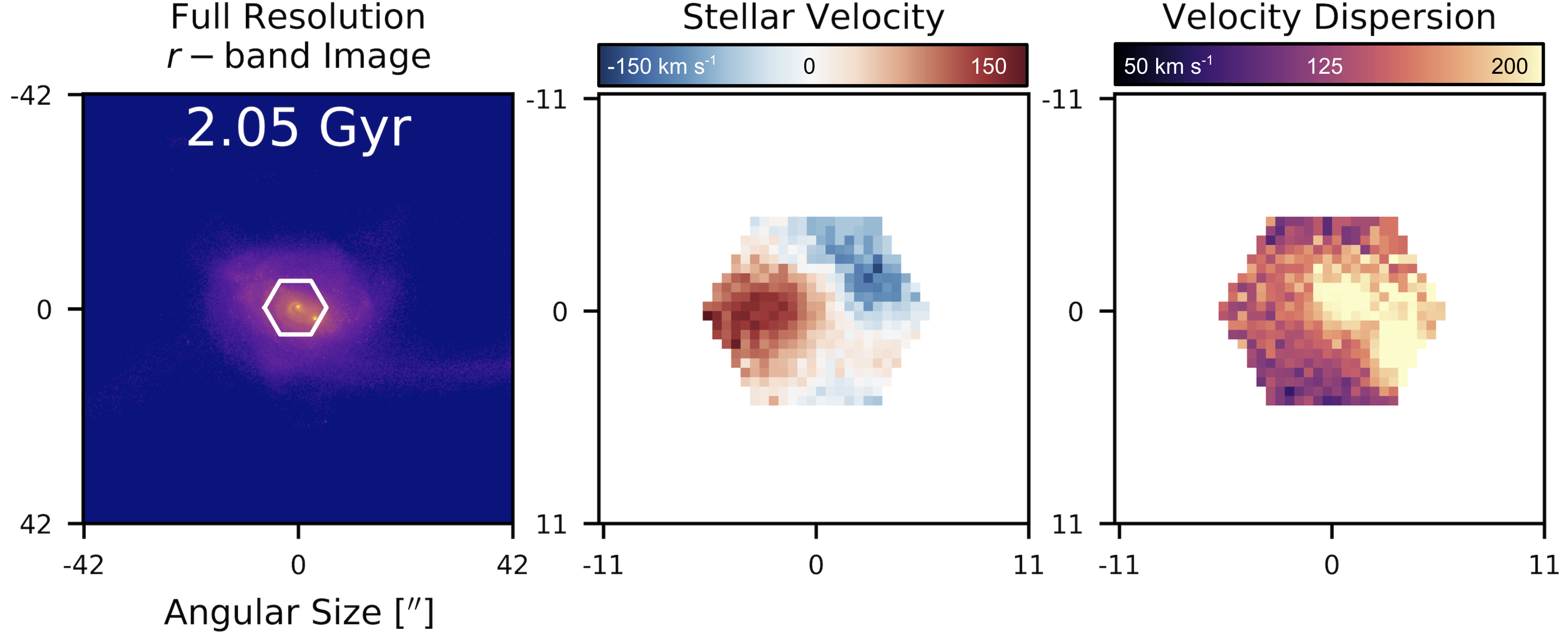}
   \includegraphics[scale=0.67, trim=1.1cm 2.7cm 1.2cm 3.0cm, clip]{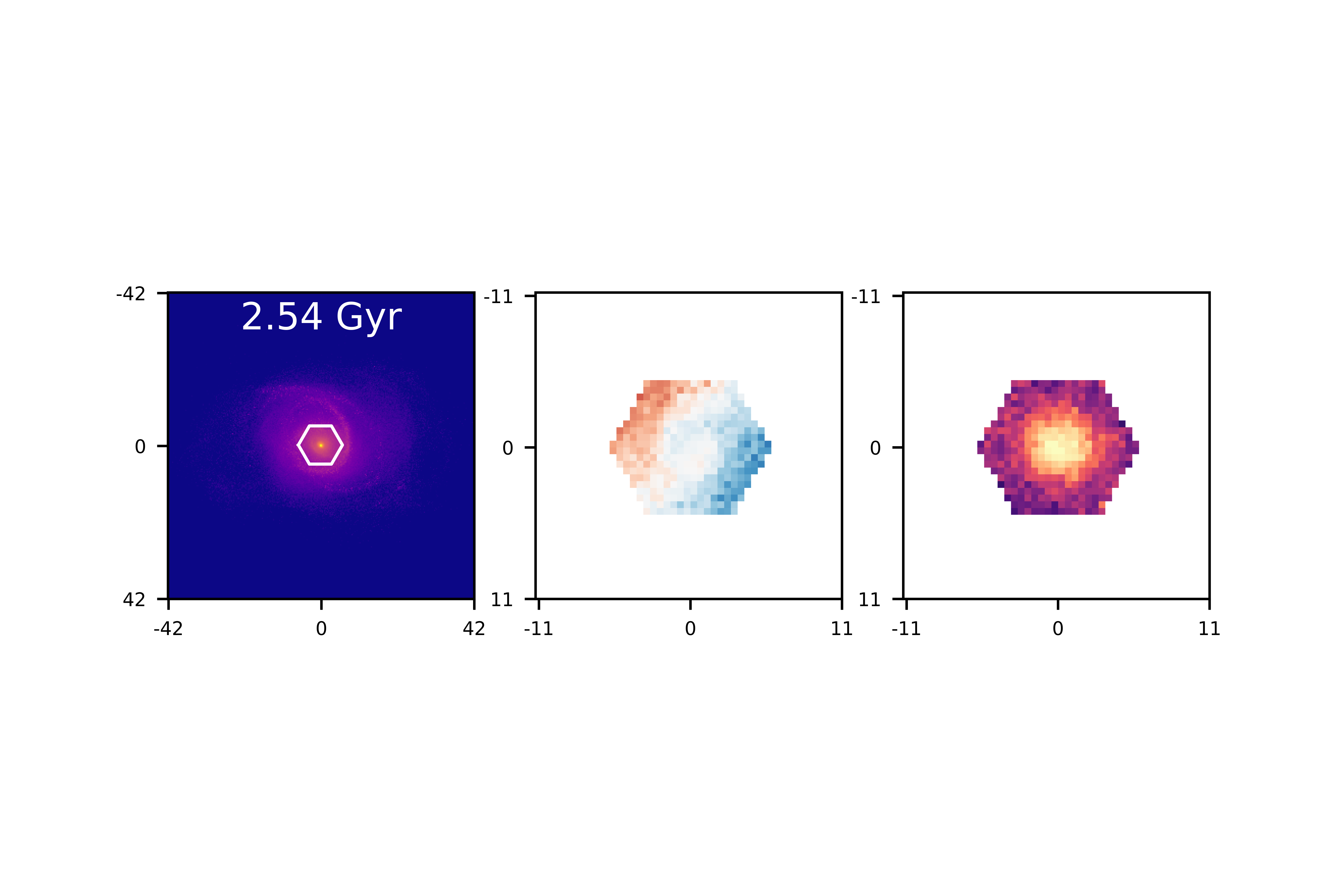}
     
     \includegraphics[scale=0.67, trim=1.1cm 2.7cm 1.2cm 3.0cm, clip]{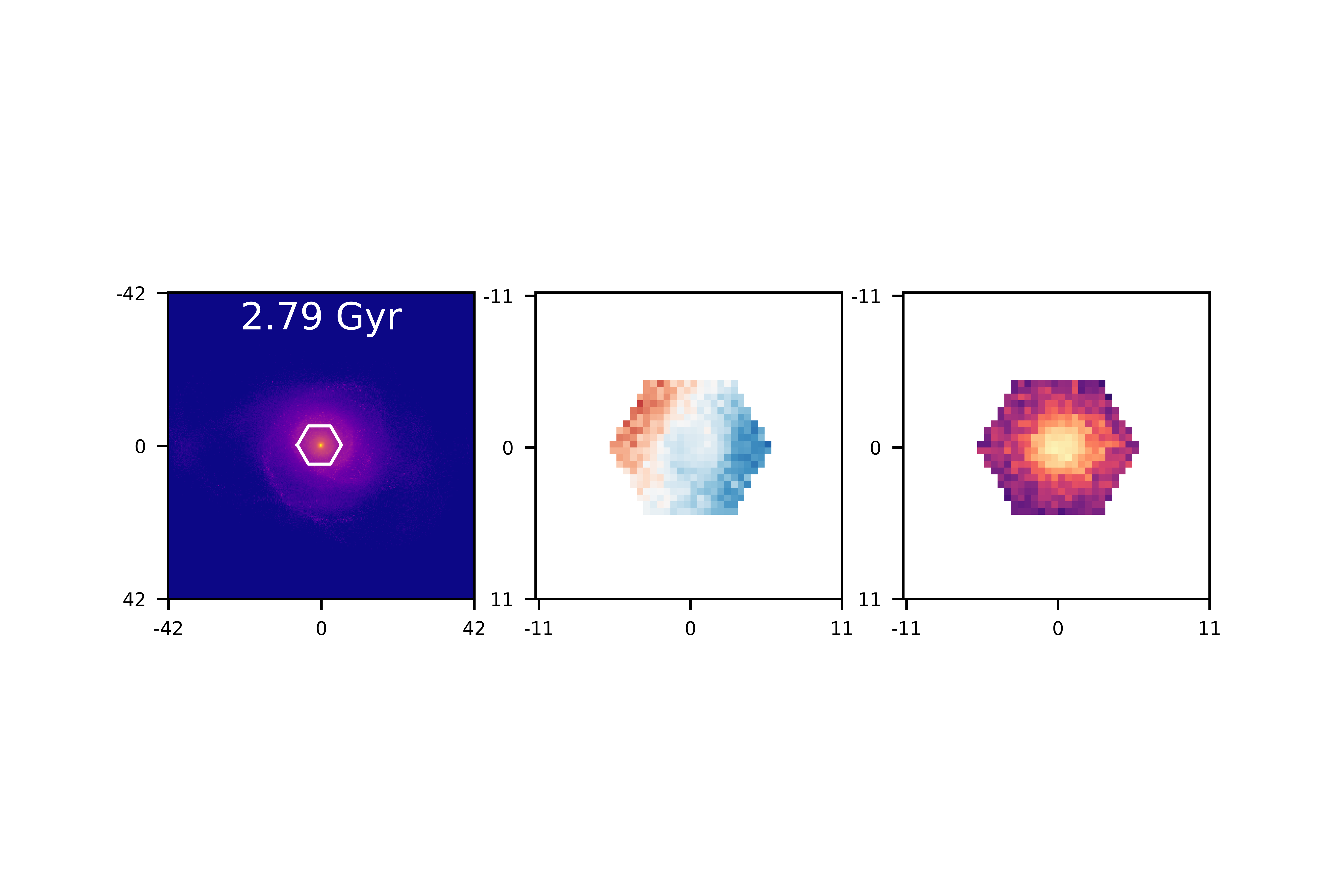}
      \includegraphics[scale=0.67, trim=1.1cm 2.7cm 1.2cm 3.0cm, clip]{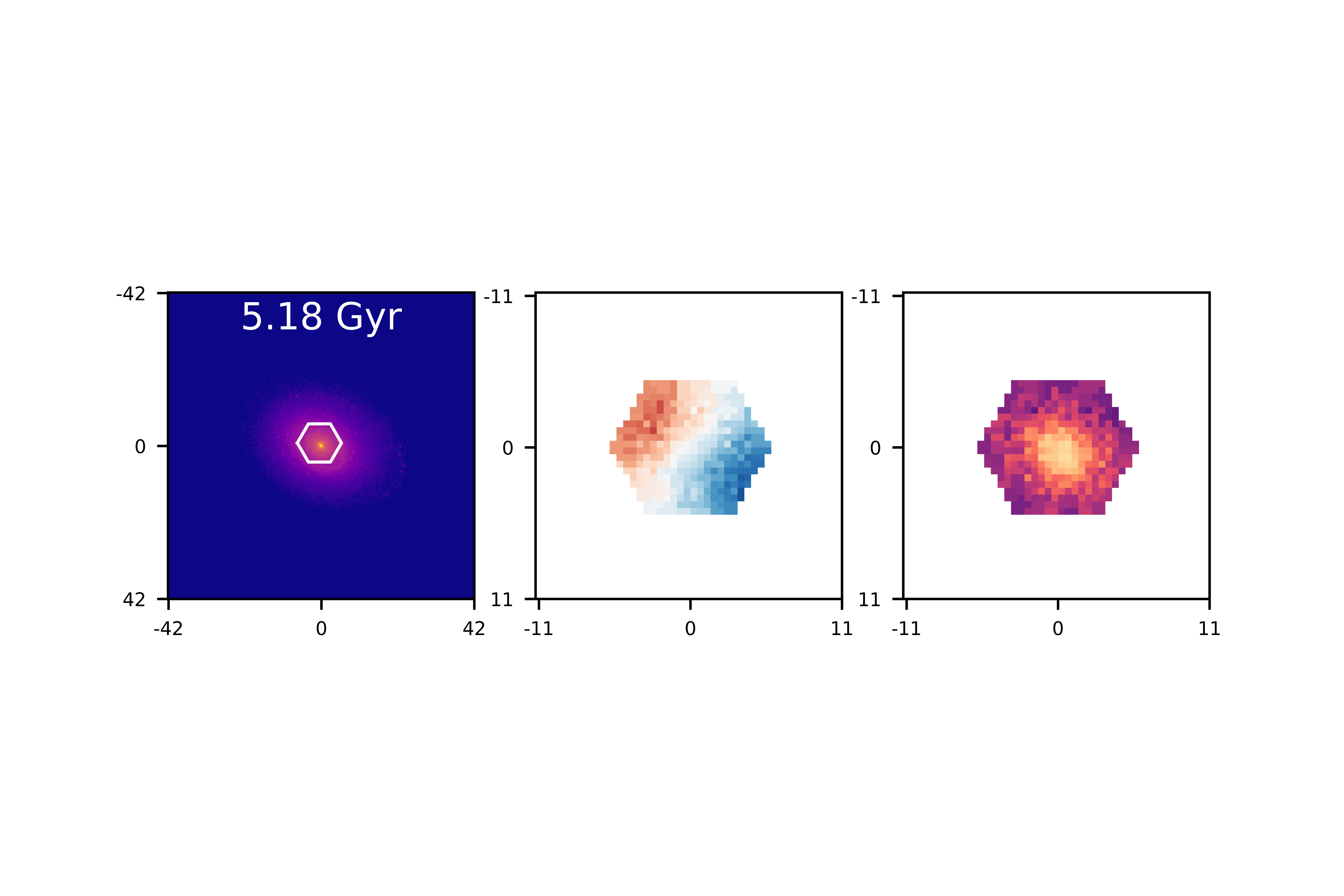}
    \caption{Evolution of the long-lived kinematically decoupled feature in the $r-$band image (left), stellar velocity map (middle), and stellar velocity dispersion map (right) of the q0.5\_fg0.3 merger. The top panel is the last snapshot before coalescence. At 2.54 Gyr the galaxy is in the post-coalescence stage and at 2.79 and 5.18 Gyrs the galaxy is in the post-merger isolated stages. The central kinematically decoupled component appears in the velocity map around 2.54 Gyr, which is $\sim$0.4 Gyr after coalescence and does not disappear until $\sim$3 Gyr after coalescence.  }
    \label{fig:endgameall}
\end{figure}

The q0.5\_fg0.3 merger has unique kinematic features in the stellar velocity map during the post-merger stages, so we focus the remainder of our discussion on this merger. We present several post-coalescence and post-merger snapshots from the q0.5\_fg0.3 merger in Figure \ref{fig:endgameall}. The stellar velocity maps are particularly intriguing because they have a distinct central component that appears in the post-coalescence phase and persists into the post-merger phase. This central feature in the stellar velocity maps is spatially coincident with the bulge-like feature in the stellar velocity dispersion maps. It is not fully counter-rotating, but is misaligned from the main stellar disk. We therefore hypothesize that we have discovered a decoupled kinematic component. 

Previous theoretical work has predicted that major mergers with mass ratios of 1:1 or 1:2 can produce this type of intriguing kinematic component (e.g., \citealt{Bendo2000,Jesseit2007,Crocker2009}). For instance, \citet{Bendo2000} and \citet{Jesseit2007} find that equal mass simulated mergers display a much wider range of kinematic features, including counter-rotating cores and global misalignments while the unequal mass mergers tend to have disk-like kinematics. The merger remnants with complex kinematics are intriguing because these decoupled central components have been discovered in observational studies as well. For example, the ATLAS$^{\mathrm{3D}}$ survey finds that a significant fraction of slow-rotating ETGs have decoupled kinematic components (e.g., \citealt{Emsellem2011}). 

Our finding of a decoupled kinematic core in the remnant of the 1:2 ($q=0.5$) mass ratio merger supports these past findings that mergers with a large mass ratio can produce dramatic central features in the kinematic maps. Furthermore, this result suggests that selecting galaxies with kinematically decoupled cores would produce a sample of post-coalescence major mergers with a mass ratio $q \gtrsim 0.5$.

\subsection{The kinematic LDA is not as good at identifying merging galaxies as the imaging technique}
\label{discussaccuracy}

The kinematic LDA technique has a significant number of false negatives, which drives down the accuracy and recall of the technique. This is partially a result of the chosen priors which skew the classification towards minimizing false positives (see \S \ref{accuracy4} for a full discussion). As we discuss in \S \ref{fails}, this is also due to the lack of identifiability of certain snapshots as mergers, meaning that they are indistinguishable from nonmerging disks and/or post-merger remnants both visually and using their kinematic predictors. 

The imaging LDA performs better for all runs, with improvements on all performance statistics. This means that fewer total galaxies will be correctly classified by the kinematic technique due to the shortcomings listed above. While the major merger combined run is slightly improved when run with imaging predictors (this improves the accuracy/recall/F1 score by 9\%/11\%/7\%), the minor merger combined run experiences significant improvement (16\%/41\%/25\%). The kinematic minor merger combined classification therefore scores lower on all performance statistics relative to both the imaging minor merger classification and the kinematic major merger classification. This reflects the particular inability of the kinematic classification to identify minor mergers.

The imaging LDA also has longer observability times on average, ranging from observability times of $2.2-2.8$ Gyr for the major mergers and $3.5-9.2$ Gyr for the minor mergers. The kinematic LDA has observability times of $0.9-2.4$ Gyr for the major mergers and $3.0-6.6$ Gyr for the minor mergers. So while the observability times for the kinematic LDA significantly improve upon the observability times from individual kinematic predictors, the mergers are still observable for slightly longer timescales by the imaging LDA.

This has important implications for the relative capability of imaging versus kinematic predictors in identifying merging galaxies. The stellar kinematics of mergers take longer to exhibit disturbance and then remain disturbed for a longer time following a merger. The imaging predictors are better contained to the duration of the merger, and better able to identify all of the different merger stages. \citet{Hung2016} recommend using kinematic predictors in combination with imaging predictors due to the frequency of false negatives in their investigation of kinematically-identified merging galaxies. We support this conclusion, finding that the kinematic predictors have failure modes that can be improved by incorporating imaging predictors. 

However, there are some advantages of kinematic predictors relative to the imaging predictors. In \S \ref{discuss:end} and \S \ref{discuss:decoupled} we find that the kinematic predictors are particularly useful for identifying the post-coalescence and post-merger stages because the kinematic disturbances persist for long after imaging predictors fade. The implication is that while the imaging predictors are more informative overall, the kinematic predictors are powerful in certain domains and provide additional information. If forced to select between the imaging and kinematic classification methods, the imaging approach is better. However, the best overall approach is to combine the two techniques into one imaging + kinematic classification. We plan to discuss this topic further in future work.

\subsection{Applying the technique to MaNGA IFS in future work}
\label{discuss:extrapolate}

In \S \ref{results:mass}, we discuss the implication of creating a kinematic classification from a suite of simulations with a narrow range in stellar mass ($3.9\times10^{10} < \mathrm{M}_{\odot}\ <4.7\times10^{10}$) and in initial B/T ratio ($0-0.2$). Many of the kinematic predictors, specifically the most important predictors for the major merger simulations, also probe the intrinsic properties of galaxies. In other words, while these predictors are useful for identifying major mergers, they also change as a function of stellar mass and morphology. The implication is that a classification created from disk-dominated intermediate mass mergers may not extrapolate well to the MaNGA sample, which spans a wider range in stellar mass ($10^8 < \mathrm{M}_{\odot} < 10^{11}$) and morphology.

At present, it is unclear if this will be a concern for just the extreme cases, i.e., the most massive bulge-dominated ETGs, or if it will also cause concern in systems with a mix of rotation and dispersion support in their kinematics. We have preliminarily investigated the distribution of bulge- versus disk-dominated galaxies in MaNGA; while \citet{Wang2020} find that MaNGA galaxies are predominantly disk-dominated, \citet{Graham2018} find that a significant fraction of MaNGA galaxies (across all masses) are bulge-dominated.

Since the MaNGA sample includes a diversity of different galaxy types, we plan to tread carefully when we apply the classification. One option could be to select MaNGA galaxies that have high values of \lambdare\ to include in the classification; in this way, we could exclude bulge-dominated galaxies. Another option could be to de-emphasize certain domains of predictor space in the classification, or to remove the kinematic predictors that are most sensitive to intrinsic galaxy properties altogether. The details of this approach will be developed in future work, since they are beyond the scope of this paper, which focuses mostly on the creation of the technique.

\section{Conclusions}
\label{conclusions4}
In this work, we build on the stand-alone imaging merger classifier in N19 to create a parallel LDA classifier that utilizes kinematic predictors to identify merging galaxies. To produce the classification, we use \sunrise\ synthetic spectra from \texttt{GADGET-3} simulated merging galaxies to create mock `MaNGA-ized' datacubes. We convolve and rebin the synthetic spectra to the spatial and spectral resolution of MaNGA, introduce noise, and implement the Voronoi binning scheme used for the MaNGA datacubes. With \texttt{ppxf}, we extract stellar velocity and stellar velocity dispersion maps from each datacube. 

We then measure a number of kinematic predictors from the velocity and velocity dispersion maps. We use a random forest regressor (RFR) followed by the LDA classifier to select the most informative kinematic predictors and to carry out the classification. The selected predictors are: the difference between the kinematic PA and the imaging PA ($\Delta$PA), the \texttt{kinemetry} residuals (resid), the approximate spin parameter (\lambdare), the asymmetry in the Radon profile ($A_2$), and the moments of the velocity and velocity dispersion distributions (\muv, \mus, \sigv, \sigs, \hthreev, \hthrees, \hfourv, and \hfours). We then run the LDA as a classifier for all simulations individually as well as for the combined major merger simulation and the combined minor merger simulation. 

We first use the LDA classification as an agnostic approach to determine the most useful kinematic predictors for identifying different types of mergers. Our main conclusions are:
    \begin{itemize}
 
    \item Many kinematic predictors that are used in previous work to identify mergers are not as useful in this work (i.e., the deviation of the velocity and velocity dispersion maps from ordered rotation (\vasym\ and \sigasym, respectively, and $\Delta$PA). These predictors are sensitive to specific stages of equal mass ratio mergers and are not as sensitive to the full range of merger parameters and stages used in this simulation suite (\S \ref{discuss:useless}).
    
    \item  The mean and variance of the values in the velocity dispersion maps (\mus\ and \sigs, respectively) are the most useful predictors for identifying mergers across all simulations, because they are sensitive to the growth of a stellar bulge component during the merger (\S \ref{discusssign}).

    \item The selected predictors differ as a function of mass ratio. The major mergers exhibit large-scale kinematic changes (i.e., a global slow-down of the rotation), so they rely more on predictors like \lambdare. The minor mergers are identified using predictors like \hfourv\ which trace the superposition of a secondary stellar nucleus (\S \ref{discussmass}).

    \end{itemize}
We also examine the performance of the LDA classification, which is measured using the four performance statistics (accuracy, precision, recall, and F1 score) as well as the observability timescale. Our main findings are: 
\begin{itemize}

    \item The LDA performance significantly improves when the interaction terms are included. These terms are capable of accounting for the non-monotonic evolution of the kinematic predictors with time (\S \ref{discuss:interaction}).
    
    \item By combining many different kinematic predictors, we create a classification where the observability timescale is a large fraction of the overall merger time (40-90\%). This corresponds to mergers that are observable for 0.9-6.6 Gyr and is an improvement on the observability timescale from any of the individual kinematic predictors (\S \ref{discussobservability}).

    \item The sensitivity of the LDA technique varies with epoch during the mergers. We find that there are more missed mergers (i.e., false negatives) during the early stage of the merger, where the stellar kinematics are disk-like. The mergers are most detectable during the late and post-coalescence stages (\S \ref{discussobservability}).
    
    \item The kinematic predictors (and the LD1 value) are long-lived and remain elevated for $\sim$2 Gyr following final coalescence. The stellar kinematics of the post-coalescence and post-merger epochs capture the formation of a stellar bulge component (\S \ref{discuss:end}). 
    
    \item For the (major, gas rich) q0.5\_fg0.3 merger, a kinematically decoupled component is visible in the stellar velocity maps (\S \ref{discuss:decoupled}).
    
    \item The imaging classification performs better than the kinematic classification and the improvement is larger ($\sim$15\% increase in accuracy, recall, and F1 score) for the minor mergers. The kinematic LDA can be improved by adding imaging predictors (\S \ref{discussaccuracy}). 
    
    \item The kinematic predictors add unique information about merging galaxies to the toolkit; for instance, the kinematic classification is better at identifying post-coalescence and post-merger galaxies relative to the imaging classification (\S \ref{discussaccuracy}). 
    
    \item The kinematic classification is created from a suite of simulations that are limited in their scope (i.e., the simulated galaxies are disk-dominated and span a range of $3.9-4.7\times10^{10} \ \mathrm{M}_{\odot}$ in stellar mass). We conclude that the results may not be applied to all MaNGA galaxies (which have a range of morphologies and an approximately flat stellar mass distribution $10^8 < \mathrm{M}_{\odot}<  10^{11}$). We plan to further address this concern in future work (\S \ref{discuss:extrapolate}).

    \end{itemize}

In Nevin et al. (2021, in prep) we will combine the kinematic classification with the imaging classification presented in N19 and apply the classifier to MaNGA galaxies. At this point, we will release the python tools for implementing these classifications. These tools are designed to be adaptable to the specifications of other imaging and/or IFS surveys, with the goal of applying the classification to other IFS surveys - i.e., SAMI, CALIFA, HECTOR.

In Nevin et al. (2021, in prep) we will further investigate whether various kinematic parameters enhance the existing imaging classifier and why, and revisit the hyperparameter tuning to determine the optimal location of the decision boundary. We also plan to investigate the possibility of splitting the classification by merger stage. Our scientific goals include identifying how the star formation histories, metallicities, and AGN activity change for these different stages as well as for different mass ratios of merging galaxies.

\section{Acknowledgements}
We thank the anonymous referee for their thorough and thoughtful comments that have improved the quality and clarity of this paper.

R. N. and J. M. C. are supported by NSF AST-1714503.

L. B. acknowledges support by NSF award \#1715413.

JAVM acknowledges support from the CONACyT postdoctoral fellowship program.

This work used the Extreme Science and Engineering Discovery Environment (XSEDE), which is supported by National Science Foundation grant number ACI-1548562. We specifically utilized Comet and Oasis through the XSEDE allocation for `An Imaging and Kinematic Approach for Improved Galaxy Merger Identifications' (TG-AST130041). We would also like to acknowledge the help of Martin Kandes, who assisted with the optimization of the LDA tool.

The authors acknowledge University of Florida Research Computing for providing computational resources and support that have contributed to the research results reported in this publication. The website for HiPerGator (the well-named supercomputer) is: http://researchcomputing.ufl.edu

This research made use of Marvin, a core Python package and web framework for MaNGA data, developed by Brian Cherinka, Jos\'{e} S\'{a}nchez-Gallego, Brett Andrews, and Joel Brownstein (\citealt{Marvin}).

Funding for the Sloan Digital Sky Survey IV has been provided by the Alfred P. Sloan Foundation, the U.S. Department of Energy Office of Science, and the Participating Institutions. SDSS-IV acknowledges support and resources from the Center for High-Performance Computing at the University of Utah. The SDSS web site is www.sdss.org.

SDSS-IV is managed by the Astrophysical Research Consortium for the Participating Institutions of the SDSS Collaboration including the 
Brazilian Participation Group, the Carnegie Institution for Science, 
Carnegie Mellon University, the Chilean Participation Group, the French Participation Group, Harvard-Smithsonian Center for Astrophysics, 
Instituto de Astrof\'isica de Canarias, The Johns Hopkins University, 
Kavli Institute for the Physics and Mathematics of the Universe (IPMU) / 
University of Tokyo, Lawrence Berkeley National Laboratory, 
Leibniz Institut f\"ur Astrophysik Potsdam (AIP),  
Max-Planck-Institut f\"ur Astronomie (MPIA Heidelberg), 
Max-Planck-Institut f\"ur Astrophysik (MPA Garching), 
Max-Planck-Institut f\"ur Extraterrestrische Physik (MPE), 
National Astronomical Observatories of China, New Mexico State University, 
New York University, University of Notre Dame, 
Observat\'ario Nacional / MCTI, The Ohio State University, 
Pennsylvania State University, Shanghai Astronomical Observatory, 
United Kingdom Participation Group,
Universidad Nacional Aut\'onoma de M\'exico, University of Arizona, 
University of Colorado Boulder, University of Oxford, University of Portsmouth, 
University of Utah, University of Virginia, University of Washington, University of Wisconsin, 
Vanderbilt University, and Yale University.

\software{astropy (\citealt{Astropy2013,Astropy2018}), matplotlib (\citealt{Matplotlib}), mangadap (\citealt{Westfall2019,Belfiore2019}), numpy (\citealt{Numpy}), openmpi (\citealt{Openmpi}), scikit-learn (\citealt{scikit-learn}), scipy (\citealt{Scipy2020}), seaborn (\citealt{Seaborn}), sdss-marvin
(\citealt{Marvin}), pandas (\citealt{Pandas})}

\bibliographystyle{apj}
\bibliography{main}

\appendix

\section{The Importance of Radiative Transfer and Dust Scattering and Absorption}
\label{scatter}

\begin{figure}
    \centering
    \includegraphics[scale=1.2]{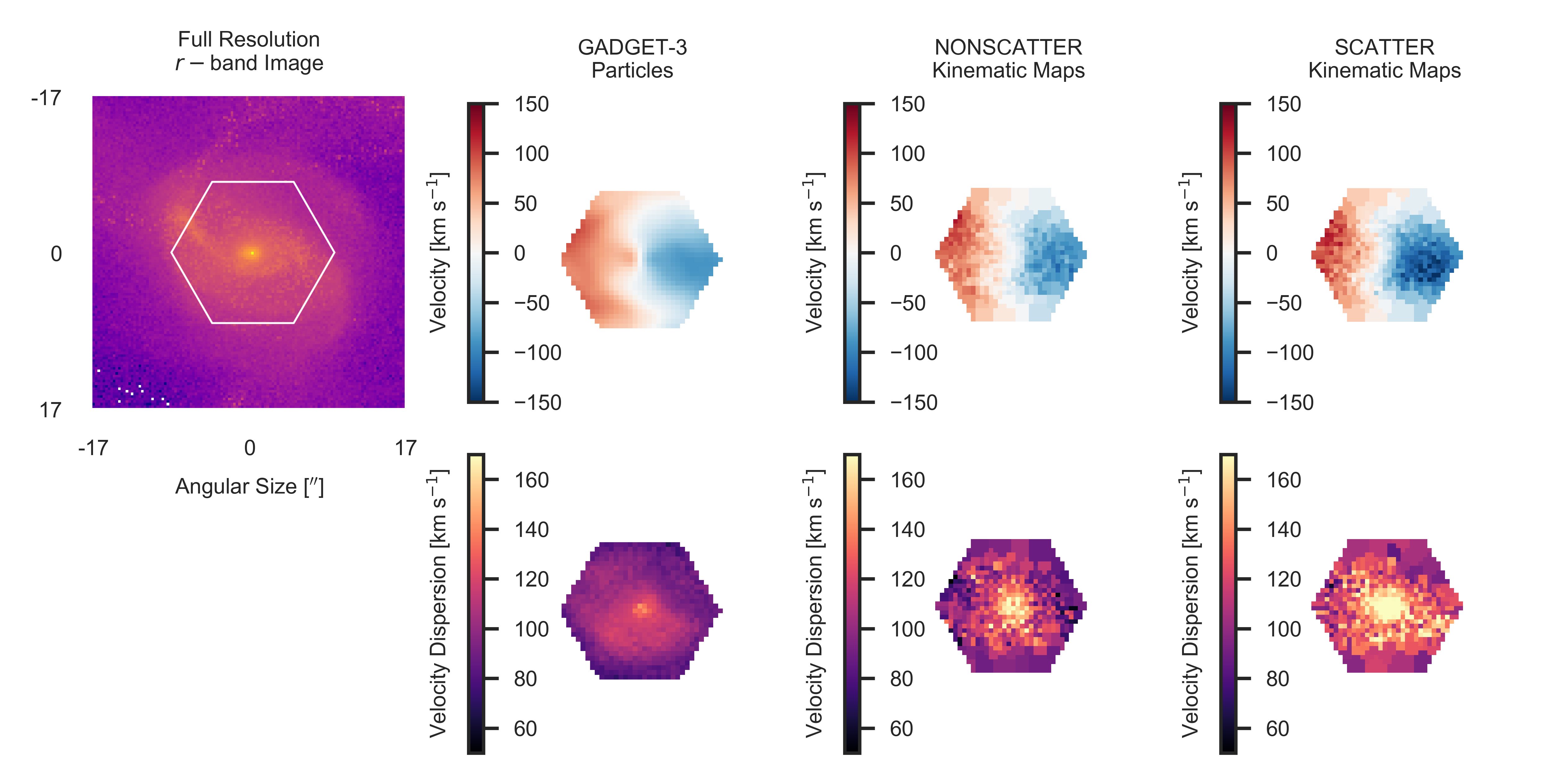}
    \caption{A comparison of the stellar velocity (top row) and velocity dispersion maps (bottom row) for an example galaxy snapshot from the q0.5\_fg0.3 merger. We include the $r-$band image for this snapshot (without observational effects, top left panel). The \texttt{GADGET-3} particle velocity and velocity dispersion map are in the middle left column, the kinematic maps derived from the NONSCATTER \sunrise\ spectra are in the middle right column, and the kinematic maps from the SCATTER \sunrise\ spectra are in the right column. The NONSCATTER and SCATTER kinematic have undergone all steps to achieve mock MaNGA observations, while the \texttt{GADGET-3} derived maps have only been convolved and rebinned. Additionally, the \texttt{GADGET-3} maps are a mass-weighted quantity, while all others are flux-weighted. The differences between the maps are therefore due to these effects in addition to the effects of radiative transfer and the proper treatment of dust. The NONSCATTER and SCATTER velocity and velocity dispersion maps have significantly larger values, which is especially apparent in the center of the velocity dispersion maps. The SCATTER maps have larger values at all locations relative to the NONSCATTER maps; this points to the importance of the proper treatment of dust when creating mock stellar kinematic maps. }
    \label{nokin}
\end{figure}

A central goal of this paper is to design a process to produce synthetic kinematic maps from simulated galaxies that are as close to full realism as possible. However, we acknowledge that running full radiative transfer (RT) and fitting the absorption lines of the resultant datacube is complex and computationally expensive. \citet{Bottrell2019} explore the concept of observational realism and which steps are necessary to produce a CNN classifier for merging galaxies that achieves high accuracy. They find that the RT step is unnecessary and that it can be avoided in order to achieve significant gains in computational time. Since the goal of our approach is to produce observations that closely mimic MaNGA kinematic maps, we carry out a full RT (`full' means that we include the effects of dust scattering and absorption) and we additionally apply observational realism, which is detailed in \S \ref{mimic}. Here we focus on the importance of radiative transfer and the proper treatment of dust.

We first explore the differences between stellar kinematics derived from the \texttt{GADGET-3} particles (no RT) and the stellar kinematics derived from the \sunrise\ spectra (RT) without dust scattering and absorption. From here on, \texttt{GADGET-3} derived kinematics are the `particle' kinematics, the \sunrise\ spectra without dust effects are the `NONSCATTER' spectra, and the \sunrise\ spectra that incorporate dust scattering and absorption are the `SCATTER' spectra\footnote{When we initially began this investigation, we found that there was a small error in the public version of \sunrise\ that affected the shape of computed emission and absorption lines (private communication, Chris Hayward and Raymond Simons). This error in \sunrise\ occurs when kinematic shifts are applied to emission and absorption lines during the dust scattering stage, and therefore applies only to SCATTER spectra generated from dust RT with kinematics. We were able to locate and fix the bug prior to running our production-run \sunrise\ simulations for this paper.}. An important caveat to our analysis on the differences between these kinematic maps is the differences in observational effects. We apply all observational effects to the NONSCATTER and SCATTER spectra, convolving, rebinning, and incorporating noise, but we create the particle kinematic maps without noise effects. Additionally, the particle kinematics are mass-weighted, so the convolution kernel that we apply to the particle velocity and velocity dispersion maps is a rough approximation of the convolution of a flux-weighted quantity (i.e., the NONSCATTER and SCATTER spectra). We apply this convolution kernel as a final step after the kinematic maps are generated.

\begin{figure}
    \centering
    \includegraphics[scale=0.36]{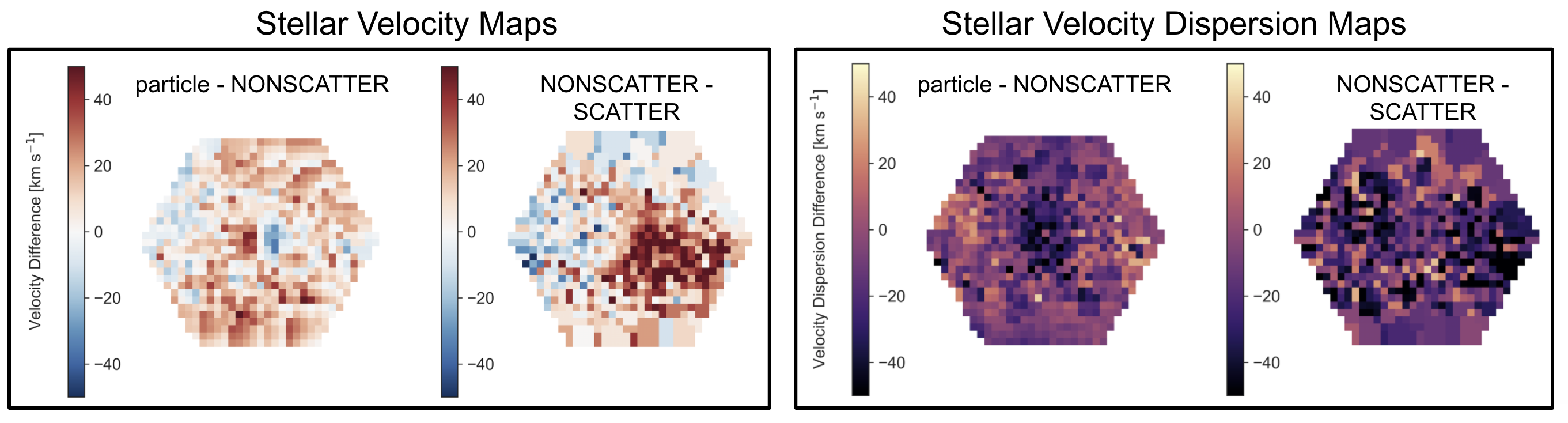}
    \caption{Difference maps for the stellar velocity (first and second panels) and stellar velocity dispersion maps (third and fourth panels). We show the difference between the particle and NONSCATTER maps and the SCATTER and NONSCATTER maps. The particle velocity map has a faster rotating core (first panel) and a lower central velocity dispersion (third panel) relative to the NONSCATTER maps, which are both characteristic beam-smearing effects. The SCATTER maps are faster rotating and have a larger velocity dispersion across the galaxy disk, which could be due to dust preferentially obscuring dynamically young stars.   }
    \label{fig:nokindifference}
\end{figure}

In Figure \ref{nokin} we present the velocity maps (top row) and the velocity dispersion maps (bottom row) produced at various stages of realism for a snapshot from the q0.5\_fg0.3 merger. We begin with the kinematic maps derived from the mass-weighted particle velocities (second column), then the kinematic maps from the NONSCATTER spectra (third column), which are produced using RT, and finally the kinematic maps from the SCATTER spectra (fourth column), which then additionally incorporates the effects of dust scattering and absorption. To first order, we find that the velocity and velocity dispersion maps are similar enough to verify that our methodology for creating stellar velocity and velocity dispersion maps from mock \texttt{SUNRISE} spectra is not failing. More specifically, significant differences between the \sunrise-derived maps and the particle-derived maps could indicate bugs in \sunrise, failures in the steps that produce mock spectral datacubes, or problems with the \texttt{ppxf} absorption line fitting that produces the velocity and velocity dispersion maps. 

We present the difference maps between the particle and NONSCATTER and SCATTER and NONSCATTER maps in Figure \ref{fig:nokindifference}. We first examine the differences between the particle kinematic maps and the NONSCATTER kinematics maps, which are the first and third panel of this figure, for the stellar velocity and velocity dispersion maps, respectively. Radiative transfer, observational effects, and the fact that the particle maps are derived from mass-weighted velocities can cause differences between these maps. For instance, the NONSCATTER maps have larger bins towards the exterior of the galaxy and additionally show more variation on a spaxel-to-spaxel basis due to noise. The NONSCATTER maps have significantly different velocity and velocity dispersion values, especially at the center of the galaxy. There is a faster rotating feature at the center of the particle velocity map, which manifests as a rotating feature in the leftmost panel of Figure \ref{fig:nokindifference}. Additionally, the NONSCATTER-derived dispersion is significantly higher at this location, which is apparent as an over subtraction in the third panel. Since the central feature is slower rotating and has a higher velocity dispersion, it is most likely related to beam-smearing. The differences in the implementation of convolution in the particle maps, which are convolved as a last step, and the NONSCATTER maps, which are convolved prior to deriving the velocity and velocity dispersion values, could contribute to the differences in these central regions.  

We next evaluate the role of dust in the kinematic maps by comparing the NONSCATTER kinematic maps to the SCATTER version in the second and fourth panels of Figure \ref{fig:nokindifference}, for the stellar velocity and stellar velocity dispersion maps, respectively. The effects of dust on the spectra of galaxies are especially important in the context of merging galaxies. For instance, nearly all ultra-luminous infrared galaxies (ULIRGs, $L_{\mathrm{IR}} > 10^{12} L_{\odot}$) are dust-dominated major mergers with buried starburst or AGNs (e.g., \citealt{Veilleux2002,Tacconi2008}). While the galaxies in this simulation suite are not ULIRGs, they still showcase kinematic effects due to the presence of dust. For instance, the NONSCATTER maps tend to underestimate the velocity dispersion relative to that of the SCATTER maps. This is apparent in the fourth panel of Figure \ref{fig:nokindifference}, which has a significant number of spaxels with negative values, indicating that the SCATTER dispersion map has larger values. In addition, the difference is apparent for the majority of spaxels in the galaxy, which could indicate that dust is important on galaxy scales. The NONSCATTER and SCATTER velocity maps also differ. The stars in the disk have higher rotational velocities for the SCATTER extension, meaning that dust may be obscuring the slower-rotating dynamically young population. We run the classification with both SCATTER and NONSCATTER extensions and find that both the individual predictors and the classification change significantly (in terms of both the selected predictors and their coefficients), so we conclude that the effects of dust are important for the stellar kinematics.

Previous work has investigated the effects of dust in \texttt{SUNRISE} spectra in depth. \citet{Stickley2016} compare measurements of the stellar velocity dispersion ($m\sigma_{*}$, from the \texttt{GADGET-3} particle velocities) to the SCATTER version of the \sunrise\ spectra ($f\sigma_{*}$), and find that the offset between the two measurements can be as large as 20-30\% for isolated galaxies and as large as 100\% for extreme cases such as actively merging systems with disturbed morphologies. They conclude that when comparing observations to simulations, it is important to measure flux-weighted kinematics and to incorporate dust. The dust acts to preferentially obscure younger stars, elevating measurements of $f\sigma_{*}$ relative to $m\sigma_{*}$. \citet{Stickley2016} also find that the distribution of the dust is more important than the total attenuation due to dust. For the case of merging galaxies, this message is especially relevant; the effects of dust on the merger kinematics cannot be ignored.

We conclude with two points: 1) The kinematic maps derived from the particle velocities are similar to the \sunrise-derived kinematic maps, indicating that the technique for creating mock stellar kinematic maps is not failing, and 2) The kinematic maps have important differences that occur over the course of RT, the proper treatment of dust, and the subsequent mock observational steps. While we do not fully propagate the particle kinematic maps through the LDA, we find that they are significantly different in appearance than the RT'ed maps and that the derived kinematic predictors are also affected. This verifies that in order to apply the classification to observed galaxies, we need to train the classification using galaxies that have undergone RT, are flux-weighted, and incorporate observational effects. It is beyond the scope of this paper to conduct a full analysis into which steps of the RT and subsequent \texttt{ppxf} derivation of the kinematic maps are most important for creating the realistic galaxies and why.

\section{Creating a Mock Noise Spectrum}
\label{noise}

\begin{figure*}
    \centering
    \includegraphics{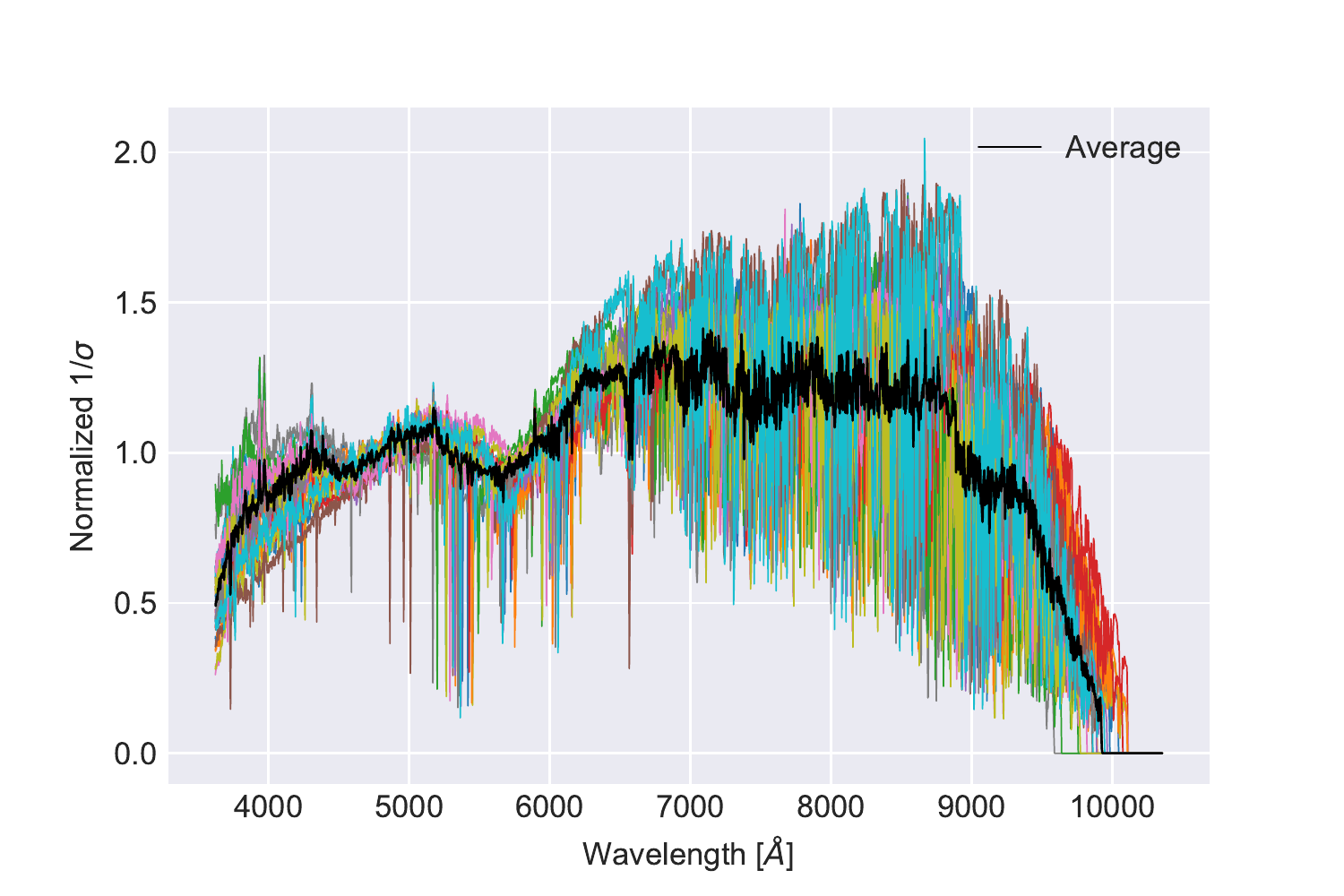}
  
    \caption{Trend of inverse noise (1/$\sigma$, or $\sqrt{ivar}$) with wavelength for the central spaxel of 20 MaNGA galaxies (color lines). The normalized median trend is black. The largest variation happens galaxy to galaxy as opposed to spaxel to spaxel, so we take the average 1/$\sigma$ spectrum (black) to use as our characteristic 1/$\sigma$, which we scale to the flux of each simulated spaxel to produce an error spectrum.}
    \label{ivar}
\end{figure*}

In order to create realistic mock datacubes we also create a mock noise datacube for each snapshot. For an IFS datacube, the typical noise trends with wavelength and so we first measure the median inverse noise (1/$\sigma$, or $\sqrt{ivar}$) trend for the central spaxel of 20 MaNGA galaxies (Figure \ref{ivar}). These 20 galaxies are composed of four randomly selected galaxies from each of the five fiber bundles.\footnote{To verify that the randomly selected galaxies are a fair representation of the MaNGA sample, we compare the average ivar spectrum in Figure \ref{ivar} with the average ivar spectra computed as a function of stellar mass. We find that the shape of the ivar spectrum varies little below a stellar mass of log M$_*\sim$ 11 and that the average spectrum presented in Figure \ref{ivar} is representative of the ivar spectra of the majority of MaNGA galaxies below this mass cutoff. } We test how the 1/$\sigma$ spectrum changes for the location of the spaxel relative to the center of the galaxy and find that there is more variation (in the shape of the 1/$\sigma$ spectrum) between different galaxies than with location in the galaxy. We therefore compute the median 1/$\sigma$ spectrum using the central spaxel for all 20 MaNGA galaxies.
 
To compute the noise spectrum for each spaxel, we multiply the normalized 1/$\sigma$ trend with wavelength by the $g-$band S/N of that spaxel to determine how S/N trends with wavelength. We then divide the mock spectra for each spaxel by the S/N trend with wavelength for that spaxel to get the error (or $\sigma$) spectrum. Finally, we multiply the error spectrum by a random normal Gaussian with a mean of zero and a standard deviation of one. This is the `realization' of the noise for that spaxel. We add the realization of the noise to the spectrum and use the error spectrum as an input to \texttt{ppxf}.

\section{AGN Contamination}
\label{AGNprobs}

\begin{figure}
    \centering
    \includegraphics[scale=0.5, trim=0cm 3cm 0cm 2cm]{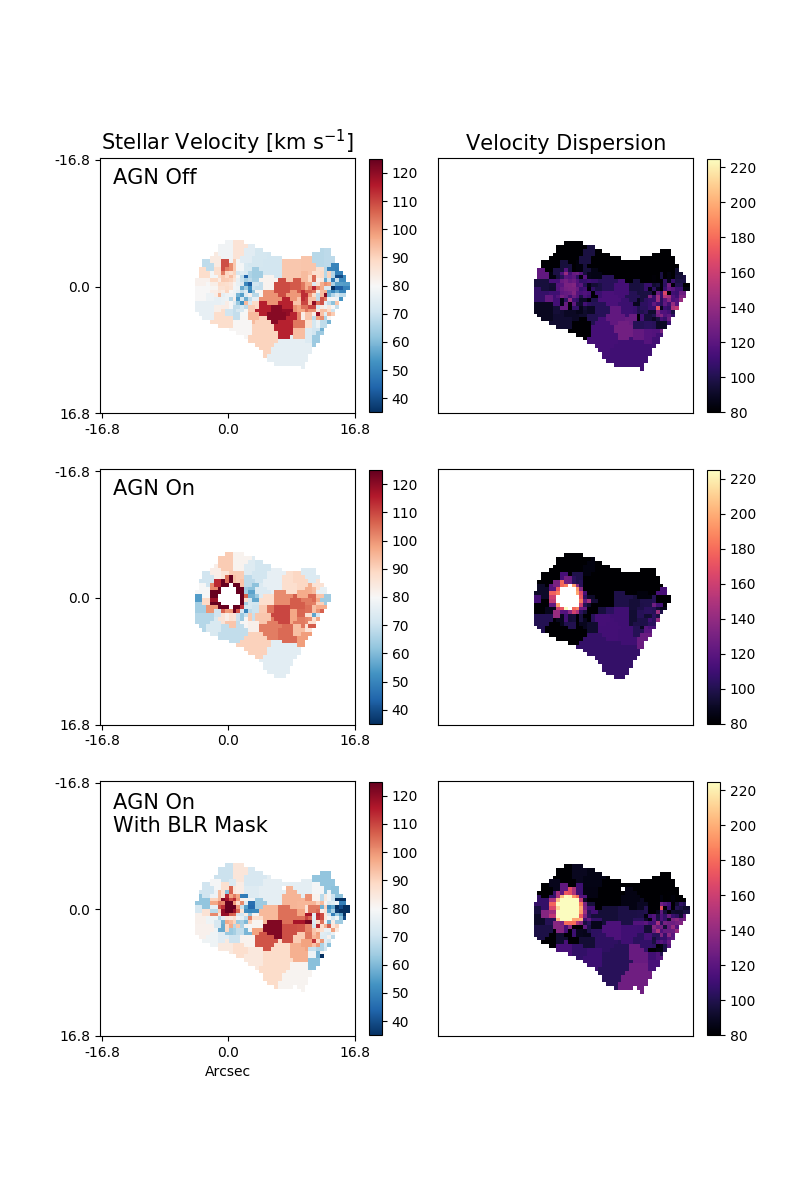}
    \caption{Stellar velocity (left) and stellar velocity dispersion (right) maps for a snapshot of the q0.5\_fg0.3 simulation with an AGN. This specific viewpoint and snapshot has the largest contribution from AGN emission and therefore showcases the most extreme example of the effects of contamination in the stellar kinematic maps by AGN emission. We present the kinematics from the snapshot with the AGN turned off in the top row, followed by the AGN turned on in the middle row, and the AGN turned on with our modified masking procedure in the bottom row. When the AGN light is not removed and no emission line mask is used (middle row), \texttt{ppxf} is unable to accurately fit the absorption lines in areas that are contaminated by the AGN light for both velocity and velocity dispersion. These snapshots are not reliable for the extraction of the kinematic predictors. When we mask the emission lines (bottom row), the kinematic predictors are slightly affected. However, this effect is only significant for a few snapshots so the LDA classification is unaffected by the slight difference between the masked emission lines and the run with the AGN emission turned off.}
    \label{agn_on_off}
\end{figure}

The framework for the simulation suite was originally created in \citet{Blecha2018} to determine the timeline of the activation and fueling of AGNs during mergers. It therefore contains broad line AGNs, which can complicate the fitting of the stellar kinematics. While all of the simulations include AGNs, the q0.5\_fg0.3 simulation hosts the brightest AGN, so we focus on this simulation to describe the effects of a BL AGN on the stellar kinematics. The AGN in the q0.5\_fg0.3 simulation achieves a maximum bolometric luminosity of $L_{\mathrm{AGN}} = 10^{46.3}$ erg s$^{-1}$, which is 90\% of the total luminosity. The duty cycle is short for moderate and high luminosity AGNs; the AGN dominates the spectra for $<$50 Myr. Therefore, even in the q0.5\_fg0.3 simulation, the AGN luminosity is subdominant for the majority of snapshots, even during the late stage of the merger.

Most MaNGA galaxies do not host AGNs, so the MaNGA DAP is not equipped to mask broad emission lines and therefore fails to fit the stellar absorption lines. Bright AGN in MaNGA galaxies are often masked by the DAP fitting procedure, leaving a hole at the center of the galaxy in the stellar kinematic maps. We demonstrate this type of failure in the middle row of Figure \ref{agn_on_off} for the snapshot of the q0.5\_fg0.3 simulation that has the largest contribution from AGN emission. When the AGN continuum is strong and the emission lines are broad, which is the case for 3-4 (out of 20 total) snapshots of the q0.5\_fg0.3 simulation, the DAP produces velocity and velocity dispersion values of 1000 km s$^{-1}$, which is the artificial convergence of \texttt{ppxf} on the maximum allowed values. These values are then masked.

Since our eventual science goals include investigating AGN fueling in mergers, we aim to include galaxies with broad line AGN in the analysis. Therefore, we explore multiple different options for removing the AGN contamination. One approach to remove the AGN contamination is to determine the PSF of the AGN and subtract this from the datacube and then fit the stellar kinematics. A different approach is to modify \texttt{ppxf} to fit broader emission lines. We find that determining the AGN PSF is nontrivial and can lead to over-subtraction of continuum light, so we pursue the other approach, modifying \texttt{ppxf} to fit broad emission lines. This approach could also be applied to MaNGA to fit the stellar kinematics of broad line AGNs.

We modify our \texttt{ppxf} fitting procedure to include a step where we use the `ELPEXT' emission line mask to mask all of the emission lines. We experiment with fitting these emission line masks to galaxies with no broad lines and we find that the returned velocity maps match that of maps without an emission line mask, so we use this procedure for all mock datacubes, not just the ones that are dominated by AGN light. We rerun the q0.5\_fg0.3 simulation with the AGN turned off in order to verify the success of our approach and we present the results in Figure \ref{agn_on_off}. We `turn the AGN off' by removing the AGN contribution to the spectrum and re-running the RT (we still include dust absorption and scattering). The GADGET-3 output is unchanged, meaning that we do not remove BH accretion and feedback from the simulated galaxy. The modified version of \texttt{ppxf} produces similar velocity and velocity dispersion maps to the simulation with the AGN turned off. In Figure \ref{agn_on_off} we demonstrate that for a few snapshots where the AGN emission is strongest, the velocity dispersion is elevated at the location of the AGN. A full multiwavelength subtraction of the PSF would be necessary to totally eliminate this effect. 

The kinematic predictors measured from the maps where the AGN is off are consistent in most cases with those measured with the modified emission line mask, which means that the classification does not change significantly with and without AGN emission. The exception is the kinematic predictors that trace the properties of the velocity dispersion map, such as \mus, \sigs, and \hthrees. These predictors all increase because the velocity dispersion distribution has higher values from the AGN region, which results in a larger mean, variance, and skew in the velocity dispersion distribution. The \lambdare\ value also increases in the presence of a bright face-on AGN because \lambdare\ is a flux-weighted measurement that is much more sensitive to the enhanced velocities in the region of the BL AGN. These predictors are significantly elevated in the presence of an AGN only in the case where the AGN is very bright and face-on. This occurs for $\sim$3 snapshots out of 20 total snapshots, and the AGN is only face-on for one of the seven viewpoints, amounting to 3/140 datapoints. We show one example of dominant AGN emission in Figure \ref{agn_on_off}). These snapshots are already classified as merging when the AGN is off, so while their LD1 value is slightly increased in the presence of an AGN, their classification is unchanged. We can conclude that the AGN alone is not determining the classification, and that it only plays a small role for only a few snapshots. In future work, we plan to more closely examine galaxies with significant AGN emission (this will be a small fraction of the MaNGA sample) to determine if the effects from AGN emission are capable of affecting the classification.

In N19, we did not consider AGN contamination, meaning that the imaging side of the technique is developed with the AGNs present. Since we did not remove the AGNs from the N19 analysis, they could be enhancing the $r-$band concentration of the gas rich major mergers. However, the most important predictors for the gas rich major mergers are asymmetry-based, which indicates that the classification relies more on the lower surface brightness features of the major mergers and are not dominated by the central AGN light.

 \section{Predictors that were not selected by the RFR}
 \label{predcont}
 
 Here we continue the description of kinematic predictors from \S \ref{kinclass} that are not selected by the LDA and therefore not introduced previously. In other words, here we describe the predictors that are not as important for identifying mergers: $A$, \vasym, \sigasym, $\Delta x_V$, and $\Delta x_{\sigma}$.

In \S \ref{kinclass} we define the Radon Transform and Radon profile. Here we describe how we determine the kinematic center of the galaxy using these as tools and we define the $A$ predictor. We follow the procedure from \citet{Stark2018} to determine the galaxy's kinematic center. We use the photometric center as the initial input, but find that the photometric center is not always the same as the kinematic center. Since an incorrect kinematic center can cause variations in the calculation of the Radon Transform, we first determine the kinematic center using the Radon profile and then extract the kinematic predictors using this kinematic center. The kinematic center is the location where the weighted kinematic asymmetry ($A$) is minimized, which is the asymmetry in the estimated $\hat{\theta}$ values, which are measured from the bounded Absolute Radon Transform. The $A$ predictor is defined as:

\begin{equation}
    A = \frac{\sum |\hat{\theta} - \hat{\theta}_{flip}|}{2 N_{i,j}}w_{i,j}
\end{equation}
where $\hat{\theta}$ is reversed to make $\hat{\theta}_{\rm{flip}}$, $N_{i,j}$ is the number of values in the $\hat{\theta}$ array at the current `center', and $w_{i,j}$ is a weight factor:
\begin{equation}
    w_{i,j} = \frac{N_{0,0}}{N_{i,j}}
    \label{eq:weight}
\end{equation}
where $N_{0,0}$ is the number of values at the photometric center.

We iteratively measure $A$ in a 3 $\times$ 3 spaxel grid centered on the photometric center, and select the spaxel with the lowest value as the kinematic center, while taking the minimized $A$ value as our $A$ predictor for each snapshot. If the spaxel with the lowest value of $A$ is consistent within errors with the photometric center, we select the photometric center as the kinematic center. If the spaxel with the lowest value is on the edge of the spaxel grid, we expand the grid by a factor of two and rerun the determination of the kinematic center. 

If the kinematic center is again at the edge of the grid, we do not expand the grid, but take the photometric center as the kinematic center. In this case, the kinematic center is not well-determined, often due to a disorganized velocity map, so the photometric center is a fair guess for the kinematic center. In \citet{Stark2018}, the galaxies where the kinematic center is not well-determined are eliminated from the analysis, but in this case, a large fraction of galaxies have disordered kinematics, so we include them in the analysis, using the photometric center as the kinematic center.

We ultimately find that $A$ is highly correlated with $A_2$ and is therefore superfluous to the analysis. As a result, it is rejected during the RFR step.

We measure the \vasym\ and \sigasym\ predictors with \texttt{kinemetry}, which we previously introduced in \S \ref{kinclass}. Our goal with these predictors is to quantify the chaotic velocity patterns expected for interacting systems in the velocity and velocity dispersion maps using the degree of kinematic asymmetries from the \texttt{kinemetry} output. The \texttt{kinemetry} output includes the $A_n$ and $B_n$ coefficients from the harmonic expansion of the best fitting ellipses for each ellipse. From these, we calculate the amplitude and phase coefficients ($k_n$ and $\phi_n$)\footnote{We do not explicitly apply a flux weighting for the calculation of the phase coefficients, as in \citet{Krajnovic2011}. However, we do use an uncertainty weighting in the \texttt{kinemetry} fitting, which indirectly incorporates a flux weighting. Additionally, \texttt{kinemetry} uses adaptively sized annuli (which increase in size with radius), which already emphasizes the high flux inner regions of the galaxy. 
}: 

\begin{equation}
    k_n=\sqrt{A_n^2 + B_n^2}
    \label{eq:kn}
\end{equation}
\begin{equation}
    \phi_n = \mathrm{arctan}\Big(\frac{A_n}{B_n}\Big)
    \label{eq:phi}
\end{equation}

The higher order $A_n$ and $B_n$ terms represent deviations from an ideal rotating disk. We quantify these perturbations with \vasym\ and \sigasym, which are measured from the $k_n$ amplitude coefficients of the model velocity and velocity dispersion maps, respectively (\citealt{Krajnovic2006,Shapiro2008,Hung2016,Bellocchi2016}):

\begin{equation}
    \rm{v}_{\mathrm{asym}} = < {\frac{\sum_{n=2}^5 k_{\mathrm{n}, v}/4}{B_{1,v}}} >_r, \sigma_{\mathrm{asym}} = < {\frac{\sum_{n=1}^5 k_{\mathrm{n}, v}/5}{A_{0,v}}} >_r
    \label{eq:vasym}
\end{equation}
where the expression is averaged over all radii, $r$. The amplitude coefficients, $k_{n,v}$, are summed and averaged for the $n$ higher order moments. We exclude the $k_{1,v}$ term from the calculation of $\rm{v}_{\mathrm{asym}}$ since it represents radial outflow, which is not associated with stars (\citealt{Shapiro2008}). We normalize $\rm{v}_{\mathrm{asym}}$ by the circular velocity term and $\sigma_{\mathrm{asym}}$ by the $A_0$ term, which is the amplitude of the velocity dispersion maps, as in \citet{Krajnovic2006}. The $\rm{v}_{\mathrm{asym}}$ and $\sigma_{\mathrm{asym}}$ predictors increase for disordered velocities and velocity dispersions.

We define $\Delta x_{\sigma}$ as the difference in spatial position between the centroid of the galaxy in $r-$band imaging and the centroid of the galaxy's velocity dispersion map. We determine the centroid (from both the $r-$band image and the velocity dispersion) by applying a 10$\times$10 pixel low pass filter, thresholding, and then identifying the center of the brightest contour. We refer to the physical distance (in kpc) between the centroid of the $r-$band image and the centroid of the velocity dispersion map as $\Delta x_{\sigma}$. We use a physical distance as opposed to a spaxel distance in order to avoid a sensitivity to galaxy redshift. We also measure $\Delta x_V$, which is the difference (in kpc) between the centroid of the galaxy in $r-$band imaging and the kinematic center.

\section{LDA stability: Is the training set biased?}
\label{fair}

Here we interrogate the biases of our simulation suite. We are specifically interested in two questions: 
\begin{enumerate}
    \item Are we justified in directly comparing the classifications from the different simulations of mergers? 
    \item Is the classification cheating? In other words, are the merging and nonmerging galaxies biased in a way that allows the classification to identify mergers using nuisance parameters?
   
\end{enumerate} 

First, we tackle the validity of comparing the different simulations. We find that the LDA differs significantly for the different merger simulations both in performance and in the selected predictors. We want to ensure that this is a reflection of the physical properties of the stellar kinematics in these different mergers and not merely a reflection of irrelevant differences in the dataset. For instance, the classification from the q0.2\_fg0.3\_BT0.2 simulation has a significantly lower accuracy than any other simulation, including the q0.1\_fg0.3\_BT0.2 simulation, which we would naively expect to have the lowest accuracy. Could this be a result of the q0.2\_fg0.3\_BT0.2 simulation having fewer datapoints than any other simulation? We randomly discard data from all simulations so that they are limited to the same amount of data as the q0.2\_fg0.3\_BT0.2 simulation. We find that both the performance and the selected predictors for all runs are stable when we decrease the number of simulated snapshots, which indicates that the number of snapshots is not biasing the classification.

Second, we carefully examine the properties of the isolated galaxies relative to the merging galaxies for each simulation. This is known as the between-class bias of the classification. Our concern is that differences between the merging and nonmerging populations (i.e., trends with inclination or size) may affect the kinematic predictors. For instance, we find that the isolated snapshots tend to be larger in size and slightly more face-on relative to the nonmerging population for all simulations. Since it is known that properties like inclination can significantly affect the observed stellar kinematics, we want to ensure that the classification is not relying upon the underlying properties of the galaxies to cheat in its classification of merging galaxies. This could happen if, for instance, the nonmerging galaxies are more face-on and therefore the kinematic predictors are picking up on this property as opposed to features that are related to the mergers themselves. We have chosen a relatively transparent classification method, so we can use our interpretation of the individual predictors to diagnose if between-class bias is a concern.

The first line of logic we use to prove that the between-class bias is not significant is the interpretation of the important predictors throughout \S \ref{discuss4}. The most important predictors track features that can be associated with merging galaxies. Additionally, the sensitivity of the method changes with time, meaning that the kinematic features are not long-lived like one could expect from a classification that relies mainly on inclination or size effects, which do not change dramatically over time in a merger.

We also investigate the between-class bias by directly introducing suspicious galaxy properties into the analysis. For instance, we expect that if the classification is instead tied to inclination or size, the kinematic predictors would be correlated with these quantities and therefore the LDA would depend most strongly on inclination or size. To test if this is the case, we introduce several parameters that quantify size and inclination into the LDA as `nuisance' predictors to determine if they are important. We use the galaxy ellipticity ($\epsilon$), the number of spaxels in the velocity map, and the effective radius $R_e$ of each galaxy as proxies for inclination, apparent size, and physical size, respectively. We expect these nuisance predictors to be important predictors if the merging and nonmerging galaxies are significantly unbalanced in properties that are measured by the nuisance predictors. We find that the nuisance predictors are only selected for a few simulations and when they are selected, they are far less important than the leading terms. Therefore, we conclude that the other predictors are not masquerading as proxies for these properties and that the isolated galaxies are not significantly biased relative to the merging galaxies.

\end{document}